\documentclass[useAMS,usedcolumn,usenatbib,usegraphicx]{mn2e}
\usepackage{times}

\usepackage{aas_macros}
\usepackage{textcmds}
\usepackage{multirow}
\usepackage{xcolor}
\usepackage{hyperref}
\usepackage{ctable}
\usepackage{amsmath,amssymb}
\usepackage{diagbox, lscape, chemfig, subcaption}
\newcommand\T{\rule{0pt}{2.6ex}}       
\newcommand\B{\rule[-1.2ex]{0pt}{0pt}} 

\DeclareSymbolFont{matha}{OML}{txmi}{m}{it}
\DeclareMathSymbol{\varv}{\mathord}{matha}{118}

\title[The sulphur connection: IRAS~16293--2422~B and 67P/C--G]{The ALMA-PILS survey: The sulphur connection between protostars and comets: IRAS~16293--2422~B and 67P/Churyumov--Gerasimenko}

\author[Maria N. Drozdovskaya et al.]{Maria~N.~Drozdovskaya$^{1}$\thanks{E-mail: maria.drozdovskaya@csh.unibe.ch}, Ewine~F.~van~Dishoeck$^{2,3}$, Jes~K.~J{\o}rgensen$^{4}$, Ursina~Calmonte$^{5}$,\newauthor
Matthijs~H.~D.~van~der~Wiel$^{6}$, Audrey~Coutens$^{7}$, Hannah~Calcutt$^{4}$, Holger~S.~P.~M\"{u}ller$^{8}$,\newauthor
Per~Bjerkeli$^{9}$, Magnus~V.~Persson$^{9}$, Susanne~F.~Wampfler$^{1}$, Kathrin Altwegg$^{5}$\\
$^{1}$~Center for Space and Habitability, Universit\"{a}t Bern, Sidlerstrasse 5, 3012 Bern, Switzerland\\
$^{2}$~Leiden Observatory, Leiden University, P.O. Box 9513, 2300 RA, Leiden, The Netherlands\\
$^{3}$~Max-Planck-Institut f\"{u}r Extraterrestrische Physik, Giessenbachstrasse 1, 85748 Garching, Germany\\
$^{4}$~Centre for Star and Planet Formation, Niels Bohr Institute \& Natural History Museum of Denmark,\\
~~~University of Copenhagen, {\O}ster Voldgade 5--7, 1350 Copenhagen K., Denmark\\
$^{5}$~Physikalisches Institut, Universit\"{a}t Bern, Sidlerstrasse 5, 3012 Bern, Switzerland\\
$^{6}$~ASTRON, The Netherlands Institute for Radio Astronomy, Postbus 2, 7990 AA Dwingeloo, The Netherlands\\
$^{7}$~Laboratoire d'Astrophysique de Bordeaux, Univ. Bordeaux, CNRS, B18N, all{\'e}e Geoffroy Saint-Hilaire, 33615 Pessac, France\\
$^{8}$~I. Physikalisches Institut, Universit{\"a}t zu K{\"o}ln, Z{\"u}lpicher Strasse 77, 50937 K{\"o}ln, Germany\\
$^{9}$~Department of Space, Earth and Environment, Chalmers University of Technology, Onsala Space Observatory, 439 92 Onsala, Sweden
}

\begin{document}

\date{Accepted xxx.  Received xxx; in original form xxx}

\pagerange{\pageref{firstpage}--\pageref{lastpage}} \pubyear{2017}

\maketitle
\label{firstpage}

\begin{abstract}
The evolutionary past of our Solar System can be pieced together by comparing analogous low-mass protostars with remnants of our Protosolar Nebula -- comets. Sulphur-bearing molecules may be unique tracers of the joint evolution of the volatile and refractory components. ALMA Band~$7$ data from the large unbiased Protostellar Interferometric Line Survey (PILS) are used to search for S-bearing molecules in the outer disc-like structure, $\sim 60$~au from IRAS~16293--2422~B, and are compared with data on 67P/C--G stemming from the ROSINA instrument aboard \textit{Rosetta}. Species such as SO$_{2}$, SO, OCS, CS, H$_{2}$CS, H$_{2}$S and CH$_{3}$SH are detected via at least one of their isotopologues towards IRAS~16293--2422~B. The search reveals a first-time detection of OC$^{33}$S towards this source and a tentative first-time detection of C$^{36}$S towards a low-mass protostar. The data show that IRAS~16293--2422~B contains much more OCS than H$_{2}$S in comparison to 67P/C--G; meanwhile, the SO/SO$_{2}$ ratio is in close agreement between the two targets. IRAS~16293--2422~B has a CH$_{3}$SH/H$_{2}$CS ratio in range of that of our Solar System (differences by a factor of $0.7-5.3$). It is suggested that the levels of UV radiation during the initial collapse of the systems may have varied and have potentially been higher for IRAS~16293-2422~B due to its binary nature; thereby, converting more H$_{2}$S into OCS. It remains to be conclusively tested if this also promotes the formation of S-bearing complex organics. Elevated UV levels of IRAS~16293-2422~B and a warmer birth cloud of our Solar System may jointly explain the variations between the two low-mass systems.
\end{abstract}


\clearpage
\newpage
\section{Introduction}
\label{intro}

Sulphur-bearing molecules have been detected in many interstellar environments from diffuse clouds to star-forming regions and rocky bodies in our Solar System. It is likely that the sulphur detected in cometary studies has its roots in the earliest diffuse phases of the interstellar medium (e.g., \citealt{LucasLiszt2002}). However, in comparison to diffuse clouds, observations show the total sulphur budget of dense cores to be depleted by several orders of magnitude \citep{Ruffle1999}. This puzzle remains unresolved; and the missing sulphur is yet to be conclusively identified (e.g., \citealt{Anderson2013}). Such depletion is unique to sulphur, making it a key element in understanding the evolution of volatile and refractory components between the diffuse and dense phases.

Some of the first detections of interstellar sulphur came in the 1970s and 1980s when molecules such as CS, OCS, H$_{2}$S, SO, H$_{2}$CS and SO$_{2}$ were observed in the gas phase towards Sgr~B2 and Orion~A \citep{Penzias1971, Jefferts1971, DrdlaKnappvanDishoeck1989, Minh1990, Pastor1991}. These detections kicked off the study of its chemistry, which suggested that in thin clouds, sulphur is predominantly in the form of S$^{+}$ ions \citep{OppenheimerDalgarno1974} and likely undergoes partial incorporation into refractories. Meanwhile, in dense clouds it is mostly neutral and gets incorporated into the observed volatiles via reactions with H$_{3}^{+}$ \citep{OppenheimerDalgarno1974} and grain-surface reactions \citep{DuleyMillarWilliams1980}. Subsequent chemical models quickly showed that the net sulphur budget as seen in volatiles is severely depleted in dense clouds \citep{PrasadHuntress1982}. Some of the sulphur is incorporated into carbon chains such as C$_{n}$S (with $n=1,2,...$; \citealt{WlodekBohmeHerbst1988, Smith1988, MillarHerbst1990, Hirahara1992}). However, since the ionization potential of sulphur is lower than that of carbon, S$^{+}$ may exist in regions where most of carbon is neutral, which may lead to the formation of S$_{2}$ (suggested early on by \citealt{DuleyMillarWilliams1980}). In photodissociation regions (PDRs), it appears that sulphur is found in the form of CS and HCS$^{+}$ \citep{Jansen1994, Goicoechea2006}.

Sulphur-bearing species are also seen in solid form: OCS ice was the first to be detected \citep{Palumbo1995, Palumbo1997, Aikawa2012} and the detection of solid SO$_{2}$ followed \citep{Boogert1997}. These ices have been seen in absorption against the bright high-mass W33A protostar and account for $<5$ per cent of the elemental sulphur abundance. Other sulphur-bearing ices are yet to be detected \citep{Boogert2015}. Such simple sulphur-bearing volatiles are likely formed via a combination of gas-phase reactions and grain-surface chemistry. It remains unclear where the remaining sulphur is at cold, dark prestellar conditions. More recently, sulphur-bearing complex organics have been detected. The S-containing methanol-analog, CH$_{3}$SH, has been detected in the gas phase towards the cold core B1 \citep{Cernicharo2012}; the hot core G327.3--0.6 \citep{Gibb2000}; Orion KL \citep{Kolesnikova2014}; the O-type protostar IRAS~16547--4247 \citep{Zapata2015}; Sgr~B2(N2) (by \citealt{Linke1979} and in the EMoCA survey by \citealt{Muller2016}); and IRAS~16293--2422 \citep{Majumdar2016}. The detection of the S-containing ethanol-analog, C$_{2}$H$_{5}$SH, has been reported towards Orion KL \citep{Kolesnikova2014}; while searches towards Sgr~B2(N2) remain uncertain \citep{Muller2016}. Such S-bearing complex organics can only form on the surfaces of grains.

Sulphur-bearing species have also been detected in numerous places in the Solar System and several comets. A large reservoir of sulphur is found on Jupiter's moon Io, which has an atmosphere dominated by SO$_{2}$, as a result of active volcanic eruptions (e.g., \citealt{Jessup2007, Moullet2008, Moullet2013}). Contrary to the interstellar medium (ISM), the majority of cometary detections of sulphur-bearing molecules belong to H$_{2}$S and S$_{2}$ \citep{AHearn1983, MummaCharnley2011}. Towards the brightest comet -- Hale--Bopp, a greater diversity has been observed, including OCS, SO$_{2}$ and H$_{2}$CS. The brighter comets C/2012 F6 (Lemmon) and C/2014 Q2 (Lovejoy) have also been shown to contain CS \citep{Biver2016}. Currently, some of the most unique in-situ data are available from the \textit{Rosetta} mission on comet 67P/Churyumov--Gerasimenko (67P/C--G hereafter; \citealt{Glassmeier2007}). With the Rosetta Orbiter Spectrometer for Ion and Neutral Analysis (ROSINA; \citealt{Balsiger2007}) aboard the orbiter, the coma has been shown to contain H$_{2}$S, atomic S, SO$_{2}$, SO, OCS, H$_{2}$CS, CS$_{2}$ and S$_{2}$ (and tentatively CS, as the mass spectrometer cannot distinguish it from CO$_{2}$) gases \citep{LeRoy2015}. Furthermore, S$_{3}$, S$_{4}$, CH$_{3}$SH, and C$_{2}$H$_{6}$S have now been detected \citep{Calmonte2016} and information on isotopologues is available \citep{Calmonte2017}. It seems that $\sim80$ per cent of sulphur is in refractories (dust) with only $\sim20$ per cent hidden in volatiles (ice; see Appendix~\ref{sulphur67P} for the details). It is likely that surface sniffing of 67P/C--G by COSAC did not reveal any sulphur-bearing species \citep{Goesmann2015} due to a lack of mass resolution \citep{Altwegg2017b}. By piecing together the sulphur puzzle from the earliest diffuse phases to the oldest cometary probes, it may be possible to disentangle the history of volatiles and refractories simultaneously, as they are formed and assembled into larger bodies.

An important parameter for the gas-phase chemistry of sulphur-bearing molecules is the initial elemental C/O ratio at the time molecules start to form, which sets the ratio between oxygen- and carbon-containing S-bearing species (e.g., as seen in the ratio between SO and CS; \citealt{WattCharnley1985}). Additional volatiles can be formed and/or enhanced via the passage of shocks, including HS, H$_{2}$S, S$_{2}$, SO$^{+}$ \citep{Mitchell1984, PineaudesForetsRoueffFlower1986, LeenGraff1988, Turner1992}. Species such as SO and SO$_{2}$ that are produced in the gas phase via reactions with OH upon the liberation of S via sputtering, have become traditional shock tracers. Sulphur-bearing molecules have also been used to study discs via CS (e.g., \citealt{Hasegawa1984, Blake1992}) and as tracers of the centrifugal barrier (disc-envelope interface) via SO (e.g., \citealt{Sakai2014}).

The possible formation routes in ices of species such as S$_{2}$ have been investigated by \citet{GrimGreenberg1987}. In their experiments, ice mixtures containing H$_{2}$S are irradiated via ultraviolet (UV) photons and the production of sulphur chains is indirectly inferred. More recently, \citet{Chen2015} have shown that energetic processing with UV of H$_{2}$S-CO ice mixtures leads to the formation of OCS and CS$_{2}$, and of H$_{2}$S-CO$_{2}$ mixtures to OCS and SO$_{2}$. It is thought that H$_{2}$S forms via the hydrogenation of atomic S and serves as a parent species for further synthesis of sulphur-bearing ices. It is suggested that SO forms on grain surfaces via oxygen addition to HS, and SO$_{2}$ forms via oxygen addition to SO and/or via the association of two SO molecules. OCS can potentially form via the addition of oxygen to CS, the addition of sulphur to CO and/or the association of HS and CO. Experiments also predict that H$_{2}$S$_{2}$ should be made on the grains via the association of two HS molecules; however, recent observational upper limits place it at least an order of magnitude lower in abundance than the laboratory works suggest (towards IRAS~16293--2422; \citealt{Martin-Domenech2016}). Neither H$_{2}$S$_{2}$ nor HS$_{2}$ have been detected with ROSINA on 67P/C--G \citep{Calmonte2016}.

This work is an attempt at piecing together the sulphur trail by comparing its budget in the warm gas on Solar System-scales around a low-mass protostar to that in our own Protosolar Nebula. For this purpose, the paper will focus on the solar-analogue IRAS~16293--2422~B, as investigated by ALMA, and the best available sample of the innate Solar Nebula -- 67P/C--G, as unraveled by ROSINA measurements. IRAS~16293--2422 is an embedded low-mass Class $0$ protostellar binary with a separation of $5\farcs{1}$ (or $610-750$~au assuming a distance of $120-147$~pc; \citealt{Loinard2008, OrtizLeon2017}), a combined luminosity of $21\pm5$~L$_{\sun}$ and disc-like structures around both sources, A and B. A full overview of the physical and chemical properties of the source are presented in \citet{Jorgensen2016}. The single dish survey with the Caltech Submillimeter Observatory (CSO) and the James Clerk Maxwell Telescope (JCMT) detected sulphur-bearing warm ($T_{\text{kin}} \gtrsim 80$~K) dense ($\sim 10^{7}$~cm$^{-3}$) gas tracers such as SO, $^{34}$SO, S$^{18}$O, SO$_{2}$, OCS, OC$^{34}$S, O$^{13}$CS, $^{18}$OCS, o- and p-H$_{2}$CS, H$_{2}$C$^{34}$S, H$_{2}^{13}$CS, SiS, $^{29}$SiS; and colder ($T_{\text{kin}} \sim 40$~K) envelope species ($10^{6} \sim 10^{7}$~cm$^{-3}$) such as CS, C$^{34}$S, HCS$^{+}$, HDS (\citealt{Blake1994, vanDishoeck1995} and reanalysed by \citealt{Schoier2002}). The single dish TIMASSS survey with IRAM-$30$~m and JCMT-$15$~m facilities expanded the list with $^{34}$SO$_{2}$, $^{13}$CS, C$^{33}$S, HDCS and C$_{2}$S \citep{Caux2011}; and CH$_{3}$SH \citep{Majumdar2016}. Interferometric observations with the Submillimeter Array (SMA) revealed the spatial distribution of CS, C$^{34}$S, $^{13}$CS, H$_{2}$S, H$_{2}$CS, H$_{2}$C$^{34}$S, HCS$^{+}$, OCS, O$^{13}$CS, SO, $^{33}$SO, $^{34}$SO, SO$_{2}$, $^{33}$SO$_{2}$, $^{34}$SO$_{2}$, SO$^{18}$O and SO$^{17}$O around the binary system on scales of $\sim 190-380$~au, shedding light on the fact that source A is significantly richer in sulphur-bearing species than B \citep{Jorgensen2011}. IRAS~16293--2422~B has been targeted with ALMA in Band~$9$: lines of $^{34}$SO$_{2}$, $^{33}$SO$_{2}$ and SO were detected in emission from a warm region near source B; and a line of H$_{2}$S was found in absorption originating from the cold foreground gas \citep{Baryshev2015}. Based on ALMA Band~$6$ data at $\sim 0\farcs{6} \times 0\farcs{5}$ spatial resolution, \citet{Oya2016} revealed that it is possible to derive the kinematic envelope structure around source A via OCS emission, meanwhile H$_{2}$CS traces both the envelope and the disc-like structure.

This paper presents the full inventory of sulphur-bearing molecules towards IRAS~16293--2422~B, based on ALMA Band~$7$ data \citep{Jorgensen2016}. Such interferometric observations make it possible to get away from the large scale outflow- and circumbinary envelope-dominated emission and to spatially resolve the thermally desorbed molecules close to the central source. The choice to focus on source B has been made, because its lines are much narrower than those observed towards source A, hence there is less line blending. This makes it ideal for studies of isotopologues and minor species. Subsequently, ratios between various molecules are compared to those deduced for the coma gases of 67P/C--G, as measured with the ROSINA instrument \citep{Calmonte2016}. Both sets of data are some of the best available for an extrasolar analogue of our Solar System and an innate Solar Nebula tracer -- a comet. The differences and similarities between the two have implications for the formation history of our Solar System. Observational details are presented in Section~\ref{observations} and the results are found in Section~\ref{results}. Molecular ratios are computed and compared to cometary values in Section~\ref{discussion} and the conclusions are given in Section~\ref{conclusions}.

\section{IRAS~16293--2422 observations}
\label{observations}

This work is based on the large unbiased Protostellar Interferometric Line Survey (PILS\footnote[2]{\url{http://youngstars.nbi.dk/PILS/}}; project-id: 2013.1.00278.S, PI: Jes K. J{\o}rgensen) of IRAS~16293--2422 carried out with ALMA in the $329-363$~GHz frequency range (Band~$7$) with a spectral resolution of $0.2$~km~s$^{-1}$ and a beam size of $0\farcs{5}$ (or $60-74$~au in diameter, assuming a distance of $120-147$~pc; \citealt{Loinard2008, OrtizLeon2017}). The data used here are continuum subtracted based on the statistical method described in \citet{Jorgensen2016}. The observations are a combination of the $12$- and the $7$-m dish arrays, thereby ensuring that the emission on scales up to $\sim13\arcsec$ is recovered, while also spatially resolving the target. The root-mean-square (RMS) noise of the combined dataset is $7-10$~mJy~beam$^{-1}$~channel$^{-1}$ (or $4-5$~mJy~beam$^{-1}$~km~s$^{-1}$ with beam sizes in the $ 0\farcs{34}-0\farcs{87}$ range; see table~$1$ of \citealt{Jorgensen2016}). Hereafter, $\sigma=10$~mJy beam$^{-1}$ channel$^{-1}$ or $5$~mJy~beam$^{-1}$~km~s$^{-1}$ is adopted. Here, the dataset convolved with a uniform circular restoring beam of $0\farcs{5}$ is used. All further details on the PILS survey, including calibration, are available in \citet{Jorgensen2016}.

The spectral analysis presented in the subsequent section is carried out towards a single position of the dataset -- one beam ($\sim60$~au) offset from source B in the SW direction, which has also been the focal point of \citet{Coutens2016}, \citet{Lykke2017} and \citet{Ligterink2017}; and lies twice as far as the position studied in \citet{Jorgensen2016} in the same direction. The position lies within the high density inner regions of the disc-like structure of source B (called ``disc'' for simplicity, hereafter), hence maximizing emission; while being sufficiently far from the source to avoid absorption against the strong dust continuum (Fig.~\ref{fgr:maps}). IRAS~16293--2422 is associated with one collimated pair of outflow lobes in the NW-SE direction and one less collimated in the E-W direction. The studied offset point lies in the most outflow-free direction of the region to avoid any additional sources of heating. The spectrum at this position is rich in numerous narrow ($\sim 1$~km~s$^{-1}$) lines from various species (and at abundances a factor of $2$ lower than at those seen in fig.~$5$ of \citet{Jorgensen2016} for the half-beam offset position in the same direction at $\sim30$~au from the source, corresponding to the comet-forming zone). The recent analysis of complex organic emission in this direction from species such as glycolaldehyde, ethylene glycol, ethylene oxide, acetone and propanal indicates that the temperatures at this point are $>100$~K, thus it is likely that the chosen position is probing hot inner envelope or face-on disc material heated by the protostellar B source.

\section{Results and analysis}
\label{results}

\ctable[
 caption = {Best-fitting parameters for detected species at the one beam offset position from source B of IRAS~16293--2422\tmark},
 label = {tbl:bestfit},
 star = 1
 ]{@{\extracolsep{\fill}}llrrrl}{
\tnote{assuming a source size of $0\farcs{5}$, FWHM of $1$~km~s$^{-1}$ and $T_{\text{ex}}=125$~K}
\tnote[b]{blended}
\tnote[c]{optically thick}
\tnote[d]{assuming $^{12}$C/$^{13}$C~$=69$ \citep{Wilson1999}}
\tnote[e]{assuming $^{32}$S/$^{34}$S~$=22$ \citep{Wilson1999}}
\tnote[f]{no estimate for the local ISM is available, so a solar ratio of $^{32}$S/$^{33}$S~$=125$ is used \citep{Asplund2009}}
\tnote[g]{assuming $^{16}$O/$^{18}$O~$=557$ \citep{Wilson1999}}
\tnote[h]{no estimate for the local ISM is available, so a solar ratio of $^{32}$S/$^{36}$S~$=4747$ is used \citep{Asplund2009}}
\tnote[i]{assuming D/H~$=0.05$, as measured with single dish observations of HDS/H$_{2}$S (table~$11$ of \citealt{vanDishoeck1995})\footnotemark[4]}
\tnote[j]{assuming D/H~$=0.005$, as recent studies suggest that single dish observations may be overestimating deuteration as a result of underestimating optical depth}
 }{
 \hline
 Species & \# of clean lines      & CDMS entry & $E_{\text{up}}$ (K) & $N$ (cm$^{-2}$)     & Derived $N$ (cm$^{-2}$)\T\\
         & (\# of lines in range) &            &                     &                     & of isotopologues\B\\
 \hline
 SO$_{2}$, $\varv=0$     & 13 (101)  & 64502 & $43-276$   & $1.5\times10^{15}$                 & \T\\
 $^{34}$SO$_{2}$         & 12 (107)  & 66501 & $35-185$   & $4.0\times10^{14}$                 & $N(\text{SO}_{2})=8.8\times10^{15,\text{e}}$\\
 SO, $\varv=0$           & 0 (8)     & 48501 & $81-87$    & $\leq5.0\times10^{14,\text{b}}$    & \\
 OCS, $\varv=0$          & 0 (2)     & 60503 & $237-254$  & $\geq2.0\times10^{16,\text{b, c}}$ & \\
 OCS, $\varv_{2}=1$      & 4 (4)     & 60504 & $986-1003$ & $2.0\times10^{17}$                 & \\
 O$^{13}$CS              & 2 (2)     & 61502 & $236-253$  & $5.0\times10^{15}$                 & $N(\text{OCS})=3.5\times10^{17,\text{d}}$\\
 OC$^{34}$S              & 3 (3)     & 62505 & $231-265$  & $1.0\times10^{16}$                 & $N(\text{OCS})=2.2\times10^{17,\text{e}}$\\
 OC$^{33}$S              & 2 (3)     & 61503 & $234-268$  & $3.0\times10^{15}$                 & $N(\text{OCS})=3.8\times10^{17,\text{f}}$\\
 $^{18}$OCS              & 3 (3)     & 62506 & $238-272$  & $5.0\times10^{14}$                 & $N(\text{OCS})=2.8\times10^{17,\text{g}}$\\
 C$^{34}$S, $\varv=0,1$  & 1 (2)     & 46501 & $65$       & $2.0\times10^{14}$                 & $N(\text{CS})=4.4\times10^{15,\text{e}}$\\
 C$^{33}$S, $\varv=0,1$  & 0 (2)     & 45502 & $65$       & $8.0\times10^{13}$                 & $N(\text{CS})=1.0\times10^{16,\text{f}}$\\
 C$^{36}$S               & 1 (1)     & 48503 & $64$       & $1.4\times10^{13}$                 & $N(\text{CS})=6.6\times10^{16,\text{h}}$\\
 H$_{2}$CS               & 9 (22)    & 46509 & $102-419$  & $1.5\times10^{15}$                 & \\
 HDCS                    & 6 (23)    & 47504 & $98-322$   & $1.5\times10^{14}$                 & $N(\text{H}_{2}\text{CS})=1.5\times10^{15}$\\
 HDS                     & 1 (10)    & 35502 & $35$       & $1.6\times10^{16}$                 & $N(\text{H}_{2}\text{S})=1.6\times10^{17^{\text{i}}-18^{\text{j}}}$\\
 HD$^{34}$S              & 1 (7)     & 37503 & $35$       & $1.0\times10^{15}$                 & $N(\text{HDS})=2.2\times10^{16,\text{e}}$\\
                         &           &       &            &                                    & $N(\text{H}_{2}\text{S})=2.2\times10^{17^{\text{i}}-18^{\text{j}}}$\\
 CH$_{3}$SH, $\varv=0-2$ & 12+ (496) & 48510 & $127-437$  & $5.5\times10^{15}$                 & \B\\
 \hline}

\ctable[
 caption = {Upper limits ($1\sigma$) on the column densities of selected species at the one beam offset position from source B of IRAS~16293--2422$^{\text{a}}$},
 label = {tbl:upperlim},
 star = 1
 ]{@{\extracolsep{\fill}}lrrr}{
\tnote[k]{not possible to derive an accurate upper limit based on one blended line, so the value from \citet{Martin-Domenech2016} is adopted, ignoring the difference in beam sizes}
\tnote[l]{b-type (CDMS entry 59504) has two lines in range, which cannot be detected due to very high $E_{\text{up}} \sim 1700$~K}
 }{
 \hline
 Species & Catalogue and entry \# & \# of lines in range & $N$ (cm$^{-2}$)\T\B\\
 \hline
 gauche-C$_{2}$H$_{5}$SH      & CDMS 62523  & 1329 & $\leq3.0\times10^{15}$\T\\
 anti-C$_{2}$H$_{5}$SH        & CDMS 62524  & 470  & $\leq6.0\times10^{14}$\\
 gauche-C$_{2}$H$_{5}^{34}$SH & CDMS 64517  & 1106 & $\leq1.0\times10^{15}$\\
 S$_{2}$                      &  JPL 64001  & 1    & $\lesssim 2.2\times10^{16, \text{k}}$\\
 S$_{3}$                      &  JPL 96002  & 303  & $\leq4.0\times10^{15}$\\
 S$_{4}$                      &  JPL 128001 & 952  & $\leq1.0\times10^{16}$\\
 HS$_{2}$                     & CDMS 65509  & 227  & $\leq5.0\times10^{14}$\\
 H$_{2}$S$_{2}$               & CDMS 66507  & 33   & $\leq9.0\times10^{14}$\\
 S$_{2}$O, $\varv=0$          & CDMS 80503  & 378  & $\leq1.0\times10^{15}$\\
 cis-S$_{2}$O$_{2}$           & CDMS 96501  & 567  & $\leq1.0\times10^{14}$\\
 HCS                          & CDMS 45507  & 6    & $\leq1.0\times10^{15}$\\
 HSC                          & CDMS 45508  & 29   & $\leq1.0\times10^{14}$\\
 HCS$^{+}$                    & CDMS 45506  & 1    & $\lesssim 2.5\times10^{13}$\\
 DCS$^{+}$                    & CDMS 46505  & 1    & $\lesssim 5.0\times10^{12}$\\
 HC$^{34}$S$^{+}$             & CDMS 47502  & 1    & $\lesssim 6.0\times10^{12}$\\
 H$_{2}$C$^{34}$S             & CDMS 48508  & 20   & $\leq7.0\times10^{13}$\\
 D$_{2}$CS                    & CDMS 48507  & 32   & $\leq1.0\times10^{14}$\\
 CCS                          & CDMS 56502  & 14   & $\leq4.0\times10^{13}$\\
 H$_{2}$C$_{2}$S              & CDMS 58501  & 63   & $\leq5.0\times10^{14}$\\
 c-C$_{2}$H$_{4}$S            & CDMS 60509  & 191  & $\leq1.0\times10^{14}$\\
 NS, $\varv=0$                & CDMS 46515  & 20   & $\leq2.0\times10^{13}$\\
 NCS                          & CDMS 58504  & 12   & $\leq4.0\times10^{14}$\\
 HNCS, a-type$^{\text{l}}$    & CDMS 59503  & 10   & $\leq1.0\times10^{14}$\\
 HSCN                         & CDMS 59505  & 75   & $\leq1.0\times10^{14}$\B\\
 \hline}

The one beam offset position from source B of IRAS~16293--2422 in the PILS Band~$7$ dataset was searched for lines of all known sulphur-bearing species, including those detected in 67P/C--G, and several that have been hypothesized to be present. The initial line searching and local thermal equilibrium (LTE) modelling have been carried out with \textsc{CASSIS}\footnote[3]{\textsc{CASSIS} has been developed by IRAP-UPS/CNRS, \url{http://cassis.irap.omp.eu}}. Thereafter, synthetic spectra have been generated with custom \textsc{IDL} routines. LTE is a good assumption in this case, because the densities are high at the chosen position and thus the molecules are expected to be thermalised. This assumption has also been quantified for the case of methanol (CH$_{3}$OH) in section~$5.1$ of \citet{Jorgensen2016}. It was shown that for the high densities probed by this dataset ($\gtrsim 3 \times 10^{10}$~cm$^{-3}$), the deviation of excitation temperatures under the assumption of LTE from kinetic temperatures of non-LTE calculations is expected to be lower than $15$ per cent. The explored grid of column densities corresponds to an uncertainty of $\sim10$ per cent. Moreover, for the molecules studied in this work, all levels are well-described by a single excitation temperature.

The detected species and the best-fitting parameters, assuming a constant $T_{\text{ex}}=125$~K for all species, are tabulated in Table~\ref{tbl:bestfit}. This value for $T_{\text{ex}}$ has been determined as best-fitting by eye based on a grid of synthetic spectra with $25$~K steps for all these species. Hence, the uncertainty on $T_{\text{ex}}$ is $\sim20$ per cent. There is no indication among this set of molecules for a different $T_{\text{ex}}$. A value of $125$~K has also been derived for some complex organics (e.g., acetaldehyde, ethylene oxide, dimethyl ether and ketene; \citealt{Lykke2017, Jorgensen2017}); while other species show significantly higher excitation temperatures of $\sim300$~K (e.g., glycolaldehyde, methyl formate, formamide; \citealt{Coutens2016, Jorgensen2016, Jorgensen2017}). These variations in excitation temperatures are likely due to different molecules tracing regions of different temperatures, which is related to the different binding energies of the species and short infall time-scales of these inner regions \citep{Jorgensen2017}. Observations at an even higher spatial resolution are necessary in order to explore this in detail. It is assumed that the source is extended and $0\farcs{5}$ in size. If, both, the beam and the source distributions are Gaussian, then the observed emission is diluted by:
\begin{equation}
 \text{beam dilution} = \text{source size}^{2} / \left( \text{source size}^{2} + \text{beam size}^{2} \right) = 0.5.
\end{equation}
A full width half-maximum (FWHM) of $1$~km~s$^{-1}$ is taken for all species. No evidence for any deviation from this value is seen at this offset position for the sulphur-bearing molecules in this dataset. The lines are spectrally resolved into about five bins. There is no indication of any additional kinematic signature such as outflows in the lines at the spectral resolution of these data at this position. The disc-like structure around source B is face-on, so Keplerian rotation is not seen. For source A, a rotating-infalling structure is seen \citep{Pineda2012}. For sources A and B, the assumed local standard of rest (LSR) velocities are $3.2$ and $2.7$~km~s$^{-1}$, respectively \citep{Jorgensen2011}. There may be a small ($0.1-0.2$~km~s$^{-1}$) shift in the best-fitting LSR velocities between molecules tracing somewhat warmer/cooler regions \citep{Jorgensen2017}. Subsequent subsections describe the detections on a molecule by molecule basis. Isotopic ratios for the local ISM have been taken from \citet{Wilson1999}, which are an update of \citet{WilsonRood1994}, where available ($^{32}$S/$^{34}$S~$=22$). Otherwise, solar ratios from \citet{Asplund2009} have been employed ($^{32}$S/$^{33}$S~$=125$, $^{32}$S/$^{36}$S~$=4747$). A selection of the detected lines and fitted synthetic spectra are shown in Appendix~\ref{linesspectra}.

\subsection{SO$_{2}$}
Sulphur dioxide (SO$_{2}$) in the $\varv=0$ state is detected based on 13 clean, non-blended lines at a column density of $1.5\times10^{15}$~cm$^{-2}$(Fig.~\ref{fgr:SO2v=0}). The $\varv_{2}=1$ state is not detected due to the covered lines having very high upper energy levels ($E_{u} \gtrsim 800$~K). $^{34}$SO$_{2}$ is detected with 12 clean lines at a column density of $4.0\times10^{14}$~cm$^{-2}$ (Fig.~\ref{fgr:S-34-O2}). For a ratio of $^{32}$S/$^{34}$S~$=22$, the column density of SO$_{2}$ should be $8.8\times10^{15}$~cm$^{-2}$. This value is a factor of $\sim4$ higher than the column density derived from the lines of SO$_{2}$ itself. This may indicate that SO$_{2}$ is marginally optically thick at this position; however, the synthetic fits to the data do not indicate this (a higher column density leads to an overproduction of the synthetic flux in comparison to the data). Alternatively, the sulphur isotopic ratio may differ at this position and/or for this molecule from that of the local ISM, specifically $^{32}$S/$^{34}$S~$=3.8\pm0.5$ instead of $22$. Finally, it may also be that the emission of SO$_{2}$ stems from a non-uniform emitting area, such as optically thick clumps. Such distributed clumps would still yield a smooth (beam-diluted) emission map on large scales (such as that shown in Fig.~\ref{fgr:maps}). ALMA Band~$9$ emission maps of $^{34}$SO$_{2}$ at $0\farcs{2}$ resolution do indeed suggest that the emission is not homogeneously distributed within the $0\farcs{5}$ PILS Band~$7$ beam (fig.~$13$ of \citealt{Baryshev2015}). None of these possibilities can be firmly ruled out. The one beam offset position has also been searched for $^{33}$SO$_{2}$, S$^{18}$OO and S$^{17}$OO; however, all these lines are very weak, with predicted emission at the noise level of the dataset.

\subsection{SO}
Sulphur monoxide (SO) in the $\varv=0$ state has 8 lines in range. Only 3 are strong enough to be detected (the other 5 are weaker than $1\sigma$) and are blended with emission from cyclopropenylidene (c-C$_{3}$H$_{2}$), methanol, ethanol (C$_{2}$H$_{5}$OH), vinyl cyanide (C$_{2}$H$_{3}$CN), glycolaldehyde (HOCH$_{2}$CHO), methyl formate (HCOOCH$_{3}$) and acetaldehyde (CH$_{3}$CHO). Therefore, the LTE fit for this molecule is less certain and only an upper limit can be derived on its column density at a value of $5.0\times10^{14}$~cm$^{-2}$ (Fig.~\ref{fgr:SOv=0}). This value reflects the maximal contribution SO can have in the observed blended lines. In the $\varv=1$ state, 7 lines are in range. However, all are predicted to be weaker than $1$~mJy~beam$^{-1}$ and, thus, weaker than the noise in the dataset. The position has also been searched for $^{34}$SO, $^{33}$SO, $^{36}$SO, S$^{18}$O, S$^{17}$O and SO$^{+}$; however, all these lines are very weak, with predicted emission below the noise level of the dataset.

\subsection{OCS}
Carbonyl sulfide (OCS) in the $\varv=0$ state has 2 lines in range that are slightly blended with lines of glycolaldehyde and ethanol, respectively (Fig.~\ref{fgr:OCSv=0}). The $\varv_{2}=1$ state is detected with 4 clean lines (Fig.~\ref{fgr:OCSv2=1}). Multiple isotopologues are also detected: O$^{13}$CS, OC$^{34}$S, OC$^{33}$S and $^{18}$OCS (Fig.~\ref{fgr:SO2v=0}-~\ref{fgr:O-18-CS}). This is the first time OC$^{33}$S has been detected towards this source. The presence of $^{33}$S has been inferred previously via $^{33}$SO$_{2}$ \citep{Jorgensen2012} and C$^{33}$S \citep{Caux2011}. All the other isotopologues of OCS have been detected before \citep{Blake1994, Schoier2002, Caux2011}. The dataset was also searched for $^{17}$OCS, OC$^{36}$S, $^{18}$OC$^{34}$S, $^{18}$O$^{13}$CS, O$^{13}$C$^{34}$S, O$^{13}$C$^{33}$S; however, the predicted emission from these species is at the noise level of the dataset.

The best-fitting column density of OCS is $2.0\times10^{17}$~cm$^{-2}$, if inferred from the 4 lines of the $\varv_{2}=1$ state; and is at most $2.0\times10^{16}$~cm$^{-2}$, if approximated from the 2 blended lines of the $\varv=0$ state. This immediately indicates that the $\varv=0$ state is optically thick and this derived column density is a lower limit. The column density of OCS can also be inferred indirectly via its optically thin isotopologues (O$^{13}$CS, OC$^{34}$S and $^{18}$OCS) and the respective local ISM isotopic ratios. The derived numbers are in Table~\ref{tbl:bestfit} and are within a factor of two from the column density calculated from the $\varv_{2}=1$ state. They are an order of magnitude higher than that based on the $\varv=0$ state confirming its optical thickness. The average of these four optically thin estimates is $\left( 2.6\pm0.9 \right)\times10^{17}$~cm$^{-2}$. The average of the values based on just the three isotopologues is $\left( 2.8\pm0.7 \right)\times10^{17}$~cm$^{-2}$ and is the best estimate of the column density of OCS at this position. As an additional check, the solar ratio of $^{32}$S/$^{33}$S~$=125$ can be used to derive the column density of OCS from its OC$^{33}$S isotopologue, since no estimate for the local ISM is available. This yields a value of $3.8\times10^{17}$~cm$^{-2}$, which is consistent within a factor of $1.4$ with the average value derived from three other isotopologues. Alternatively, the observed column density of OC$^{33}$S and the estimated average column density of OCS can be used to calculate $^{32}$S/$^{33}$S for IRAS~16293--2422~B to be $93\pm11$ (with $1\sigma$ significance), which is marginally sub-solar compared with the solar value of $125$ at the highest end of the error bars.

\subsection{CS}
Carbon monosulfide (CS) in all the $\varv=0,1,2,3,4$ vibrational states has 5 rotational lines in range. Only one of them ($\varv=0$, $J=7-6$) is expected to be strong enough to be detected ($E_{u} = 66$~K and $A_{ij} = 8.40\times10^{-4}$~s$^{-1}$), but is blended with ethyl cyanide (C$_{2}$H$_{5}$CN) and ethylene glycol ((CH$_{2}$OH)$_{2}$). The C$^{34}$S isotopologue in the $\varv=0,1$ state has 2 lines in range. One is cleanly detected, while the second is predicted to be too weak for a detection due to a high value of $E_{u}\sim1880$~K. The best-fitting column density is $2.0\times10^{14}$~cm$^{-2}$ (Fig.~\ref{fgr:CS-34v=01}). The C$^{33}$S isotopologue in the $\varv=0,1$ also has 2 lines in range. One is cleanly detected, while the second is again too weak to detect as $E_{u}\sim1887$~K. The best-fitting column density is $8.0\times10^{13}$~cm$^{-2}$ (Fig.~\ref{fgr:CS-33v=01}). The fourth S-isotopologue C$^{36}$S has 1 line in range; and it is cleanly detected ($E_{u} = 64$~K and $A_{ij} = 7.66\times10^{-4}$~s$^{-1}$), giving a best-fitting column density of $1.4\times10^{13}$~cm$^{-2}$ (Fig.~\ref{fgr:CS-36}). This is a tentative first-time detection of $^{36}$S towards a low-mass protostar. C$^{36}$S has been previously detected towards high-mass hot cores \citep{Mauersberger1996} as a first-time detection of interstellar $^{36}$S. More lines are required to confirm this detection; however, the isotopic ratios are in agreement with solar ratios. Other excitation states and isotopologues lack lines in the observed frequency range, specifically: CS in the $\varv=1-0, 2-1$ and $\varv=2-0$ states; C$^{34}$S in the $\varv=1-0$ state; $^{13}$CS in the $\varv=0,1$ and $\varv=1-0$ states; $^{13}$C$^{34}$S and $^{13}$C$^{33}$S. Meanwhile, $^{13}$C$^{36}$S has 1 line in range, but it is a clear non-detection. CS$^{+}$ has 2 lines in range; however, one is too weak ($A_{ij} = 7.12\times10^{-7}$~s$^{-1}$), while the other suffers from blending and absorption.

Assuming the local ISM ratio of $^{32}$S/$^{34}$S~$=22$, the column density of CS can be estimated at $4.4\times10^{15}$~cm$^{-2}$ from its C$^{34}$S isotopologue. If this column density is used to fit the detected line of CS suffering from absorption, then the synthetic line width matches the observed. Thus this is the best available estimate of the column density of CS. Based on C$^{33}$S and C$^{36}$S, the column density of CS can be derived to be $1.0\times10^{16}$~cm$^{-2}$ and $6.6\times10^{16}$~cm$^{-2}$, respectively, which is factors of $2.3$ and $15$ larger than the value derived based on C$^{34}$S. This may imply that either the C$^{36}$S column density is poorly constrained by the single line; or that the solar isotopic ratio is not applicable for IRAS~16293--2422~B. Alternatively, the observed column densities of C$^{33}$S and C$^{36}$S in conjunction with the estimated CS column density can be used to calculate $^{32}$S/$^{33}$S and $^{32}$S/$^{36}$S for IRAS~16293--2422~B to be $55\pm8$ and $314\pm44$ (both with $1\sigma$ significance), i.e., a factor of $2$ and an order of magnitude lower than the solar values of $125$ and $4747$, respectively.

\subsection{H$_{2}$CS}
Thioformaldehyde (H$_{2}$CS) is detected with 9 clean lines at a column density of $1.5\times10^{15}$~cm$^{-2}$ (Fig.~\ref{fgr:H2CS}). H$_{2}$C$^{34}$S cannot be cleanly detected due to a lack of non-blended lines, thus only an upper limit of $7.0\times10^{13}$~cm$^{-2}$ can be derived (Table~\ref{tbl:upperlim}, Fig.~\ref{fgr:H2CS-34}). The tentative assignments of the strongest lines of H$_{2}$C$^{34}$S at this maximal column density would imply a column density of $1.5\times10^{15}$~cm$^{-2}$ for H$_{2}$CS (assuming $^{32}$S/$^{34}$S~$=22$), thus testifying to H$_{2}$CS emission being (very close to) optically thin.

HDCS is detected with 6 clean lines at a column density of $1.5\times10^{14}$~cm$^{-2}$ (Fig.~\ref{fgr:HDCS}). Since H$_{2}$CS appears to be optically thin, the best-fitting column densities can be used to calculate the HDCS/H$_{2}$CS ratio to be $0.1\pm0.014$ and a D/H~$=0.05\pm0.007$\footnote[4]{A correction for the statistical factor of 2 is applied, which assumes that deuteration events are mutually exclusive}. A lower level of deuteration is ruled out based on the low quantity of H$_{2}$C$^{34}$S. This implies that thiofomaldehyde is highly singly-deuterated (5 per cent), but comparable to many oxygen-bearing complex organic species \citep{Jorgensen2017}.

D$_{2}$CS is not detected with an upper limit of $1.0\times10^{14}$~cm$^{-2}$ (Table~\ref{tbl:upperlim}, Fig.~\ref{fgr:D2CS}). This yields a D$_{2}$CS/HDCS ratio of $<0.67$, and a D$_{2}$CS/H$_{2}$CS ratio of $<0.067$. \citet{Persson2017} derived the ratios between the deutrated isotopologues of formaldehyde (H$_{2}$CO) to be: HDCO/H$_{2}$CO~$=0.065\pm0.01$, D$_{2}$CO/HDCO~$=0.128^{+0.033}_{-0.041}$, and D$_{2}$CO/H$_{2}$CO~$=0.0064\pm0.001$. This means that in comparison to formaldehyde, the sulphur-bearing analog (thioformaldehyde) is a factor of $\sim1.5$ more singly deuterated relative to its main isotopologue, at most a factor of $\sim5$ more doubly-deuterated relative to its singly-deuterated isotopologue and at most a factor of $\sim10$ more doubly-deuterated relative to its main isotopologue. This may suggest that thiofomaldehyde forms under even more deuterium-rich conditions than formaldehyde, or that it undergoes fewer chemical reactions that would lower its level deuteration from the time of initial synthesis. Other isotopologues have not been detected: H$_{2}$C$^{33}$S and H$_{2}^{13}$CS. Higher sensitivity data would be useful for ascertaining the column density of D$_{2}$CS and firmly verifying the degree of deuteration of thioformaldehyde.

\subsection{H$_{2}$S}
Hydrogen sulfide (H$_{2}$S) has 1 line in the observed frequency range; however, it is not detected due to a high upper energy level ($E_{u}=758$~K) and line weakness ($A_{ij} = 6.28\times10^{-9}$~s$^{-1}$). On the other hand, its isotopologue HDS has 10 lines in range. One of those is a clear detection at the $65\sigma$ level, while the others have upper energy levels that are too high ($E_{u}>537$~K) and thus, are not expected to be strong. The best-fitting column density is $1.6\times10^{16}$~cm$^{-2}$ (Fig.~\ref{fgr:HDS}). HD$^{34}$S has 7 lines in range. One is strong-enough and blend-free to be detected, while all others have upper energy levels that are too high ($E_{u}>536$~K) to be stronger than the noise level. The best-fitting column density is $1.0\times10^{15}$~cm$^{-2}$ (Fig.~\ref{fgr:HDS-34}). Assuming that $^{32}$S/$^{34}$S~$=22$, gives a column density of $2.2\times10^{16}$~cm$^{-2}$ for HDS, which closely (a factor of $1.4$) agrees with the value derived based on its single observed line. Other isotopologues are not detected. H$_{2}^{34}$S and H$_{2}^{33}$S do not have any lines in the observed frequency range. D$_{2}$S has 4 lines in range; however, only 2 are above the noise level and both suffer from blending, thus no clear detection is possible. D$_{2}^{34}$S has 2 lines in range, but one is blended with a strong absorption feature; and the other with emission lines of acetaldehyde and the $^{13}$C-isotopologue of ethyl cyanide (C$_{2}$H$_{5}^{13}$CN); thus, no confident detection can be claimed.

Assuming D/H~$=0.05$, as measured with single dish observations of HDS/H$_{2}$S (table~$11$ of \citealt{vanDishoeck1995}\footnotemark[4]), the column density of H$_{2}$S can be estimated from HDS and HD$^{34}$S (Table~\ref{tbl:bestfit}). The average of these two values of $\left( 1.9\pm0.3 \right)\times10^{17}$~cm$^{-2}$ is the best estimate of the column density of H$_{2}$S at this position. It is possible that single dish observations may be overestimating deuteration, either because of sampling colder material or as a result of underestimating optical thickness of the main species. The lowest D/H ratio seen towards IRAS~16293-2422 is $0.01$ ($1$ per cent) for HNCO \citep{Coutens2016}. To account for this other extreme for the case of H$_{2}$S, a D/H~$=0.005$ may be assumed, increasing the best estimate to $\left( 1.9\pm0.3 \right)\times10^{18}$~cm$^{-2}$.

The previously obtained SMA observations of IRAS~16293--2422 have detected a line of H$_{2}$S around $\sim 216.71$~GHz (see fig.~$6$ of \citealt{Jorgensen2011}). LTE modelling assuming a beam size of $3\arcsec$, a spectral resolution of $0.56$~km~s$^{-1}$ (as given in table~$1$ of \citealt{Jorgensen2011}), FWHM of $1$~km~s$^{-1}$ and a source size of $0\farcs{5}$ (as assumed for the ALMA observations) shows that the line is optically thick. In order to match the observed line intensity of $\sim2$~Jy~beam$^{-1}$, a larger source size is necessary. Upon the assumption of a source $1\arcsec$ in size, the lower limit on the column density of H$_{2}$S is $4.0\times10^{16}$~cm$^{-2}$. This illustrates the uncertainty in the emitting area of this molecule. Alternatively, a non-detection of the sole line in range of the SMA observations of H$_{2}^{34}$S at $\sim226.70$~GHz does not yield a strongly constraining column density estimate. Assuming the same parameters, in order to be weaker than the SMA $3\sigma$ noise level (where $\sigma=0.24$~Jy~beam$^{-1}$~channel$^{-1}$; table~$1$ of \citealt{Jorgensen2011}), the column density of H$_{2}^{34}$S must not exceed $4.0\times10^{19}$~cm$^{-2}$. Assuming $^{32}$S/$^{34}$S~$=22$, the upper limit on the column density of H$_{2}$S is $8.8\times10^{20}$~cm$^{-2}$. This is consistent with the limits obtained with the uncertain limit obtained from HDS. Dedicated high resolution ALMA observations are needed in order to constrain the spatial distribution and column density of H$_{2}$S.

\subsection{CH$_{3}$SH and other species}
Methyl mercaptan (CH$_{3}$SH, also known as methanethiol) in the $\varv=0-2$ state is detected with more than $12$ clean lines at a column density of $5.5\times10^{15}$~cm$^{-2}$ (Fig.~\ref{fgr:CH3SHv=02}), which is comparable to the column density of H$_{2}$CS. Thioformaldehyde is a likely precursor for methyl mercaptan's grain-surface formation pathways, analogously to formaldehyde being a precursor for methanol.

No other sulphur-bearing species were detected towards the one beam offset position; however, selected $1\sigma$ upper limits have been derived and are given in Table~\ref{tbl:upperlim} (Fig.~\ref{fgr:g-C2H5SH}-~\ref{fgr:HSCN}). The values correspond to the maximal column densities at the fixed $T_{\text{ex}}=125$~K that the species can have in order to not exceed the observed flux at the frequencies of their emission lines. The obtained upper limits for all three versions of ethyl mercaptan (C$_{2}$H$_{5}$SH) are lower than the observed column density of methyl mercaptan by at most an order of magnitude, which is consistent with it being a step-up in chemical complexity.

The derived upper limits for the column densities of HS$_{2}$ and H$_{2}$S$_{2}$ are more constraining than those obtained by \citet{Martin-Domenech2016} owing to the much smaller beam size of the PILS Band~$7$ data. It is not possible to obtain an accurate upper limit for S$_{2}$ as only one line is covered in this dataset; however, the upper limit derived by \citet{Martin-Domenech2016} in a much larger beam size based on two lines is consistent with the observed non-detection. Carbon-sulphur chains larger than CS, e.g., C$_{2}$S, are not detected.

Upper limits on other likely carriers of sulphur in conjunction with H, C, O and N atoms are low (almost all $\lesssim 10^{15}$~cm$^{-2}$, compared with $\gtrsim 10^{17}$~cm$^{-2}$ for OCS and H$_{2}$S), which indicates that the dominating volatile sulphur reservoirs have been accounted for. The only remaining untapped sulphur reservoir would be the refractory/dust component. All other species that were not detected at this position are listed in Appendix~\ref{nondetspecies}.

\subsection{Integrated intensity maps}
\label{sec:intmaps}

\begin{figure*}
 \centering
  \includegraphics[width=0.95\textwidth,height=0.8\textheight,keepaspectratio]{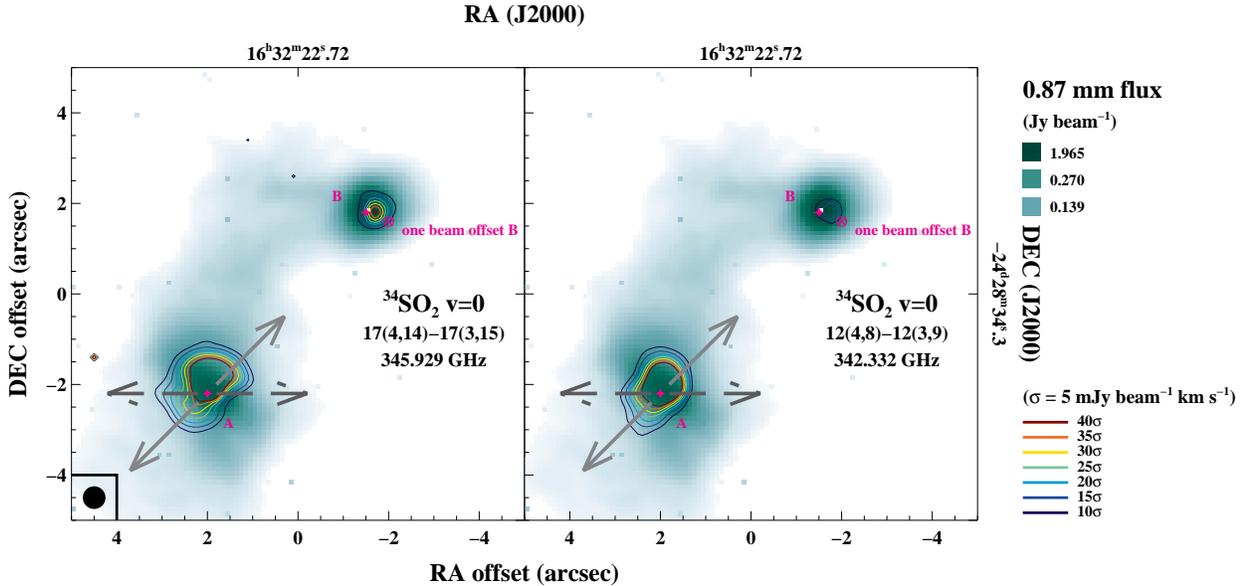}
 \caption{Integrated intensity maps of two transition of $^{34}$SO$_{2}$, see Section~\ref{sec:intmaps} for details. The centre of the pixel of the one beam offset position from IRAS~16293--2422~B corresponds to (RA, Dec)~J2000~$= (16^{\text{h}}32^{\text{m}}22.^{\text{s}}58, -24^{\circ}28\arcmin32\farcs{80})$. The directions of the outflows are indicated with the arrows.}
 \label{fgr:maps}
\end{figure*}

In order to gain insight into the spatial distribution of the emission of sulphur-bearing species, integrated intensity maps of two $^{34}$SO$_{2}$, $\varv=0$ transitions (at $342.332$ and $345.929$~GHz) are displayed in Fig.~\ref{fgr:maps}. This molecule has been chosen as it is firmly detected with many lines; and its emission is optically thin, without a doubt, at the one beam offset position from IRAS~16293--2422~B. The pixels in the maps are $0\farcs{1}\times0\farcs{1}$ in order to comfortably oversample the spatial resolution of $0\farcs{5}$ of the synthesized beam. The integration is performed over $9$ spectral bins (one containing the line frequency and $4$ to each side of that bin). Hence, the integrated maps cover $\pm0.9$~km~s$^{-1}$ around the source velocity of B of $2.7$~km~s$^{-1}$ (i.e., $v_{\text{lsr}}$ from $+1.9$ to $+3.6$~km~s$^{-1}$ at the spectral resolution of the data of $0.244$~MHz or $0.2$~km~s$^{-1}$). This implies that the maps are appropriate for the emission stemming from quiescent gas near source B; however, near source A they are less representative even though the systemic velocity of $3.2$~km~s$^{-1}$ of source A is covered. All emission lines near source A are likely wider than $2$~km~s$^{-1}$; thus, some of the flux from the line wings is missed when integrating over just $9$ bins. Moreover, there is a steep velocity gradient around source A. Integration over a wider velocity range will pollute the maps with emission from other species. In fact, even with integration over this velocity range already leads to contamination by emission from broadened neighbouring lines. Bearing in mind these limitations, Fig.~\ref{fgr:maps} clearly shows that source A has stronger emission in $^{34}$SO$_{2}$ and also on larger scales. The $346$~GHz line shows a higher flux than the $342$~GHz line around source B, while little difference is seen around source A. Since emission in two lines from the same molecule differs spatially, the physical conditions exciting the $^{34}$SO$_{2}$ molecule must also differ around the two sources.

These maps show convincingly that the $^{34}$SO$_{2}$ emission near source B is dominated by material on small scales either from the disc or the inner envelope (barely no difference is seen when analyzing the $12$-m data instead of the combination with ACA). For the $342$~GHz line of $^{34}$SO$_{2}$, the line peaks at $\sim38$~mJy~beam$^{-1}$ (Fig.~\ref{fgr:S-34-O2}). Integrating over $9$ spectral bins yields the total flux of this line at the one beam offset position of $\sim26$~mJy~beam$^{-1}$~km~s$^{-1}$. The total flux of this line integrated over the entirety of source B (Fig.~\ref{fgr:maps}) is $\sim150$~mJy~km~s$^{-1}$ (for RA$\in\left[ -1\farcs{0},-2\farcs{4} \right]$ and Dec$\in\left[ 1\farcs{3},2\farcs{2} \right]$). Taking the ratio of these two numbers implies that $\sim17$ per cent of the total $342$~GHz line emission of $^{34}$SO$_{2}$ towards source B is picked up in a beam solid angle at the one beam offset position. For the purposes of this paper, which are to compare and contrast the chemical composition of the smallest disc-scales around source B to those of comet 67P/C-G, it has been chosen to focus on this single position rather than the entire emitting area (which also differs per molecule). Emission from other species, such as CS and its isotopologues, is likely a superposition of extended (envelope) and the more compact (disc) components. However, these ALMA observations are marginally sensitive to envelope emission, since any structure that is smooth on scales of more than $\sim13\arcsec$ is filtered, even by the ACA.

\section{Discussion}
\label{discussion}

\subsection{Comparison with single dish observations}

\ctable[
 caption = {Molecular ratios relative to H$_{2}$S and OCS as measured with PILS interferometric ALMA observations at the one beam offset position from source B of IRAS~16293--2422 and with previous single dish (SD) work$^{\text{m}}$},
 label = {tbl:SDcomp},
 star = 1
 ]{@{\extracolsep{\fill}}llllllll}{
\tnote[m]{tables~$5$ and~$6$ of \citet{Schoier2002}}
 }{
 \hline
 \multirow{2}{*}{Species} & \multirow{2}{*}{N (cm$^{-2}$)} & \multicolumn{2}{c}{Molecular ratios relative to H$_{2}$S (\%)} & \multicolumn{4}{c}{Molecular ratios relative to OCS (\%)}\T\\
 & & ALMA B & SD constant & ALMA B & SD constant & SD inner & SD outer\B\\
 \hline
 H$_{2}$S  & $1.9\times10^{17,\text{i}-18,\text{j}}$ & $100$       & $100$ & $68-679$ & $23$  & -     & -\T\\
 OCS       & $2.8\times10^{17}$                      & $147-15$   & $438$ & $100$    & $100$ & $100$ & $100$\\
 SO        & $5.0\times10^{14}$                      & $0.3-0.03$ & $275$ & $0.2$    & $63$  & $100$ & $117$\\
 SO$_{2}$  & $1.5\times10^{15}$                      & $0.8-0.08$ & $39$  & $0.5$    & $9$   & $40$  & $15$\\
 CS        & $4.4\times10^{15}$                      & $2-0.2$    & $188$ & $2$      & $43$  & -     & -\\
 H$_{2}$CS & $1.5\times10^{15}$                      & $0.8$      & $13$  & $0.5$    & $3$   & $1$   & $3$\B\\
\hline}

\ctable[
 caption = {Molecular ratios relative to H$_{2}$S and OCS as measured with these interferometric ALMA observations at the one beam offset position from source B of IRAS~16293--2422 in comparison to those of 67P/C--G$^{\text{n}}$ and ISM ices towards W33A$^{\text{o}}$},
 label = {tbl:67Pcomp},
 star = 1
 ]{@{\extracolsep{\fill}}llllclll}{
\tnote[n]{see Section~\ref{comp67P} for the details on the ROSINA data}
\tnote[o]{ice ratios in the cold outer protostellar envelope of the high-mass protostar W33A, which are assumed to be representative of ISM ices \citep{Boogert1997, vanderTak2003}}
\tnote[p]{using the average of the upper limits on the column densities of gauche-C$_{2}$H$_{5}$SH and anti-C$_{2}$H$_{5}$SH}
 }{
 \hline
 \multirow{2}{*}{Species} & \multirow{2}{*}{N (cm$^{-2}$)} & \multicolumn{2}{c}{Molecular ratios relative to H$_{2}$S (\%)} &  & \multicolumn{3}{c}{Molecular ratios relative to OCS (\%)}\T\B\\
\cline{3-4}
\cline{6-8}
 & & ALMA B & 67P/C--G &  & ALMA B & 67P/C--G & ISM ices W33A\T\B\\
 \hline
 H$_{2}$S         & $1.9\times10^{17,\text{i}-18,\text{j}}$ & $100$      & $100$            &  & $68-679$  & $2257$    & $<30-90$\T\\
 OCS              & $2.8\times10^{17}$                      & $147-15$   & $4.43\pm0.15$    &  & $100$     & $100$     & $100$\\
 SO               & $5.0\times10^{14}$                      & $0.3-0.03$ & $7.06\pm0.17$    &  & $0.2$     & $159$     & -\\
 SO$_{2}$         & $1.5\times10^{15}$                      & $0.8-0.08$ & $12.5\pm0.3$     &  & $0.5$     & $282$     & $3-7$\\
 S$_{2}$          & $\leq2.2\times10^{16}$                  & $\leq12$   & $0.25\pm0.05$    &  & $\leq8$   & $6$       & -\\
 S$_{3}$          & $\leq4.0\times10^{15}$                  & $\leq2$    & $\sim1$          &  & $\leq1$   & $\sim23$  & -\\
 HS$_{2}$         & $\leq5.0\times10^{14}$                  & $\leq0.3$  & $\leq0.01$       &  & $\leq0.2$ & $\leq0.2$ & -\\
 H$_{2}$S$_{2}$   & $\leq9.0\times10^{14}$                  & $\leq0.5$  & $\leq0.057$      &  & $\leq0.3$ & $\leq1$   & -\\
 H$_{2}$CS        & $1.5\times10^{15}$                      & $0.8$      & $2.25\pm1.38$    &  & $0.5$     & $51$      & -\\
 CH$_{3}$SH       & $5.5\times10^{15}$                      & $3-0.3$    & $3.5\pm1.0$      &  & $2$       & $79$      & -\\
 C$_{2}$H$_{5}$SH & $\leq1.8\times10^{15, \text{p}}$        & $\leq0.9$  & $0.0338\pm0.018$ &  & $\leq0.6$ & $0.08$    & -\B\\
\hline}
\ctable[
 caption = {Molecular ratios as measured with these interferometric ALMA observations at the one beam offset position from source B of IRAS~16293--2422 and with the ROSINA instrument for 67P/C--G$^{\text{n}}$},
 label = {tbl:comprat},
 star = 1
 ]{@{\extracolsep{\fill}}lrl}{
 }{
 \hline
 Ratio & ALMA B & 67P/C--G\T\B\\
 \hline
 SO/SO$_{2}$                 & $0.33$                & $0.4-0.7$ (section~$4.4$ of \citealt{Calmonte2016})\T\\
 CH$_{3}$SH/H$_{2}$CS        & $3.7$                 & $0.69-5.2$ (based on values in Table~\ref{tbl:67Pcomp})\\
 C$_{2}$H$_{5}$SH/CH$_{3}$SH & $\leq0.33^{\text{p}}$ & $0.0010-0.021$ (based on values in Table~\ref{tbl:67Pcomp})\B\\
\hline}

The presented interferometric observations spatially resolve the hot inner regions near the source B continuum peak, meaning that column densities can be derived and compared on the same scales unambiguously. Moreover, it is possible to detect weaker lines from multiple isotopologues, thereby allowing a better determination of the optical depth. By comparing molecular ratios of interferometric and single dish observations, it is possible to disentangle whether molecules primarily emit on large scales or small inner scales near the source, or whether they are associated with outflows. For this purpose, Table~\ref{tbl:SDcomp} has been compiled with molecular ratios of the detected species relative to H$_{2}$S and OCS (where available), as obtained with the PILS survey with ALMA at the one beam offset position and as compiled from previous single dish observations of \citet{Schoier2002}. Since single dish observations cannot spatially resolve the emission, three different sets of abundances are typically provided: one assuming a constant abundance profile and one assuming a jump abundance profile, giving an inner (hot) abundance and an outer (cold) one.

Under the assumption of a constant abundance profile, relative to H$_{2}$S, the interferometric ratios are lower than those from single dish observations: by a factor of $3-30$ for OCS, by $3-4$ orders of magnitude for SO, and by $2-3$ orders of magnitude for SO$_{2}$, CS, and H$_{2}$CS (Table~\ref{tbl:SDcomp}). Relative to OCS, the interferometric ratio is a factor of $3-30$ higher than the single dish value for H$_{2}$S. All other interferometric ratios remain lower than the single dish: by $2$ orders of magnitude for SO, by a factor of $\sim 18$ for SO$_{2}$, by a factor of $\sim 27$ for CS and by a factor of $\sim 6$ for H$_{2}$CS. Under the assumption of a jump abundance profile, relative to OCS, the differences are either comparable or exacerbated further. The inner ratio is expected to yield the most meaningful comparison with interferometric data, as it is the closest estimate for the hot inner regions studied in this work.

The derived lower interferometric ratios mean that sulphur-bearing species emit on small and large scales. Single dish observations are, therefore, a combination of beam-diluted disc-scale emission (isotopologues detected) and large-scales emission from the envelope and outflows. The largest differences are seen for SO, which is a well-known shock tracer that shows larger line widths in the single dish data \citep{Blake1994}. Some SO$_{2}$ emission likely originates from outflows as well. Meanwhile, OCS, H$_{2}$S and H$_{2}$CS emit also from the envelope. In fact, H$_{2}$CS is one of the few species detected at large distances from source A (appendices~C1 and~C2 of \citealt{Murillo2017}), definitively pointing to its origins in the extended envelope. However, the detection of all these molecules with interferometric data testifies to their presence on small scales of the disc and/or inner envelope as well. Unfortunately, it is not meaningful to quantify these differences, since single dish observations also encompass source A. The emission from sulphur-bearing species is seen to be brighter towards source A, which may correlate with A powering stronger outflows. It is beyond the scope of this work to quantify the individual contributions of the two sources in single dish observations.

\subsection{Comparison with interstellar ices}
\label{compISM}
The interstellar ices in clouds and cores prior to the onset of star formation (i.e., prestellar ices) can only be observed in absorption against a bright background source. Observations towards the high-mass protostar W33A also show deep absorption features corresponding to ices in a protostellar envelope, which are water-dominated, but also contain trace species, including CH$_{3}$OH \citep{Boogert1997, vanderTak2003}. The ices in the outer cold protostellar envelope have been suggested to be chemically representative of prestellar ices (just prior to star formation) based on observations of non-sulphur bearing molecules \citep{Boogert2015}. Table~\ref{tbl:67Pcomp} contains the ice molecular ratios relative to OCS as observed towards W33A in its outer cold protostellar envelope and are assumed to be representative of ISM ices, thus called as such. It is interesting to compare ISM ices to the warm ($>100$~K) gas near the forming B protostar of IRAS~16293--2422 to assess the degree of processing between the cold outer protostellar envelope ices and this earliest warm embedded protostellar phase; assuming that the ices have just thermally desorbed, and no gas-phase chemistry has occurred. The ISM SO$_{2}$/OCS ratio is roughly an order of magnitude higher, which means that in the cold outer protostellar envelope there is more SO$_{2}$ or less OCS in comparison to the warm protostellar phase of IRAS~16293--2422~B. The H$_{2}$S/OCS ratio is poorly constrained due to a lack of data on H$_{2}$S ice and gas. Based on the current limits, the ISM ratio can be lower (by an order of magnitude) or higher (by a factor of $\sim1.3$) than that of source B. If the ISM H$_{2}$S/OCS is indeed higher than that towards source B, then there may indeed be less OCS in the cold outer protostellar envelope than in the warm protostellar phase.

\subsection{Comparison with 67P/C-G and ISM ices}
\label{comp67P}

\begin{figure}
 \centering
  \includegraphics[width=0.45\textwidth,height=0.8\textheight,keepaspectratio]{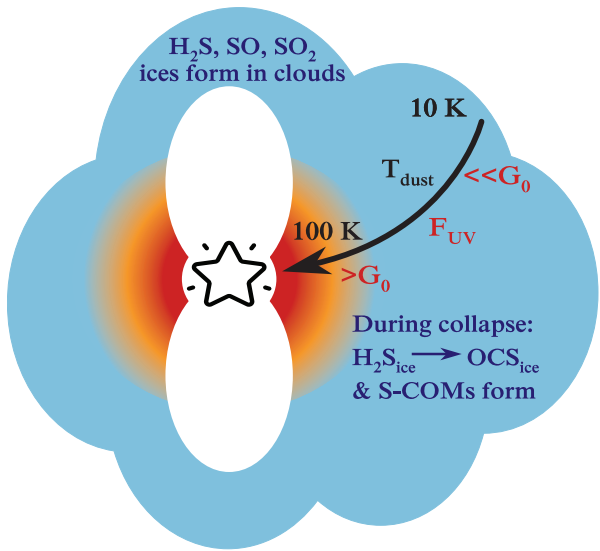}
 \caption{A schematic summarizing the chemical evolution of S-bearing species during the earliest phases of low-mass star formation. S-COMs stands for S-bearing complex organic molecules. G$_{0}$ is the interstellar FUV radiation field of $1.6\times10^{-3}$~erg~cm$^{-2}$~s$^{−1}$ \citep{Habing1968}.}
 \label{fgr:summary_schematic}
\end{figure}

IRAS~16293--2422 is thought to be analogous to our innate Solar Nebula, as it is one of the most chemically rich low-mass protostellar systems. Thanks to its proximity and the face-on orientation of the disc around source B, ALMA has been able to start precisely quantifying the chemistry of the disc-neighbourhood of B. Meanwhile, it has also been postulated that comets are the most pristine tracers of the innate cloud and protoplanetary disc that evolved into the Solar System that we have today. Therefore, it is interesting to compare the chemical inventories of comets to those of IRAS~16293--2422~B in order to quantify the chemical budgets of solar and extrasolar building blocks of planetary systems \citep{Schoier2002, Bockelee-Morvan2015}. The emission seen in the hot inner regions of the B protostellar core could potentially be tracing solid species that are just undergoing thermal desorption and thus, is in fact probing the hidden solid reservoir of planet-building material. The chemical composition of such building blocks will be shaped by many processes at different radii; however, this location likely uniquely probes the solid volatiles prior to modification in the gas-phase.

Table~\ref{tbl:67Pcomp} contains the molecular ratios relative to H$_{2}$S and OCS for IRAS~16293--2422~B in comparison to those of 67P/C--G. The cometary values correspond to bulk coma quantities measured with DFMS/ROSINA and corrected for photodissociation and ionization by the radiation from the Sun (more important when in high proximity to the Sun and when the comet-spacecraft distance is large, e.g., $\sim200$~km; table~$3$ of \citealt{Calmonte2016}), as obtained between equinox in May 2015 and perihelion in August 2015 (heliocentric distances between $1$ and $2$~au). The upper limits for HS$_{2}$ and H$_{2}$S$_{2}$ are based on October 2014 data, when H$_{2}$S abundance was maximized (sections~$4.1.3$ and $4.3.5$ of \citealt{Calmonte2016}). The value given for S$_{3}$ is an estimate based on the sole measurement obtained in March 2015 via a direct comparison of the measured ion current (largely uncertain, section~$4.1.2$ of \citealt{Calmonte2016}). The value tabulated for H$_{2}$CS is based upon the sole clear signal at $46$~$u/e$ during a flyby in February 2015, when the CO$_{2}$ abundance was very low and the overlap with C$^{16}$O$^{18}$O could be avoided (priv. comm. and section~$4.2.3$ of \citealt{Calmonte2016}). The value quoted for CH$_{3}$SH is the mean of the four periods between the start of the mission and perihelion with a signal at $48$~$u/e$ corrected for photodissociation and ionization (section~$4.3.4$ of \citealt{Calmonte2016}). The value quoted for C$_{2}$H$_{5}$SH is the mean of four different periods between the start of the mission and perihelion with a signal at $62$~$u/e$ not corrected for photodissociation and ionization (sections~$4.1.1$ and $4.3.4$ of \citealt{Calmonte2016}), which is highly uncertain due to the assumption of the entire C$_{2}$H$_{6}$S peak being associated with ethyl mercaptan and the close overlap with the C$_{5}$H$_{2}$ fragment. The errors in Table~\ref{tbl:67Pcomp} for the DFMS/ROSINA measurements stem from instrumental uncertainties (section~$3.3$ of \citealt{Calmonte2016}).

From the table, it can be seen that for OCS the ratio relative to H$_{2}$S for IRAS~16293--2422~B is a factor of $3 \sim 33$ higher than that for 67P/C--G, for SO a factor of $2 \sim 24$ lower, for SO$_{2}$ a factor of $2 \sim 16$ lower and for CH$_{3}$SH a factor of $1.2 \sim 12$ lower\footnote[5]{This contradicts section~$5.2$ of \citet{Calmonte2016} due to different assumptions on the quantity of H$_{2}$S and the use of abundances versus column densities.}. Relative to OCS, the IRAS~16293--2422~B ratios for SO and SO$_{2}$ are $\sim3$ orders of magnitude lower, and for CH$_{3}$SH is $\sim 1$ order of magnitude lower. Such large differences are consistent with the fact that H$_{2}$S has not yet been detected in interstellar ices, but OCS has (Section~\ref{compISM}).

These results indicate that in the case of IRAS~16293--2422~B there is significantly more OCS available, while for 67P/C--G more H$_{2}$S is present. Potentially, this has to do with the amount of solid carbon monoxide (CO) available in these systems. OCS may be more easily produced via grain-surface chemistry when CO ice is abundant, as seen in laboratory experiments of \citet{Ferrante2008}. If so, then once CO undergoes thermal desorption into the gas-phase around $20$~K, the production of OCS would be inhibited (excluding any CO that maybe trapped in other ices). This could be an initial indication of our Solar System being born in a somewhat warmer, $>20$~K, environment rather than in the $10-15$~K regime, which would lead to a deficiency of CO ice on the grains for the synthesis of OCS. This scenario is also favoured from the point of view of oxygen chemistry and the detection of O$_{2}$ on 67P/C--G \citep{Bieler2015, Taquet2016b}. In the case of IRAS~16293--2422~B, observations have shown its surrounding core to be very cold ($\sim12$~K; \citealt{Menten1987}), thus making CO ice plentiful in its birth cloud. It has been suggested that regions with elevated temperatures lead to CO ice-poor conditions, for example, the R~CrA region with several low-mass protostars under the irradiation of an intermediate-mass protostar \citep{Lindberg2014a}, and Orion hosting intermediate-mass protostars \citep{Jorgensen2006}.

A warmer environment would imply less efficient hydrogenation reactions, as the residence time of hydrogen atoms on the grains is reduced. This would make it difficult to produce H$_{2}$S through grain-surface atom-addition reactions. This molecule is predominantly formed via the associations of H and S, followed by the reaction of HS with H (e.g, \citealt{Furuya2015, Vidal2017}). Some H$_{2}$S gas can also be produced via gas-phase channels at warmer temperatures. Temperature differences between birth clouds should also be noticeable in terms of complex organics. An initially warmer cloud would enhance mobility of heavy radicals on grain surfaces, such as CH$_{3}$ and SH, thereby boosting the abundance of complex species, such as CH$_{3}$SH (although it is not exclusively formed via grain-surface chemistry). The CH$_{3}$SH/H$_{2}$CS ratio for IRAS~16293-2422~B is in range of the ratio for 67P/C--G given the errors in ROSINA data (differences by a factor of $0.7-5.3$; Table~\ref{tbl:comprat}). Hence, based the complex S-bearing species CH$_{3}$SH relative to H$_{2}$CS, it is not possible to conclusively say which target is richer in complex organic molecules. Given the large differences in the quantities of H$_{2}$S and OCS between the two data sets, ratios relative to them are less meaningful. As it has only been possible to derive an upper limit for C$_{2}$H$_{5}$SH, the C$_{2}$H$_{5}$SH/CH$_{3}$SH ratio cannot yet be used to compare the budgets of S-bearing complex species between IRAS~16293--2422~B and 67P/C--G. A rigorous comparison between IRAS~16293--2422~B and 67P/C--G in terms of a larger set of other O- and N-bearing complex organics is the subject of a dedicated paper.

Additional physics could be at play here - the degree of irradiation that the two targets are subjected to. Our Sun is an isolated star that is thought to have formed in a medium-sized stellar cluster of $10^{3}-10^{4}$ members \citep{Adams2010}. Laboratory experiments of \citet{Chen2015} have shown rapid conversion of H$_{2}$S into OCS upon VUV and EUV irradiation. If IRAS~16293--2422~B experienced higher UV fluxes, perhaps due to its binary A companion, then that may have stimulated the conversion of its H$_{2}$S reservoir into OCS. In addition, higher levels of UV would explain the higher fraction of complex S-bearing species relative to H$_{2}$CS of IRAS~16293--2422~B, since UV enhances the availability of radicals in the ice and boosts grain-surface chemistry. This scenario would also correlate with OCS forming during the collapse phases and there, thus, being less OCS in ISM ices (Section~\ref{compISM}; although, the presence of H$_{2}$S ice is yet to be demonstrated and quantified in prestellar sources).

Another diagnostic independent of the H$_{2}$S and OCS quantities, is the SO/SO$_{2}$ ratio. Based on the presented PILS Band~$7$ data, it is $0.33$ for IRAS~16293--2422~B; and based on the ROSINA data, it is in the $0.4-0.7$ range for 67P/C--G (Table~\ref{tbl:comprat}). Such excellent (a factor of $\sim2$) agreement could point to both datasets sampling comparable materials. If the quantities of SO and SO$_{2}$ are set before and/or during the prestellar stage, as supported by the detection of SO$_{2}$ as an ISM ice (Section~\ref{compISM}), then this again points to IRAS~16293--2422~B and 67P/C--G having formed from similar birth clouds. Unfortunately, even with JWST it may be impossible to quantify the ISM SO/SO$_{2}$ ice ratio due to the overlap of their symmetric stretching modes. Gas-phase routes to SO and SO$_{2}$ via reactions of OH and O$_{2}$ with S at warm temperatures may also skew the ratio.

The ROSINA instrument determined that $\sim27$ per cent of the sulphur is in the atomic form and constitutes the second largest sulphur reservoir after H$_{2}$S (fig.~$16$ of \citealt{Calmonte2016}). It was also argued that five times more H$_{2}$S than detected would be necessary to account for the measured quantity of atomic sulphur, hence it is indeed stored in its atomic form in 67P/C-G \citep{Calmonte2016}. Unfortunately, atomic sulphur does not have any lines in the frequency range covered by the PILS Band~$7$ data. 67P/C--G has been observed to contain sulphur-chain molecules like S$_{2}$, S$_{3}$ and S$_{4}$, which could, in part, be fragments of even heavier S$_{n}$ species (up to $n=8$). The upper limits derived in this work are less constraining than the quantities measured for 67P/C--G (Table~\ref{tbl:67Pcomp}). ROSINA did not (conclusively) detect HS$_{2}$ and H$_{2}$S$_{2}$ (section~$4.1.3$ of \citealt{Calmonte2016}). These species were not detected towards IRAS~16293--2422~B with the PILS Band~$7$ data either. It was not possible to derive the abundance of CS with ROSINA, due to interference with CO$_{2}$; and due to it being a common fragment of other sulphur-bearing species. Hence, it cannot be compared in a meaningful fashion with the data in this work. Finally, as CS$_{2}$ is a linear symmetric molecule without strong rotational lines, it is not possible to search for it in the PILS Band~$7$ data; even though it has been clearly detected with mass spectrometry on 67P/C--G.

The findings of this work can be summarized in Fig.~\ref{fgr:summary_schematic}. It seems that the abundances of SO, SO$_{2}$ and H$_{2}$S are set in the translucent cloud and prestellar core phases, prior to the onset of collapse. This is supported by the similar relative ratios between the PILS data on IRAS~16293--2422~B, the ISM ices towards W33A (Section~\ref{compISM}) and the ROSINA data on 67P/C--G. The differences between the datasets can be explained in terms of the amount of UV radiation that the materials are exposed to, which occurs during the collapse of the system, already in the earliest embedded stages, as well as the temperature of the birth cloud. Irradiation brings about the conversion of H$_{2}$S into OCS and the formation of S-bearing complex organic molecules. This explains the richer chemistry towards IRAS~16293--2422~B and our Solar System, in comparison to cold outer protostellar envelopes. Whether IRAS~16293--2422~B is relatively chemically richer than our Solar System as a result of higher UV fluxes based on its binary nature still remains to be conclusively tested via a larger set of complex organic species. If this is shown to be the case, then this would correlate with IRAS~16293--2422 being one of the most chemically rich low-mass protostars; and imply that our Solar System is not special chemically speaking.

\section{Conclusions}
\label{conclusions}

In this paper, the sulphur inventory at the one beam offset position from source B in the SW direction in the binary protostellar system IRAS~16293--2422 has been presented. The ALMA Band~$7$ data analysed are part of the PILS survey towards this target. Sulphur is thought to be a unique simultaneous tracer of both the volatile and refractory components. By comparing molecular ratios observed towards IRAS~16293--2422~B -- a Solar System proxy, and those obtained for 67P/C--G -- a pristine tracer of the innate Solar Nebula, the chemical links between the early embedded protostellar phases and the protoplanetary building blocks can be explored. The main conclusions of this paper are as follows.

\begin{enumerate}
	\item The sulphur-bearing species previously detected towards IRAS~16293--2422~B have now been firmly detected towards the one beam offset position and spatially resolved with ALMA: SO$_{2}$ in the $\varv=0$, $^{34}$SO$_{2}$, OCS in the $\varv_{2}=1$ state, O$^{13}$CS, OC$^{34}$S, OC$^{33}$S (first-time detection towards this source), $^{18}$OCS, H$_{2}$CS, HDCS and CH$_{3}$SH in the $\varv=0-2$ state. Furthermore, several are detected tentatively due to a lack of lines and/or blending: SO in the $\varv=0$ state, OCS in the $\varv=0$ state, C$^{34}$S in the $\varv=0,1$ state, C$^{33}$S in the $\varv=0,1$ state, C$^{36}$S (tentative first time detection towards a low-mass protostar), HDS and HD$^{34}$S. All the lines from these sulphur-bearing molecules are narrow ($\sim 1$~km~s$^{-1}$) and are probing the small disc-scales of source B. Unfortunately, the unknown HDS/H$_{2}$S ratio leaves the column density of H$_{2}$S uncertain by at least an order of magnitude.
	\item In comparison to earlier single dish observations, the molecular ratios determined from interferometric data can be up to four orders of magnitude lower. On large scales, SO and SO$_{2}$ likely emit from outflows; and OCS, H$_{2}$S and H$_{2}$CS originate from the envelope. All these S-bearing species are also present on the small disc- and inner envelope-scales. Single dish data may be dominated by source A, which is bright in emission from S-bearing molecules.
	\item In comparison to ROSINA measurements of the bulk volatile composition of 67P/C--G, the molecular ratios of IRAS~16293--2422~B can differ significantly, potentially by several orders of magnitude. In particular, IRAS~16293--2422~B contains much more OCS than H$_{2}$S. However, the SO/SO$_{2}$ is in close agreement between the two targets.
	\item The agreement in terms of SO, SO$_{2}$, the disagreement in terms of H$_{2}$S and OCS, and comparable (differences by a factor of $0.7-5.3$) CH$_{3}$SH/H$_{2}$CS ratios towards IRAS~16293--2422~B in comparison to that of our Solar System (as probed by 67P/C--G) may stem from different levels of UV irradiation during the initial collapse of the systems. Potentially higher UV levels near source B, as a result of its binary structure, may lead to the conversion of H$_{2}$S into OCS and enhance the formation of S-bearing complex organic molecules. An initially warmer birth cloud may also contribute to the lower quantities of OCS in the Solar System by reducing the amount of available CO ice.
\end{enumerate}

The results also highlight the importance of the level of UV exposure and temperature of the parental clouds, in determining the physical and chemical structures of low-mass protostars. Given that these may vary significantly from region to region (see, e.g., review by \citealt{Adams2010}), it is also likely that significant source-to-source variations in chemistry may be found. Future studies will also explore the protostellar-cometary connection via isotopic ratios and a full set of complex organic species, thereby isolating chemical links formed during cold phases of evolution that are dominated by grain-surface chemistry from those occurring during warmer, irradiated phases of collapse.

\section{Acknowledgements}
\label{acknowledgements}

The authors would like to thank Dr. Catherine Walsh, Dr. Vianney Taquet, Mr. Ko-Ju Chuang and Dr. Martin Rubin for useful discussions on sulphur chemistry and ROSINA measurements. This work is supported by a Huygens fellowship from Leiden University, the European Union A-ERC grant 291141 CHEMPLAN, the Netherlands Research School for Astronomy (NOVA), a Royal Netherlands Academy of Arts and Sciences (KNAW) professor prize, the Center for Space and Habitability (CSH) Fellowship and the IAU Gruber Foundation Fellowship.

The research of JKJ and his group is supported by the European Research Council (ERC) under the European Union's Horizon 2020 research and innovation programme (grant agreement No~646908) through ERC Consolidator Grant ``S4F''. Research at the Centre for Star and Planet Formation is funded by the Danish National Research Foundation. A.C. postdoctoral grant is funded by the ERC Starting Grant 3DICE (grant agreement 336474).

This paper makes use of the following ALMA data: ADS/JAO.ALMA\#2013.1.00278.S. ALMA is a partnership of ESO (representing its member states), NSF (USA) and NINS (Japan), together with NRC (Canada), MOST and ASIAA (Taiwan), and KASI (Republic of Korea), in cooperation with the Republic of Chile. The Joint ALMA Observatory is operated by ESO, AUI/NRAO and NAOJ.

\clearpage
\bibliographystyle{mn2e}
\bibliography{mybib} 

\clearpage
\newpage
\appendix

\section{Partition of sulphur on 67P/C--G}
\label{sulphur67P}
For the moment, the data on elemental abundances in the dust of 67P/C--G has not been released by the Cometary Secondary Ion Mass Spectrometer (COSIMA) team. Therefore, the partition of sulphur between the refractory (dust) and volatile (ice) components has been derived based upon the information on the S/O ratio in volatiles as derived by the ROSINA team and the results obtained for comet 1P/Halley \citep{Jessberger1988}.

For comet 1P/Halley, it has been shown that the dust to ice mass ratio is:
\begin{equation}
m_{\text{dust 1P}}/m_{\text{ice 1P}} = 2,
\end{equation}
\noindent so then:
\begin{equation}
m_{\text{comet} 1P} = m_{\text{dust 1P}}+m_{\text{ice 1P}} = 3 m_{\text{ice 1P}} = \frac{3}{2} m_{\text{dust 1P}}.
\end{equation}
\noindent In other terms:
\begin{equation}
m_{\text{dust 1P}} = \frac{2}{3} m_{\text{comet 1P}}
\end{equation}
\noindent and
\begin{equation}
m_{\text{ice 1P}} = \frac{1}{3} m_{\text{comet 1P}}.
\end{equation}
\noindent According to \citet{Geiss1988}, $23$ per cent of all the oxygen atoms on 1P/Halley is in the refractory (dust) component by number, i.e.:
\begin{equation}
n_{\text{O}_{\text{dust 1P}}} = 0.23 n_{\text{O}_{\text{tot 1P}}},
\end{equation}
\noindent where $n$ are total numbers of atoms and is dimensionless. So:
\begin{equation}
n_{\text{O}_{\text{ice 1P}}} = 0.77 n_{\text{O}_{\text{tot 1P}}}.
\end{equation}
\noindent Then the number densities (in cm$^{-3}$) of oxygen in the refractory (dust) and volatile (ice) components of 1P/Halley are given by:
\begin{align}
\rho_{\text{O}_{\text{dust 1P}}} &= \frac{n_{\text{O}_{\text{dust 1P}}}}{m_{\text{O}_{\text{dust 1P}}}} \times P_{\text{O}_{\text{dust 1P}}} = \frac{0.23 n_{\text{O}_{\text{tot 1P}}}}{\frac{2}{3} m_{\text{O}_{\text{comet 1P}}}} \times P_{\text{O}_{\text{dust 1P}}},\\
\rho_{\text{O}_{\text{ice 1P}}} &= \frac{n_{\text{O}_{\text{ice 1P}}}}{m_{\text{O}_{\text{ice 1P}}}} \times P_{\text{O}_{\text{ice 1P}}} = \frac{0.77 n_{\text{O}_{\text{tot 1P}}}}{\frac{1}{3} m_{\text{O}_{\text{comet 1P}}}} \times P_{\text{O}_{\text{ice 1P}}},
\end{align}
\noindent where $P$ are the mass densities (in g~cm$^{-3}$). The ratio of the two is:
\begin{equation}
\rho_{\text{O}_{\text{ice 1P}}} / \rho_{\text{O}_{\text{dust 1P}}} = \frac{0.77 \times 2}{0.23} \times \frac{P_{\text{O}_{\text{ice 1P}}}}{P_{\text{O}_{\text{dust 1P}}}} \approx 6.7 \frac{P_{\text{O}_{\text{ice 1P}}}}{P_{\text{O}_{\text{dust 1P}}}}.
\end{equation}

Now let us assume that 1P/Halley and 67P/C--G have been made from the same dust and ice. This means that $\rho_{\text{O}_{\text{ice 1P}}} = \rho_{\text{O}_{\text{ice 67P}}}$, $\rho_{\text{O}_{\text{dust 1P}}} = \rho_{\text{O}_{\text{dust 67P}}}$, and that $P_{\text{O}_{\text{ice 1P}}} = P_{\text{O}_{\text{ice 67P}}}$, $P_{\text{O}_{\text{dust 1P}}} = P_{\text{O}_{\text{dust 67P}}}$. Then it also follows that:
\begin{equation}
\rho_{\text{O}_{\text{ice 67P}}} / \rho_{\text{O}_{\text{dust 67P}}} \approx 6.7 \frac{P_{\text{O}_{\text{ice 67P}}}}{P_{\text{O}_{\text{dust 67P}}}}.
\end{equation}

For comet 67P/C--G, \citet{Rotundi2015} derived that the dust to ice mass ratio is:
\begin{equation}
m_{\text{dust 67P}}/m_{\text{ice 67P}} = 4,
\end{equation}
\noindent so then:
\begin{equation}
m_{\text{comet 67P}} = m_{\text{dust 67P}}+m_{\text{ice 67P}} = 5 m_{\text{ice 67P}} = \frac{5}{4} m_{\text{dust 67P}}.
\end{equation}
\noindent In other terms:
\begin{equation}
m_{\text{dust 67P}} = \frac{4}{5} m_{\text{comet 67P}}
\end{equation}
\noindent and
\begin{equation}
m_{\text{ice 67P}} = \frac{1}{5} m_{\text{comet 67P}}.
\end{equation}

Subsequently, the numbers of oxygen atoms in the refractory (dust) and volatile (ice) components of 67P/C--G become:
\begin{align}
n_{\text{O}_{\text{dust 67P}}} &= \rho_{\text{O}_{\text{dust 67P}}} \times \frac{m_{\text{O}_{\text{dust 67P}}}}{P_{\text{O}_{\text{dust 67P}}}} =\\
&= \rho_{\text{O}_{\text{dust 67P}}} \times \frac{\frac{4}{5} m_{\text{comet 67P}}}{P_{\text{O}_{\text{dust 67P}}}},\\
n_{\text{O}_{\text{ice 67P}}} &= \rho_{\text{O}_{\text{ice 67P}}} \times \frac{m_{\text{O}_{\text{ice 67P}}}}{P_{\text{O}_{\text{ice 67P}}}} =\\
&= \rho_{\text{O}_{\text{ice 67P}}} \times \frac{\frac{1}{5} m_{\text{comet 67P}}}{P_{\text{O}_{\text{ice 67P}}}},
\end{align}
\noindent and the ratio of the two is:
\begin{align}
n_{\text{O}_{\text{dust 67P}}} / n_{\text{O}_{\text{ice 67P}}} &= 4 \times \frac{\rho_{\text{O}_{\text{dust 67P}}}}{\rho_{\text{O}_{\text{ice 67P}}}} \times \frac{P_{\text{O}_{\text{ice 67P}}}}{P_{\text{O}_{\text{dust 67P}}}} \approx\\
&\approx 4 \times \frac{P_{\text{O}_{\text{dust 67P}}}}{6.7 \times P_{\text{O}_{\text{ice 67P}}}} \times \frac{P_{\text{O}_{\text{ice 67P}}}}{P_{\text{O}_{\text{dust 67P}}}} \approx 0.60.
\end{align}
\noindent Hence:
\begin{equation}
n_{\text{O}_{\text{comet 67P}}} = n_{\text{O}_{\text{dust 67P}}} + n_{\text{O}_{\text{ice 67P}}} = 1.60 n_{\text{O}_{\text{ice 67P}}} = 2.7 n_{\text{O}_{\text{dust 67P}}}.
\end{equation}
\noindent This implies that $37$ per cent of all the oxygen atoms on 67P/C--G is in the refractory (dust) component by number; and that $63$ per cent of all the oxygen atoms on 67P/C--G is in the volatile (ice) component by number.

It has also been shown by \citet{Jessberger1988} that S/O in the dust of 1P/Halley is $8.1$ per cent (based on values in table~$1$). Assuming again that 1P/Halley and 67P/C--G have been made from the same dust, i.e.:
\begin{equation}
\left(\text{S}/\text{O}\right)_{\text{dust 67P}} = \left(\text{S}/\text{O}\right)_{\text{dust 1P}} = 8.1\%;
\end{equation}
\noindent then the overall S/O ratio in the dust and ice of 67P/C--G is:
\begin{multline}
\left(\text{S}/\text{O}\right)_{\text{comet 67P}} = \frac{n_{\text{O}_{\text{dust 67P}}}}{n_{\text{O}_{\text{comet 67P}}}} \left(\text{S}/\text{O}\right)_{\text{dust 67P}} + \frac{n_{\text{O}_{\text{ice 67P}}}}{n_{\text{O}_{\text{comet 67P}}}} \left(\text{S}/\text{O}\right)_{\text{ice 67P}}=\\
=0.37\times8.1\% + 0.63\times1.47\%\approx3.9\%,
\end{multline}
\noindent where the S/O in the ice of 67P/C--G of $1.47$ per cent has been used (as given in section~$5.6$ of \citealt{Calmonte2016}) as derived from ROSINA data.

If $\left(\text{S}/\text{O}\right)_{\text{comet 67P}} = 3.9\%$ and $\left(\text{S}/\text{O}\right)_{\text{ice 67P}} = 1.47\%$, then:
\begin{align}
\frac{\left(\text{S}/\text{O}\right)_{\text{ice 67P}}}{\left(\text{S}/\text{O}\right)_{\text{comet 67P}}} &= \frac{1.47}{3.9},\\
\frac{n_{\text{S}_{\text{ice 67P}}}}{n_{\text{S}_{\text{comet 67P}}}} \times \frac{n_{\text{O}_{\text{comet 67P}}}}{n_{\text{O}_{\text{ice 67P}}}} &= \frac{1.47}{3.9},\\
\frac{n_{\text{S}_{\text{ice 67P}}}}{n_{\text{S}_{\text{comet 67P}}}} \times 1.6 &= \frac{1.47}{3.9},\\
\frac{n_{\text{S}_{\text{ice 67P}}}}{n_{\text{S}_{\text{comet 67P}}}} &= \frac{1.47}{3.9 \times 1.6} = 0.24.
\end{align}
\noindent This means that $24$ per cent of all the sulphur atoms on 67P/C--G are in the volatile (ice) component by number; and that $76$ per cent of all the sulphur atoms on 67P/C--G are in the refractory (dust) component by number.

\section{A selection of lines and synthetic spectra of S-bearing species in the PILS Band~$7$ dataset}
\label{linesspectra}

\begin{figure*}
 \centering
 \includegraphics[width=0.95\textwidth,height=0.8\textheight,keepaspectratio]{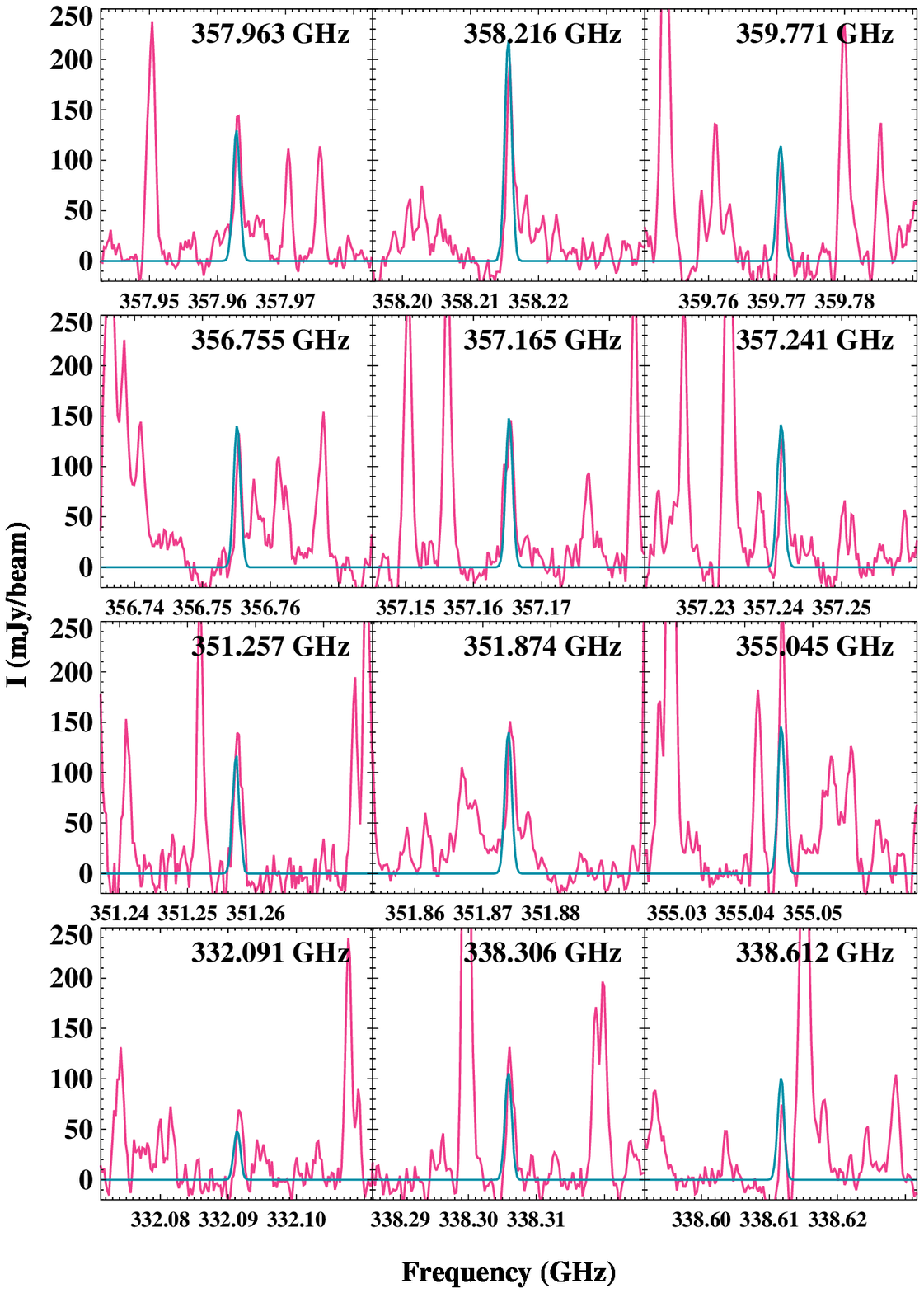}
 \caption{Twelve selected lines of SO$_{2}$ in the $\varv=0$ state. The observed ALMA Band~$7$ spectrum convolved with a uniform circular restoring beam of $0\farcs{5}$ at the one beam ($\sim60$~au) offset from source B of IRAS~16293--2422 in the SW direction is in pink. The turquoise line is the LTE fit for each displayed transition, assuming a source size of $0\farcs{5}$, FWHM of $1$~km~s$^{-1}$, $T_{\text{ex}}=125$~K and a column density as prescribed in Table~\ref{tbl:bestfit}.}
 \label{fgr:SO2v=0}
\end{figure*}

\begin{figure*}
 \centering
 \includegraphics[width=0.95\textwidth,height=0.8\textheight,keepaspectratio]{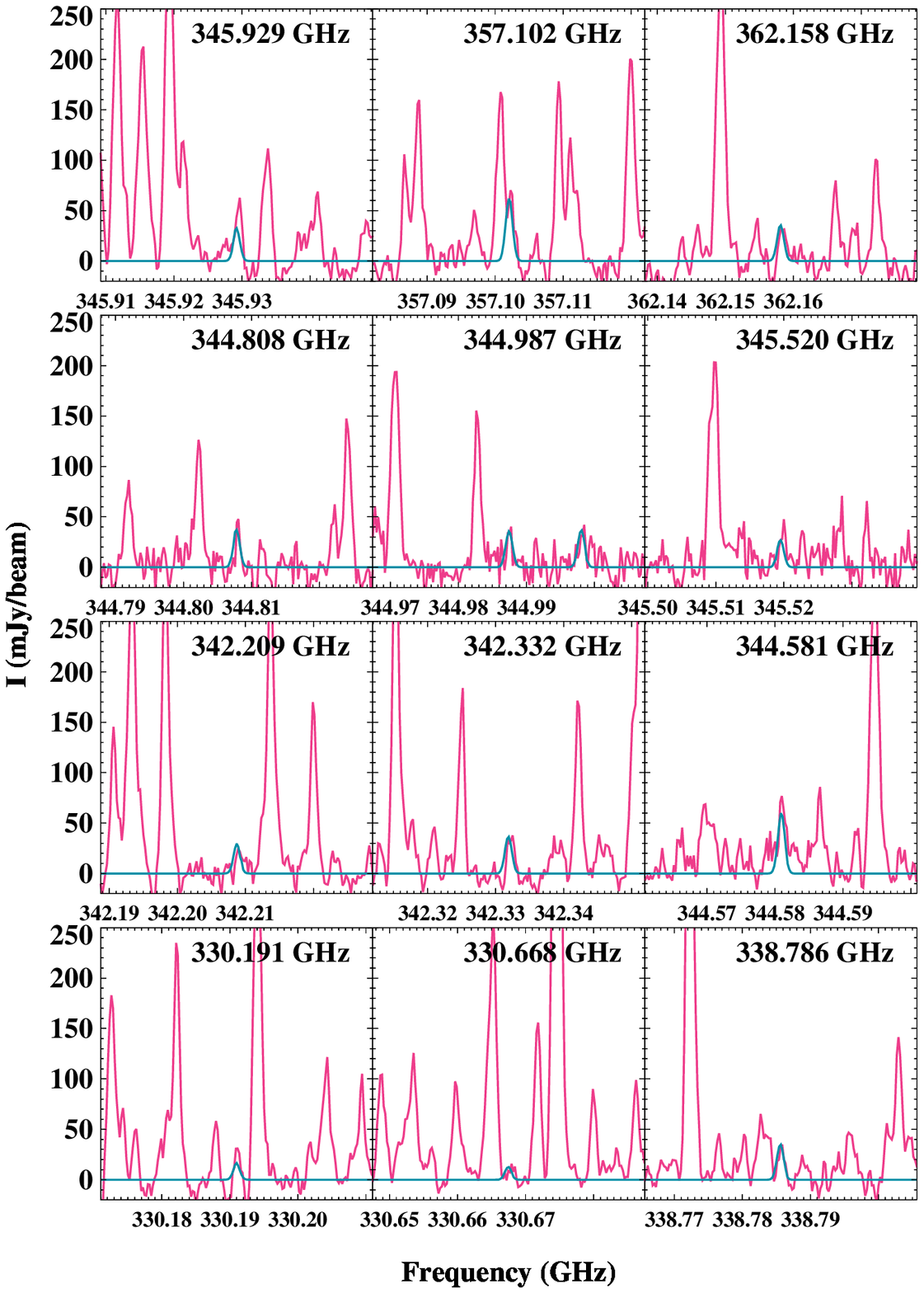}
 \caption{Twelve selected lines of $^{34}$SO$_{2}$. Idem Fig.~\ref{fgr:SO2v=0}.}
 \label{fgr:S-34-O2}
\end{figure*}

\begin{figure*}
 \centering
 \includegraphics[width=0.95\textwidth,height=0.2\textheight,keepaspectratio]{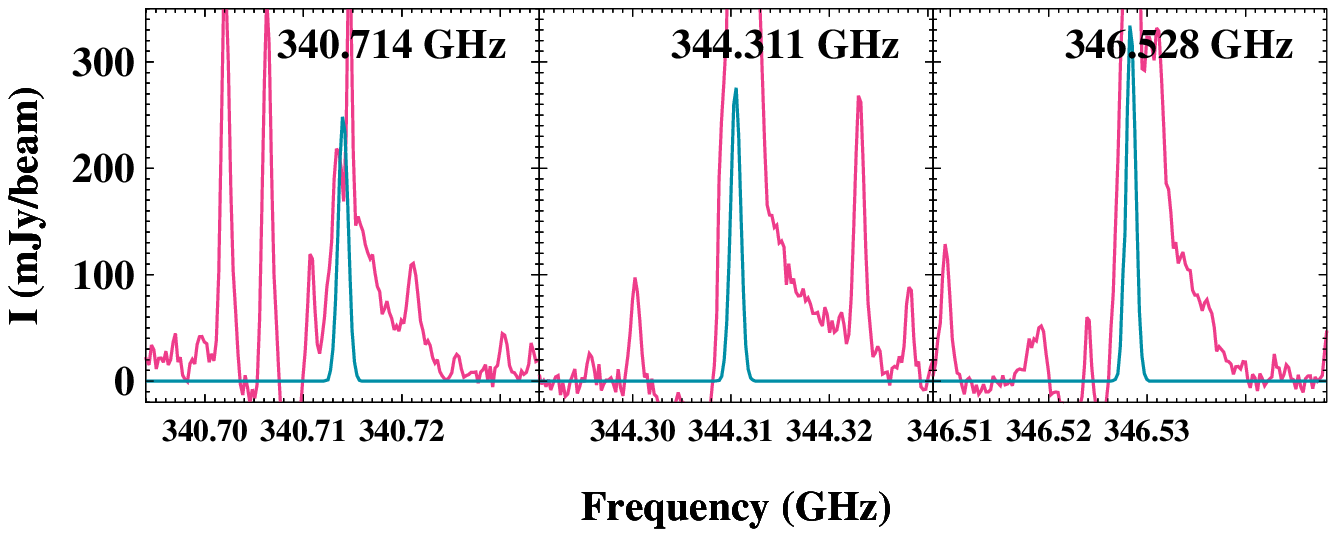}
 \caption{Three selected lines of SO in the $\varv=0$ state. Idem Fig.~\ref{fgr:SO2v=0}.}
 \label{fgr:SOv=0}
\end{figure*}

\begin{figure*}
 \centering
 \includegraphics[width=0.95\textwidth,height=0.2\textheight,keepaspectratio]{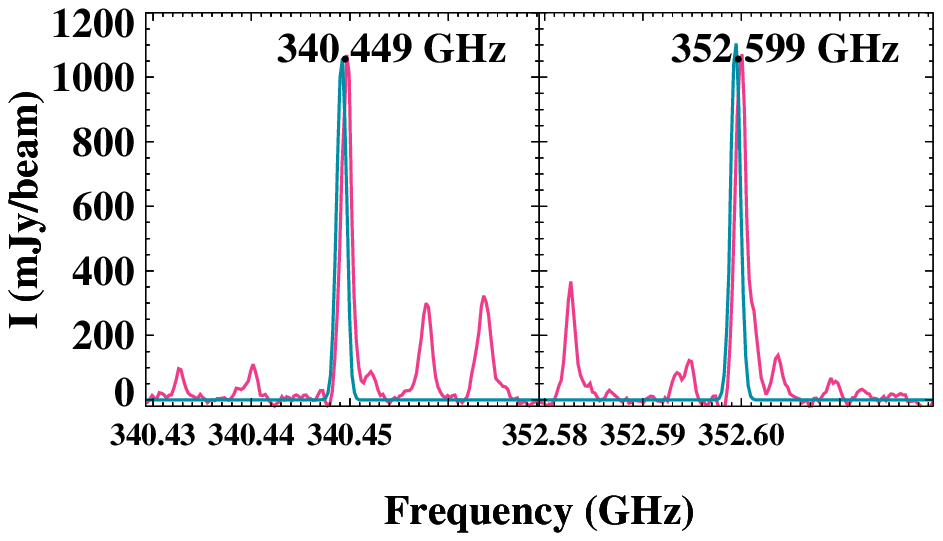}
 \caption{Two lines of OCS in the $\varv=0$ state. Idem Fig.~\ref{fgr:SO2v=0}.}
 \label{fgr:OCSv=0}
\end{figure*}

\begin{figure*}
 \centering
 \includegraphics[width=0.95\textwidth,height=0.4\textheight,keepaspectratio]{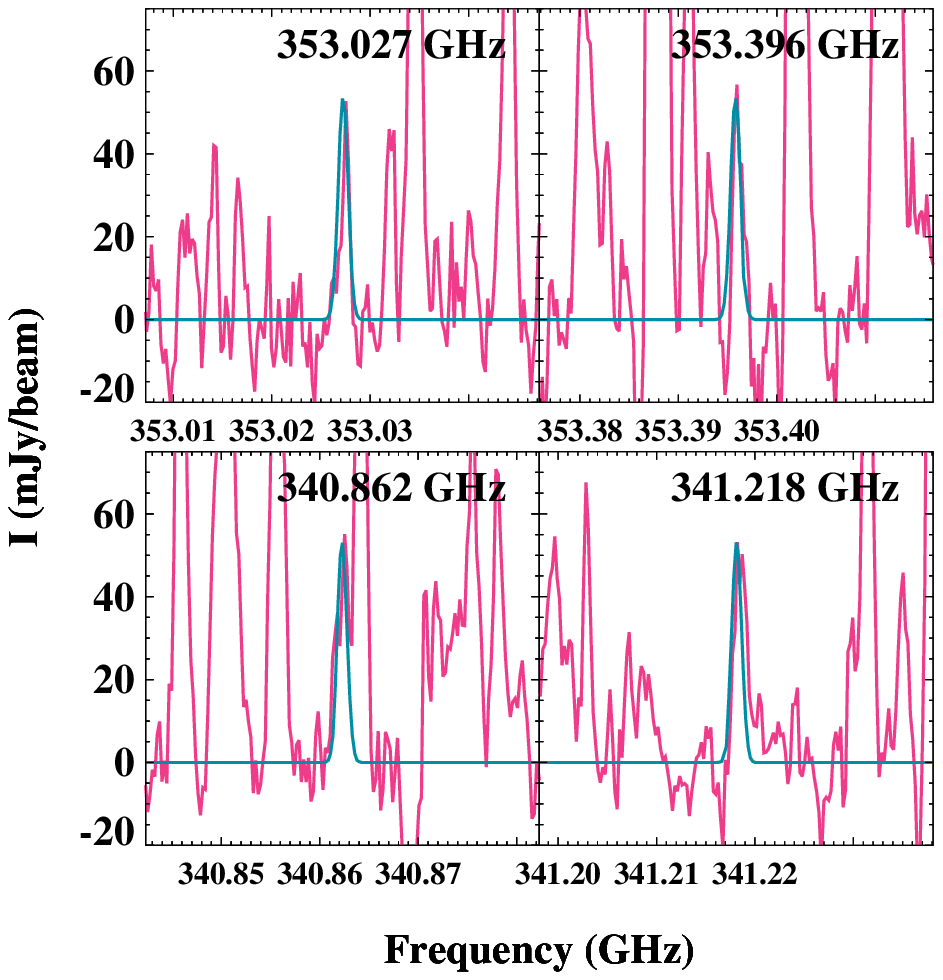}
 \caption{Four lines of OCS in the $\varv_{2}=1$ state. Idem Fig.~\ref{fgr:SO2v=0}.}
 \label{fgr:OCSv2=1}
\end{figure*}

\begin{figure*}
 \centering
 \includegraphics[width=0.95\textwidth,height=0.2\textheight,keepaspectratio]{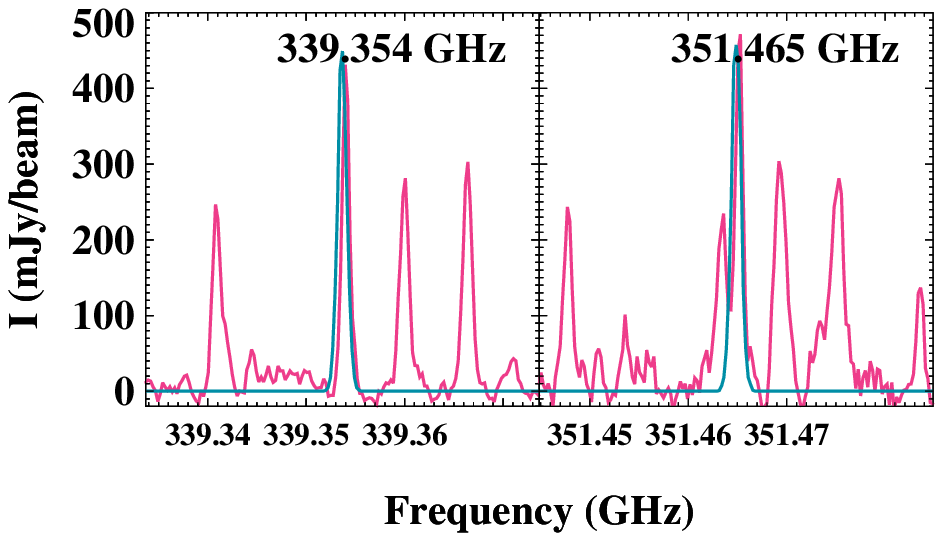}
 \caption{Two lines of O$^{13}$CS. Idem Fig.~\ref{fgr:SO2v=0}.}
 \label{fgr:OC-13-S}
\end{figure*}

\begin{figure*}
 \centering
 \includegraphics[width=0.95\textwidth,height=0.2\textheight,keepaspectratio]{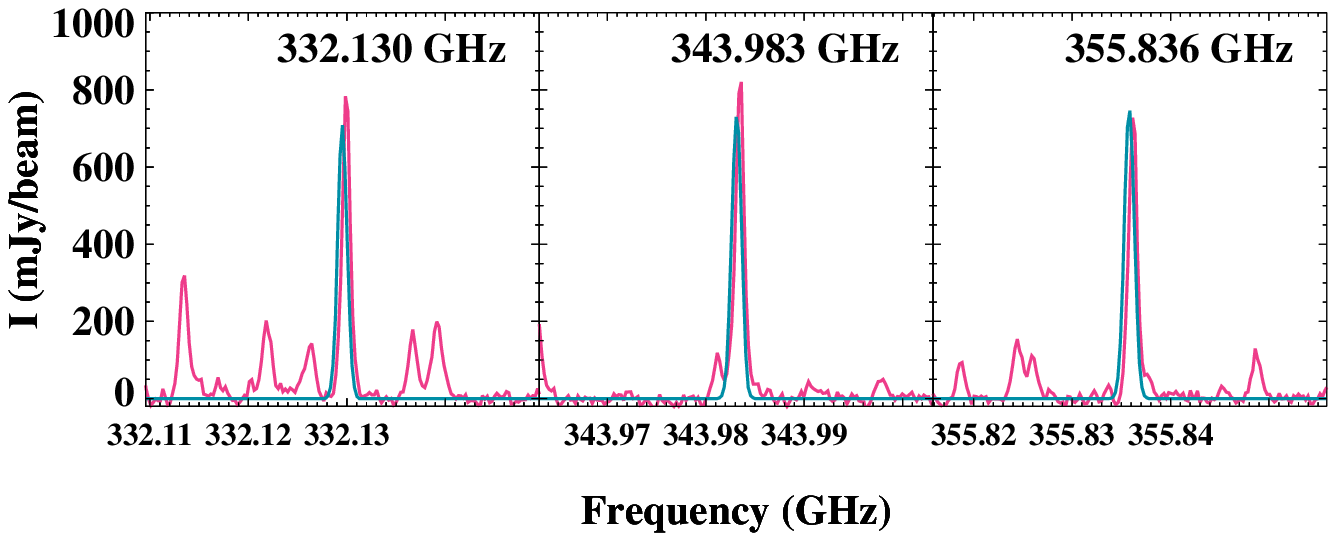}
 \caption{Three lines of OC$^{34}$S. Idem Fig.~\ref{fgr:SO2v=0}.}
 \label{fgr:OCS-34}
\end{figure*}

\begin{figure*}
 \centering
 \includegraphics[width=0.95\textwidth,height=0.2\textheight,keepaspectratio]{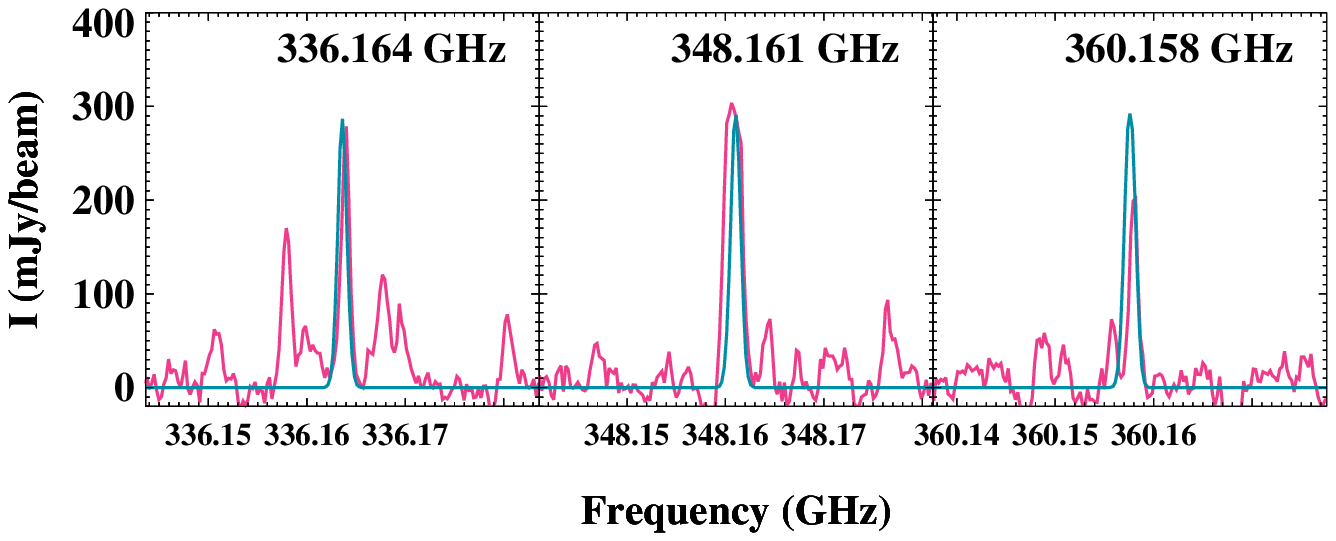}
 \caption{Three lines of OC$^{33}$S. Idem Fig.~\ref{fgr:SO2v=0}.}
 \label{fgr:OCS-33}
\end{figure*}

\begin{figure*}
 \centering
 \includegraphics[width=0.95\textwidth,height=0.2\textheight,keepaspectratio]{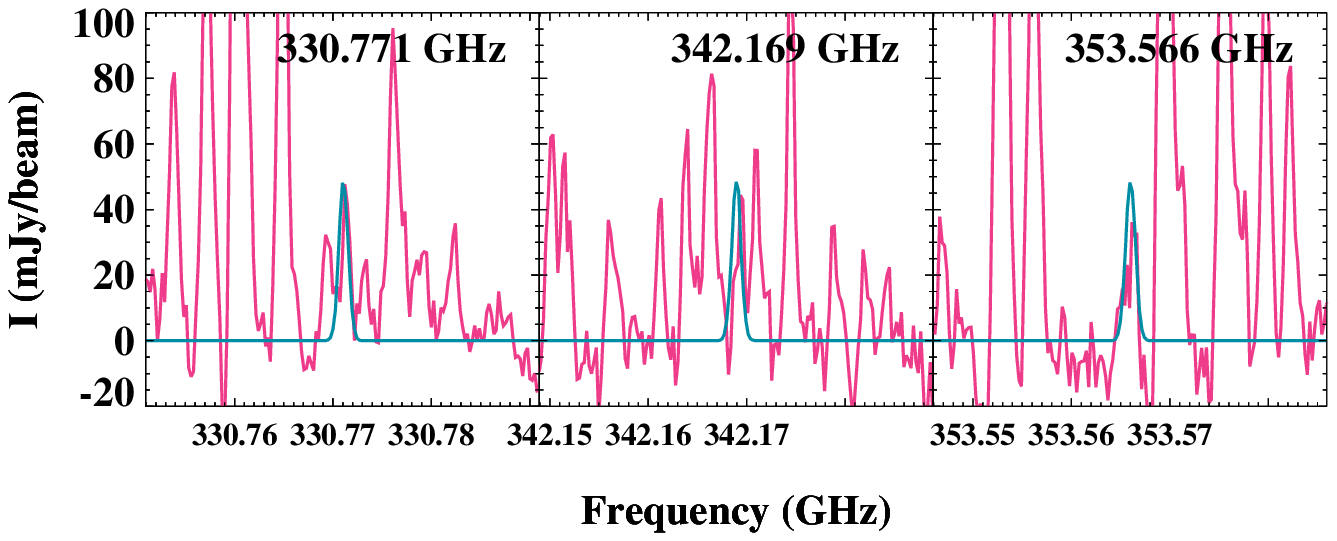}
 \caption{Three lines of $^{18}$OCS. Idem Fig.~\ref{fgr:SO2v=0}.}
 \label{fgr:O-18-CS}
\end{figure*}

\begin{figure*}
 \centering
 \includegraphics[width=0.95\textwidth,height=0.2\textheight,keepaspectratio]{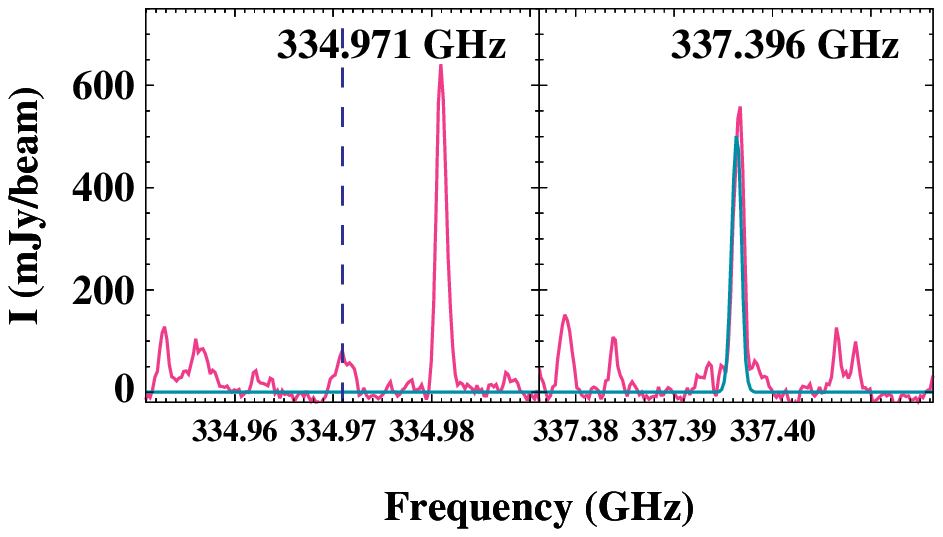}
 \caption{Two lines of C$^{34}$S in the $\varv=0,1$ state. Idem Fig.~\ref{fgr:SO2v=0}. The vertical blue dashed line in one of the panels indicates the position of the line that is too weak to generate an emission line under the assumed conditions.}
 \label{fgr:CS-34v=01}
\end{figure*}

\begin{figure*}
 \centering
 \includegraphics[width=0.95\textwidth,height=0.2\textheight,keepaspectratio]{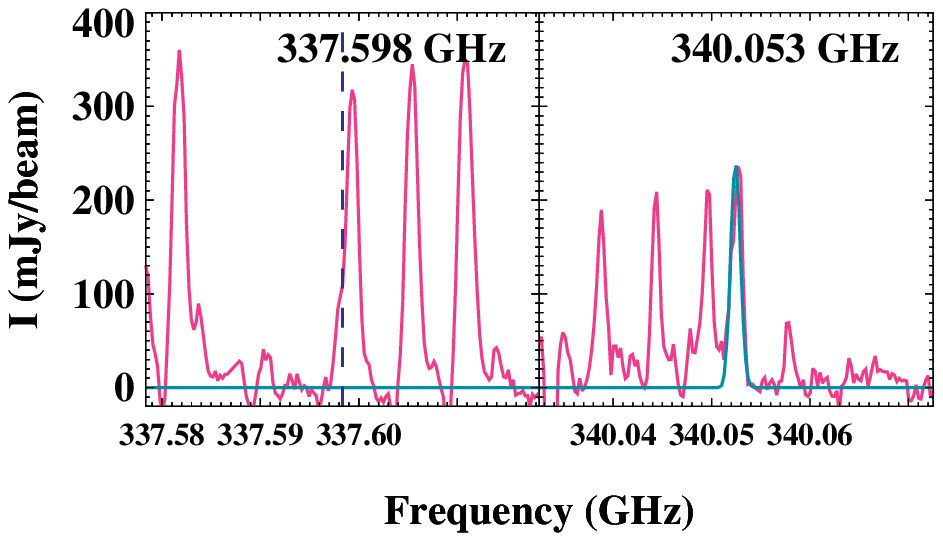}
 \caption{Two lines of C$^{33}$S in the $\varv=0,1$ state. Idem Fig.~\ref{fgr:SO2v=0}. The vertical blue dashed line in one of the panels indicates the position of the line that is too weak to generate an emission line under the assumed conditions.}
 \label{fgr:CS-33v=01}
\end{figure*}

\begin{figure*}
 \centering
 \includegraphics[width=0.95\textwidth,height=0.2\textheight,keepaspectratio]{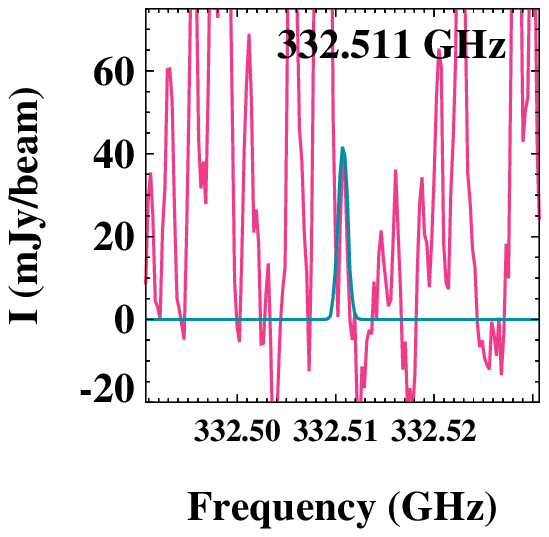}
 \caption{A line of C$^{36}$S. Idem Fig.~\ref{fgr:SO2v=0}.}
 \label{fgr:CS-36}
\end{figure*}

\begin{figure*}
 \centering
 \includegraphics[width=0.95\textwidth,height=0.6\textheight,keepaspectratio]{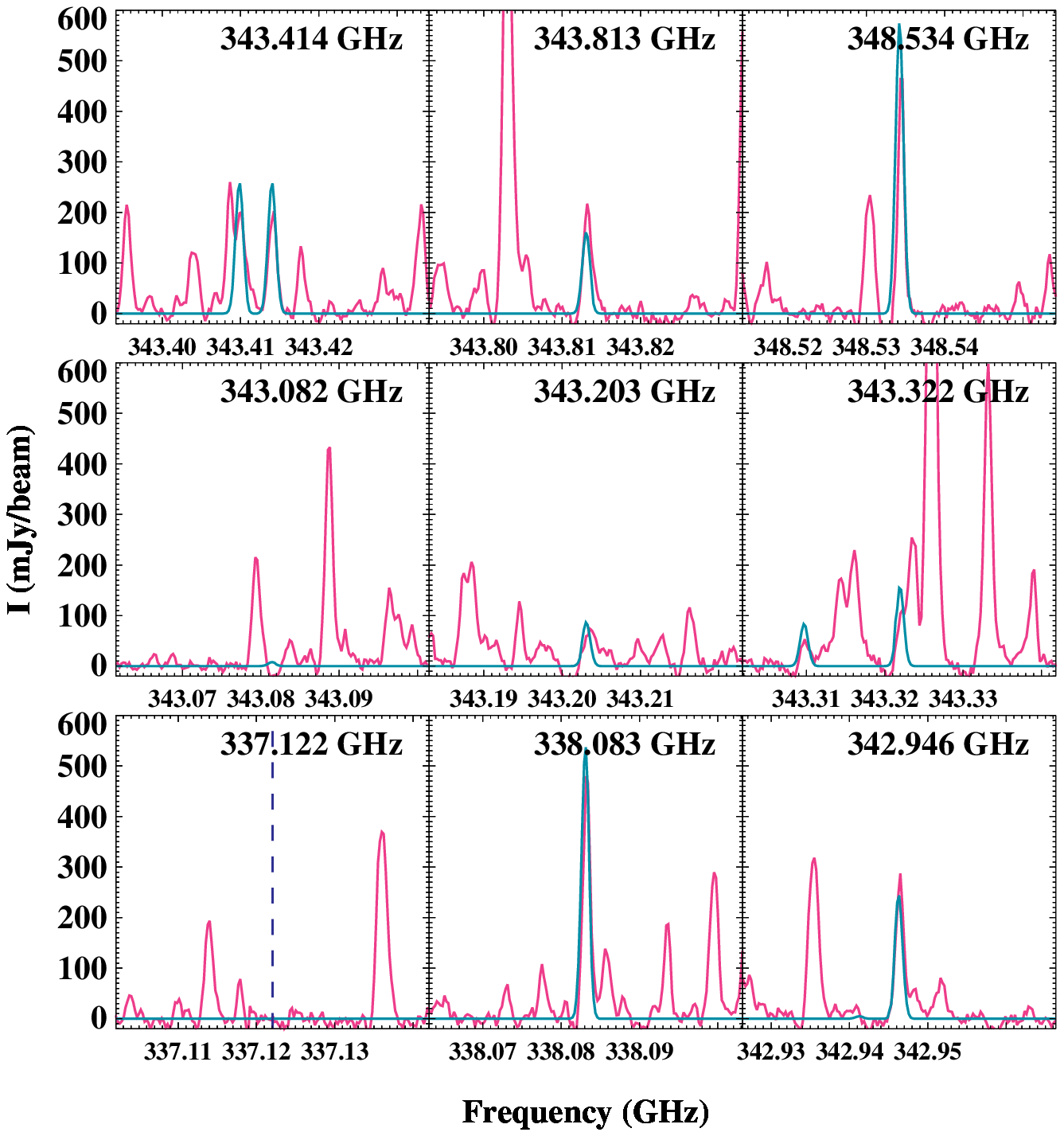}
 \caption{Nine selected lines of H$_{2}$CS. Idem Fig.~\ref{fgr:SO2v=0}. The vertical blue dashed line in one of the panels indicates the position of the line that is too weak to generate an emission line under the assumed conditions.}
 \label{fgr:H2CS}
\end{figure*}

\begin{figure*}
 \centering
 \includegraphics[width=0.95\textwidth,height=0.6\textheight,keepaspectratio]{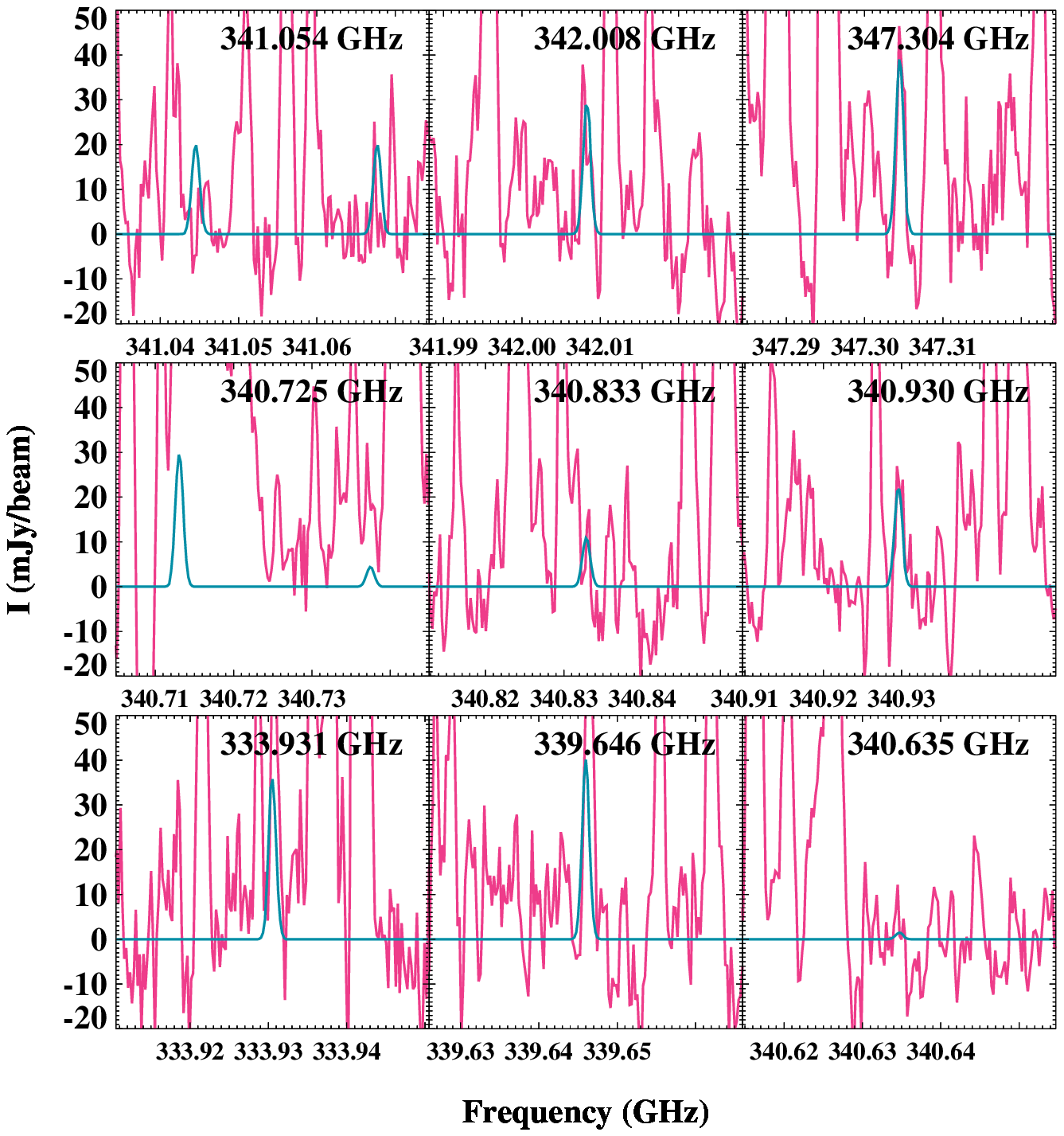}
 \caption{Nine selected lines of HDCS. Idem Fig.~\ref{fgr:SO2v=0}.}
 \label{fgr:HDCS}
\end{figure*}

\begin{figure*}
 \centering
 \includegraphics[width=0.95\textwidth,height=0.2\textheight,keepaspectratio]{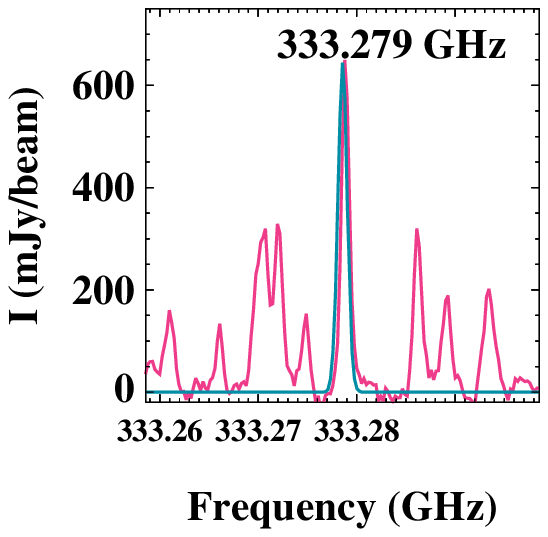}
 \caption{A selected line of HDS. Idem Fig.~\ref{fgr:SO2v=0}.}
 \label{fgr:HDS}
\end{figure*}

\begin{figure*}
 \centering
 \includegraphics[width=0.95\textwidth,height=0.2\textheight,keepaspectratio]{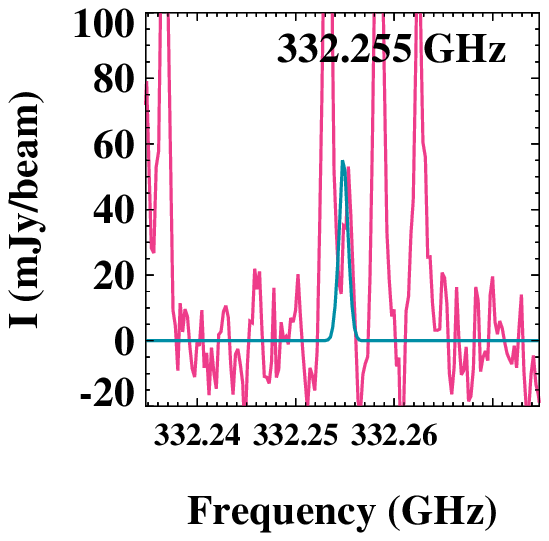}
 \caption{A selected line of HD$^{34}$S. Idem Fig.~\ref{fgr:SO2v=0}.}
 \label{fgr:HDS-34}
\end{figure*}

\begin{figure*}
 \centering
 \includegraphics[width=0.95\textwidth,height=0.8\textheight,keepaspectratio]{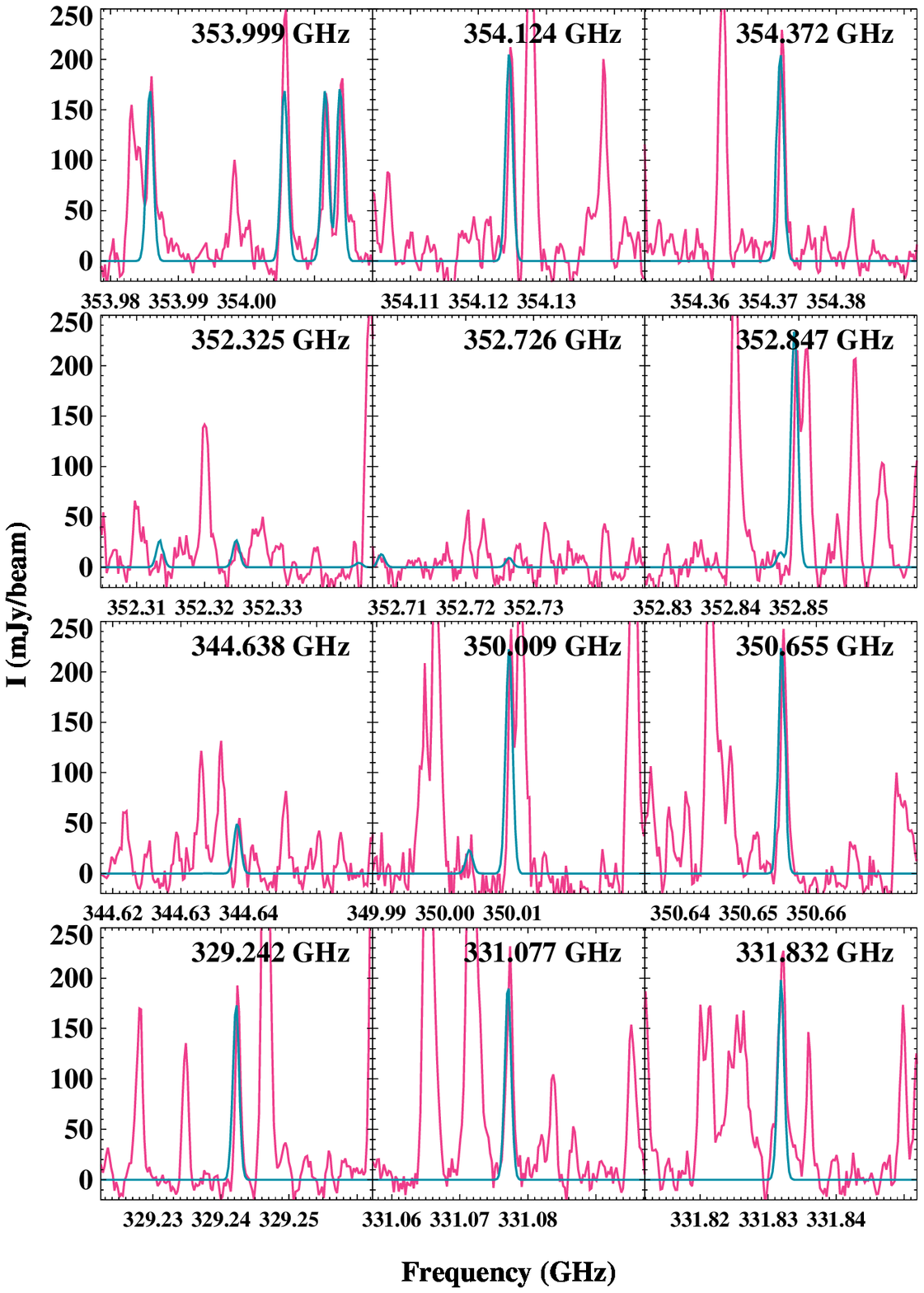}
 \caption{Twelve selected lines of CH$_{3}$SH in the $\varv=0-2$ state. Idem Fig.~\ref{fgr:SO2v=0}.}
 \label{fgr:CH3SHv=02}
\end{figure*}

\begin{figure*}
 \centering
 \includegraphics[width=0.95\textwidth,height=0.8\textheight,keepaspectratio]{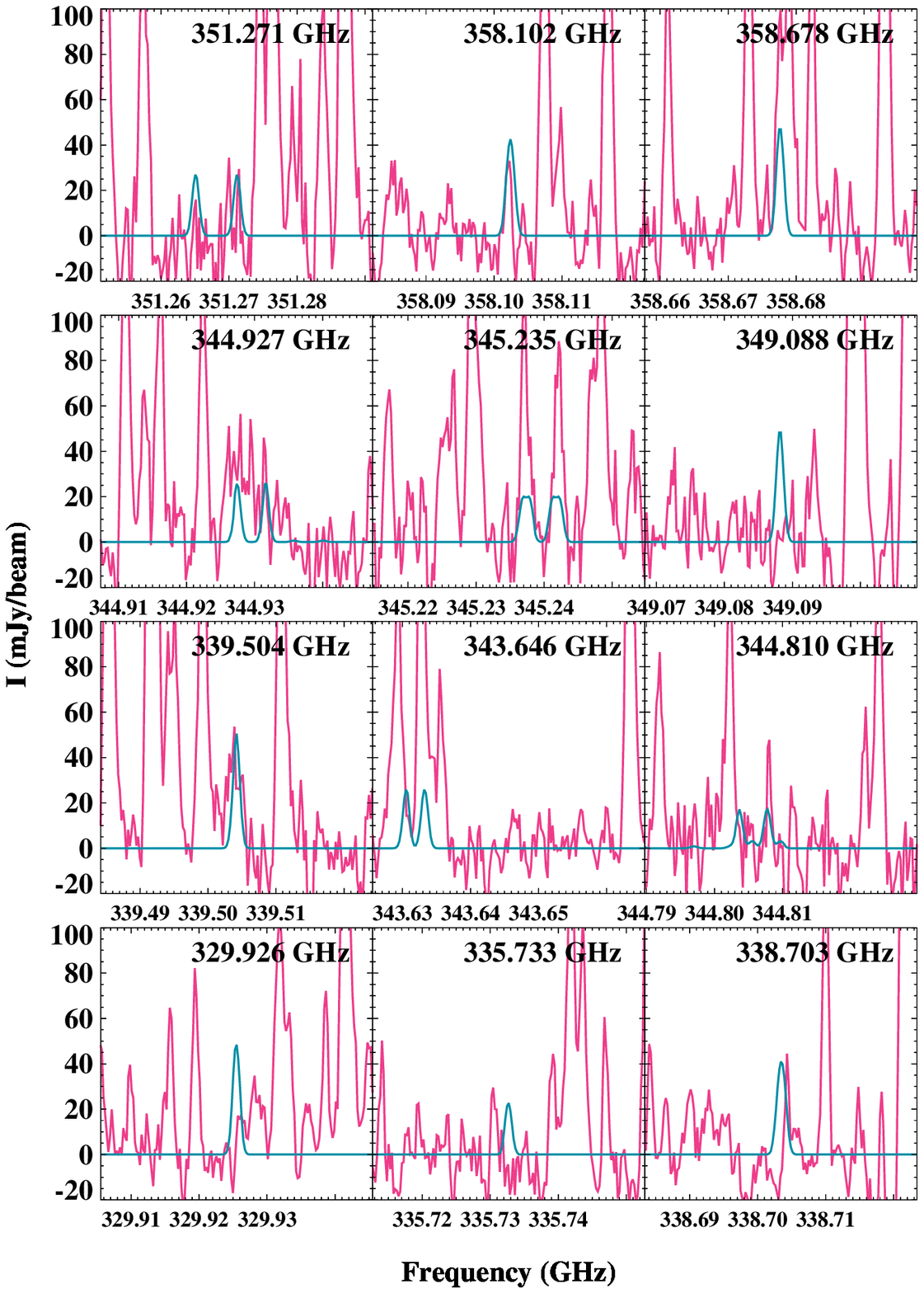}
 \caption{Twelve selected lines of gauche-C$_{2}$H$_{5}$SH. Idem Fig.~\ref{fgr:SO2v=0}.}
 \label{fgr:g-C2H5SH}
\end{figure*}

\begin{figure*}
 \centering
 \includegraphics[width=0.95\textwidth,height=0.8\textheight,keepaspectratio]{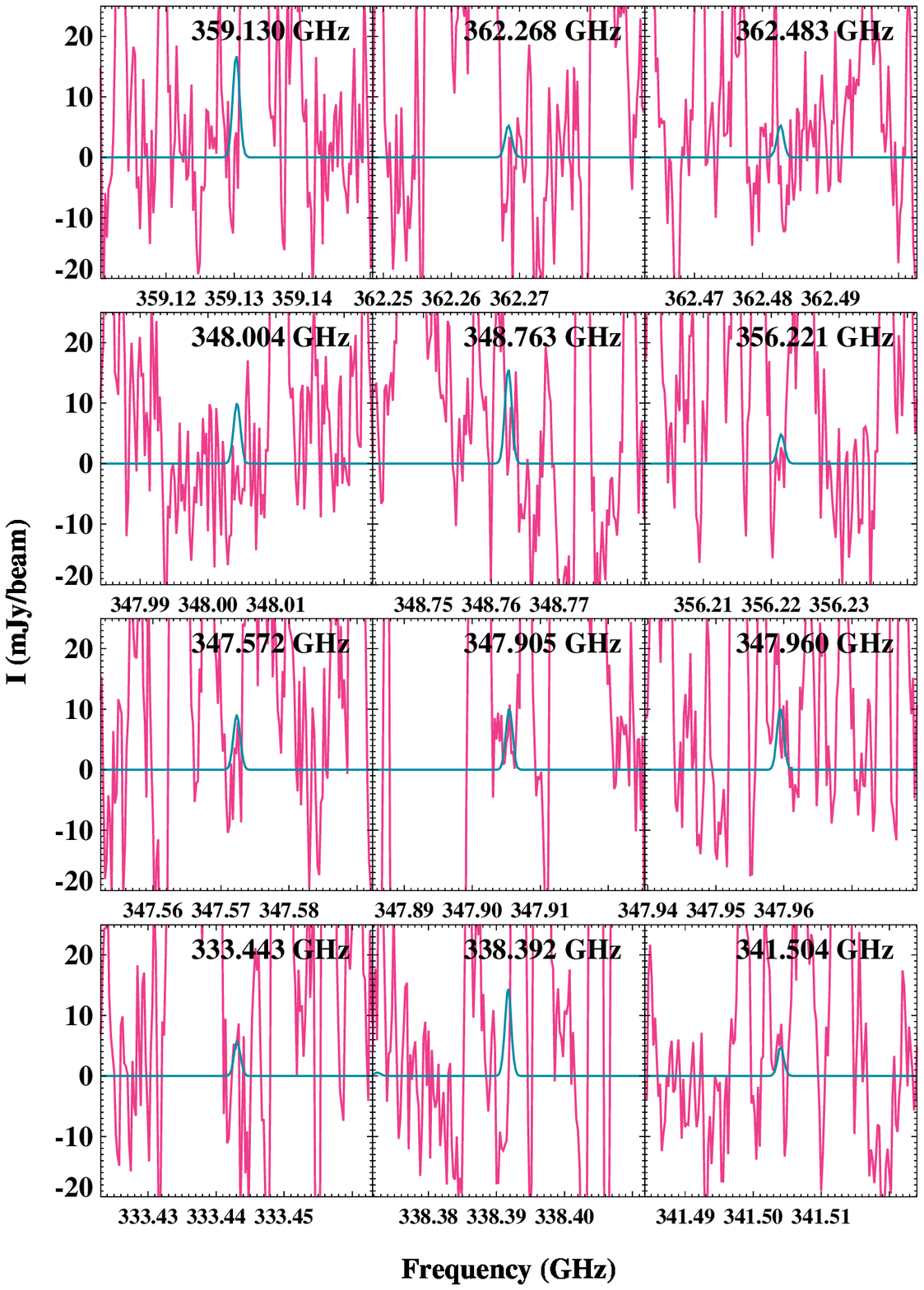}
 \caption{Twelve selected lines of anti-C$_{2}$H$_{5}$SH. Idem Fig.~\ref{fgr:SO2v=0}.}
 \label{fgr:a-C2H5SH}
\end{figure*}

\begin{figure*}
 \centering
 \includegraphics[width=0.95\textwidth,height=0.8\textheight,keepaspectratio]{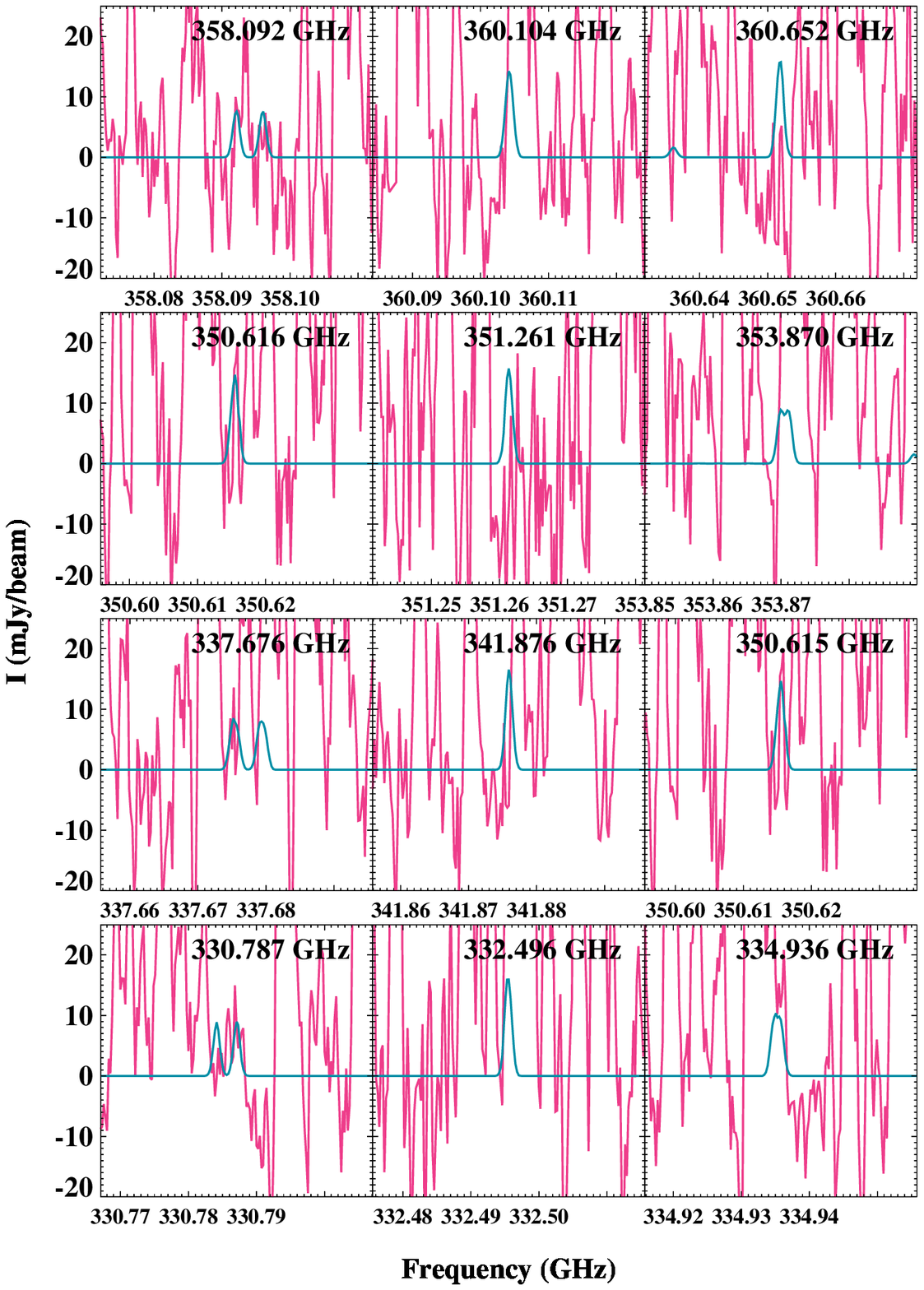}
 \caption{Twelve selected lines of gauche-C$_{2}$H$_{5}^{34}$SH. Idem Fig.~\ref{fgr:SO2v=0}.}
 \label{fgr:g-C2H5S-34-H}
\end{figure*}

\begin{figure*}
 \centering
 \includegraphics[width=0.95\textwidth,height=0.2\textheight,keepaspectratio]{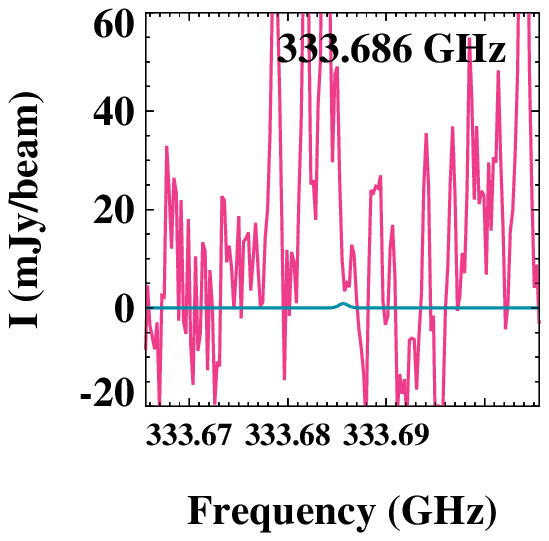}
 \caption{A line of S$_{2}$. Idem Fig.~\ref{fgr:SO2v=0}.}
 \label{fgr:S2}
\end{figure*}

\begin{figure*}
 \centering
 \includegraphics[width=0.95\textwidth,height=0.8\textheight,keepaspectratio]{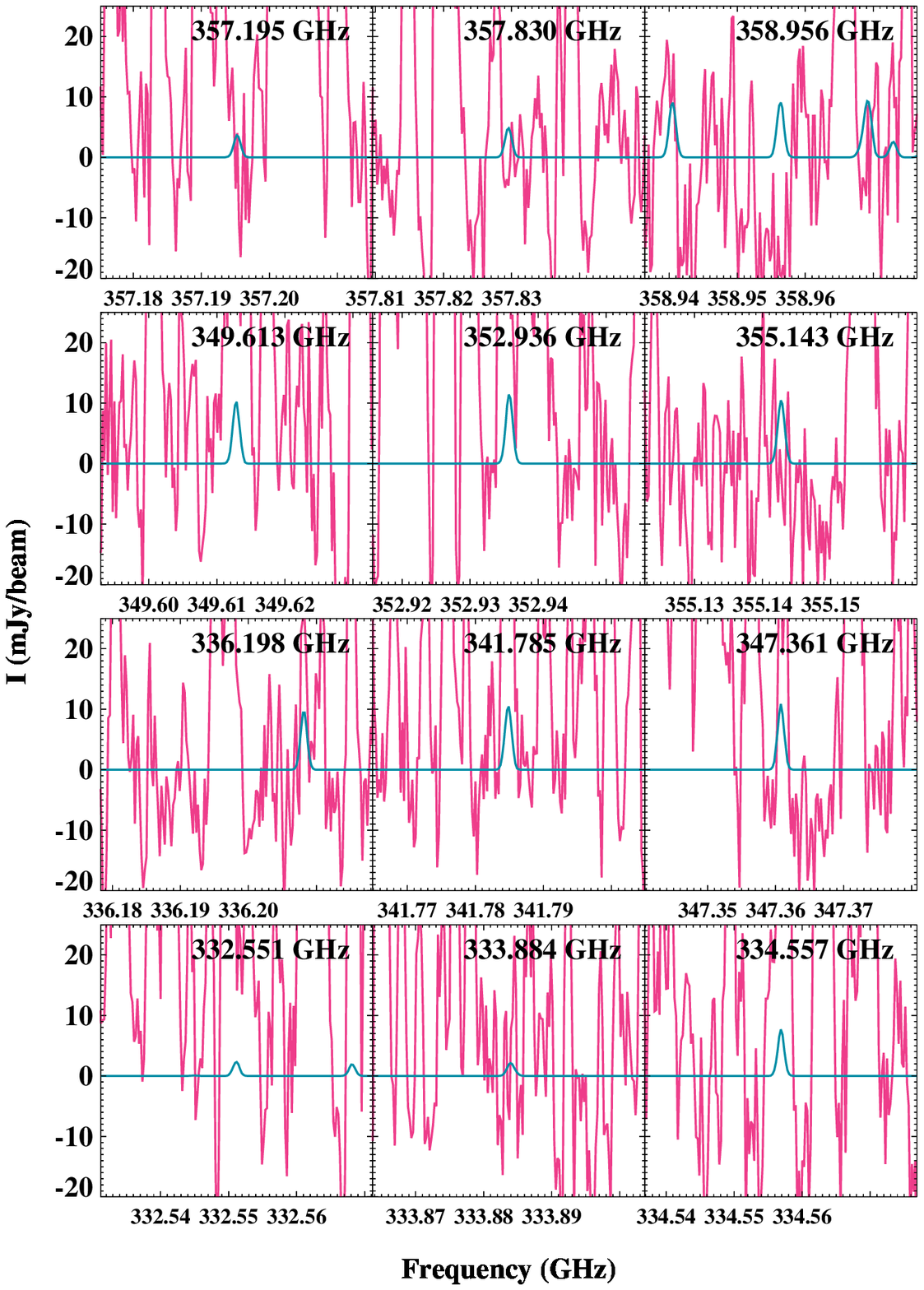}
 \caption{Twelve selected lines of S$_{3}$. Idem Fig.~\ref{fgr:SO2v=0}.}
 \label{fgr:S3}
\end{figure*}

\begin{figure*}
 \centering
 \includegraphics[width=0.95\textwidth,height=0.8\textheight,keepaspectratio]{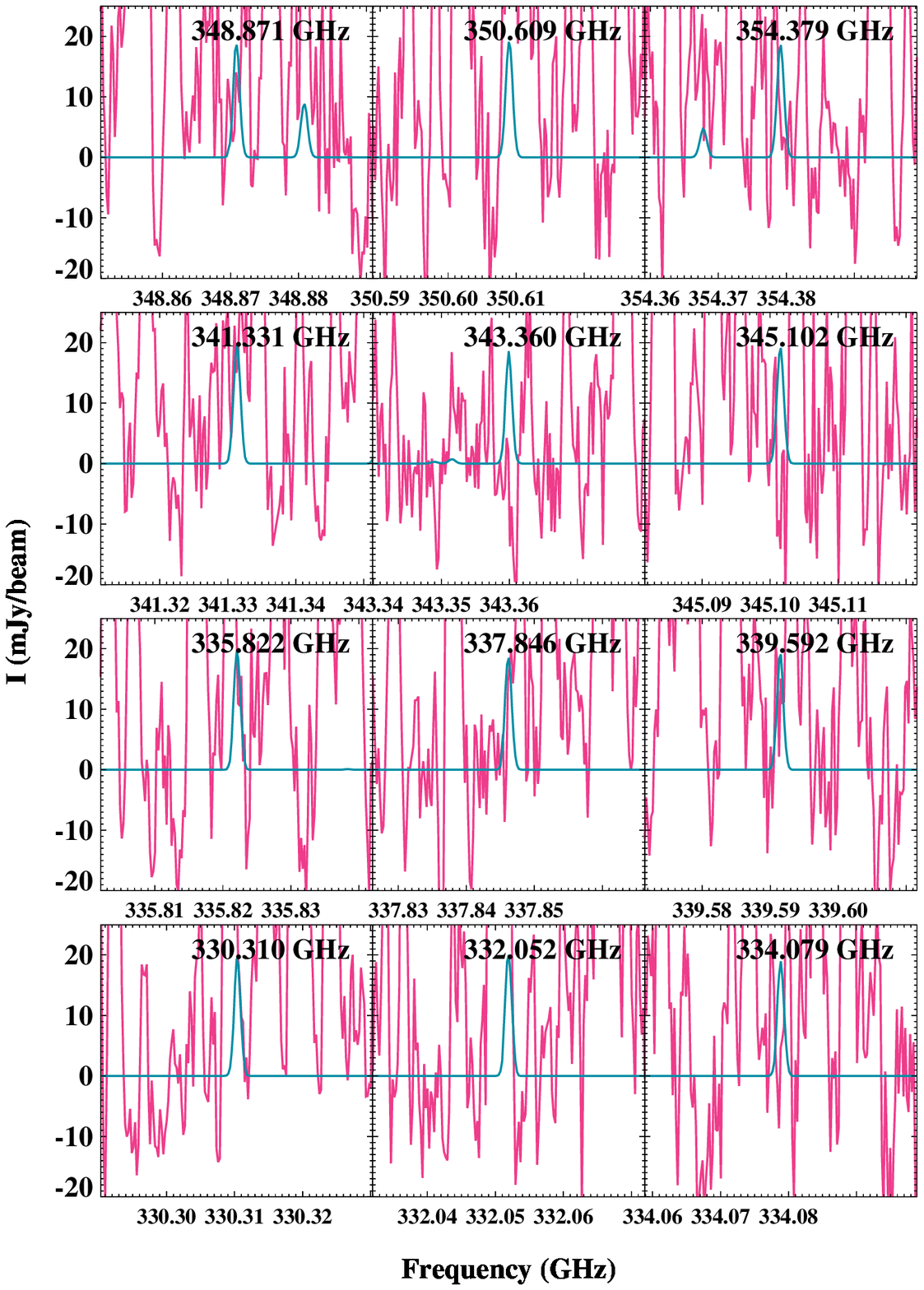}
 \caption{Twelve selected lines of S$_{4}$. Idem Fig.~\ref{fgr:SO2v=0}.}
 \label{fgr:S4}
\end{figure*}

\begin{figure*}
 \centering
 \includegraphics[width=0.95\textwidth,height=0.8\textheight,keepaspectratio]{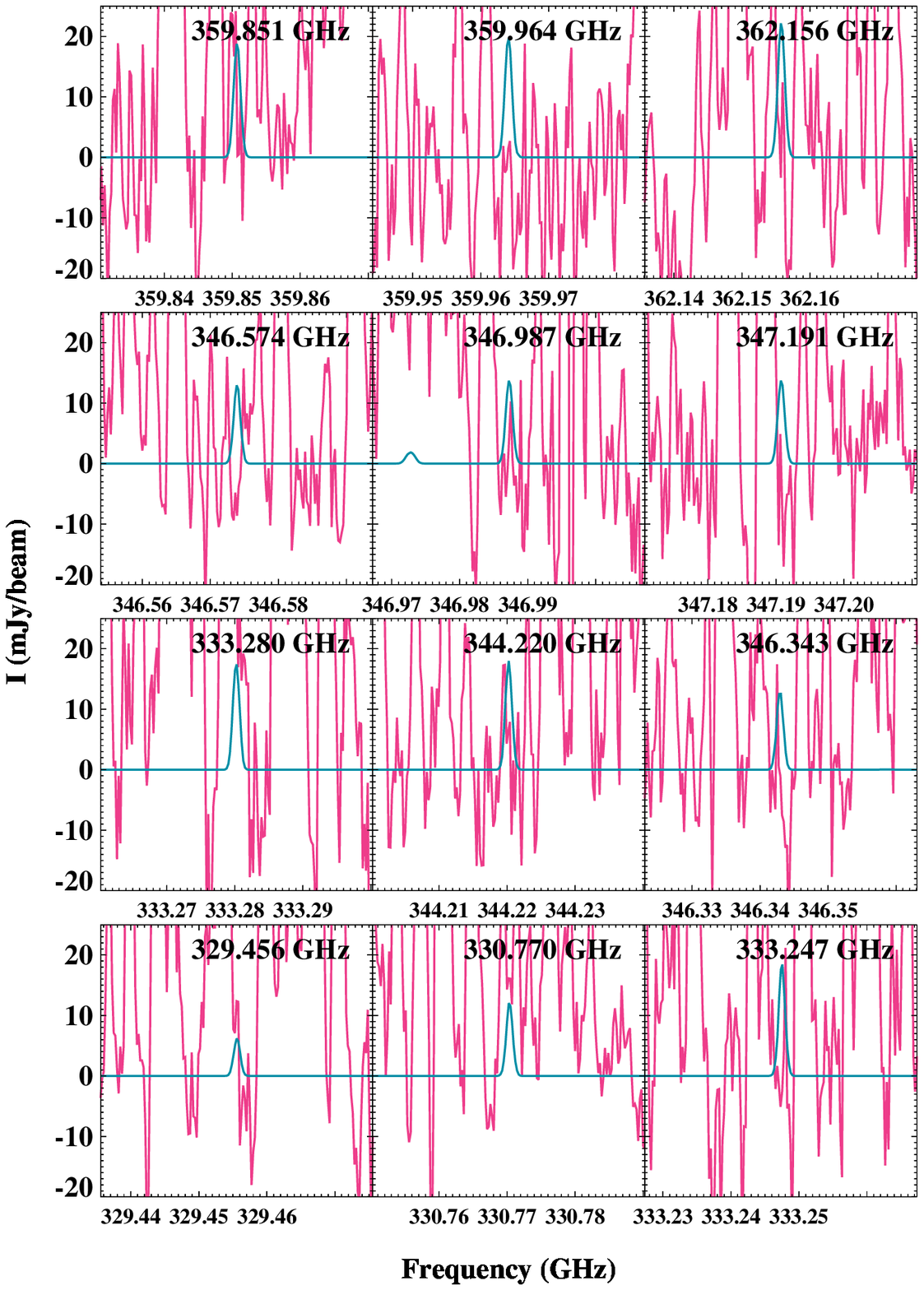}
 \caption{Twelve selected lines of HS$_{2}$. Idem Fig.~\ref{fgr:SO2v=0}.}
 \label{fgr:HS2}
\end{figure*}

\begin{figure*}
 \centering
 \includegraphics[width=0.95\textwidth,height=0.6\textheight,keepaspectratio]{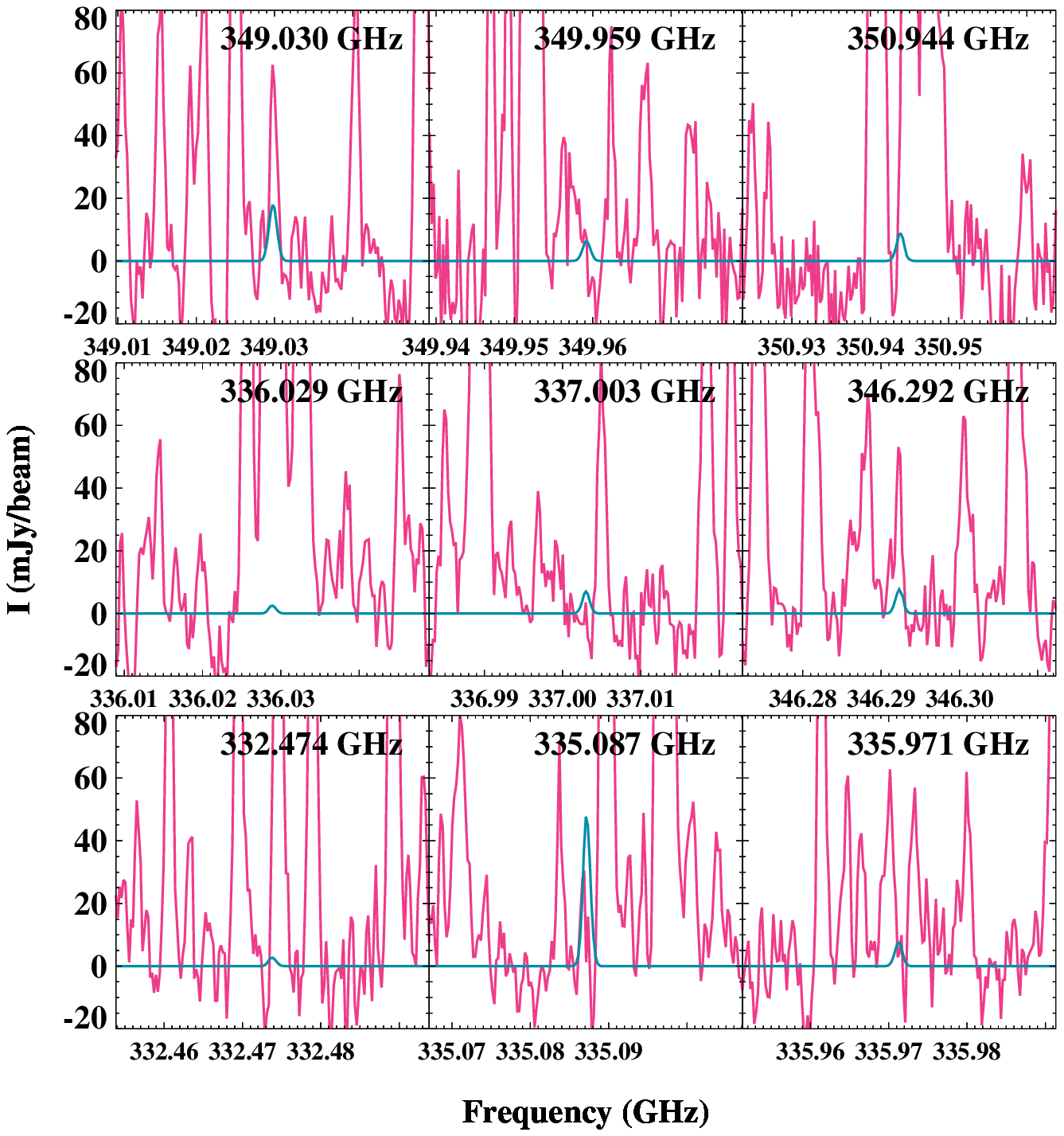}
 \caption{Nine selected lines of H$_{2}$S$_{2}$. Idem Fig.~\ref{fgr:SO2v=0}.}
 \label{fgr:H2S2}
\end{figure*}

\begin{figure*}
 \centering
 \includegraphics[width=0.95\textwidth,height=0.8\textheight,keepaspectratio]{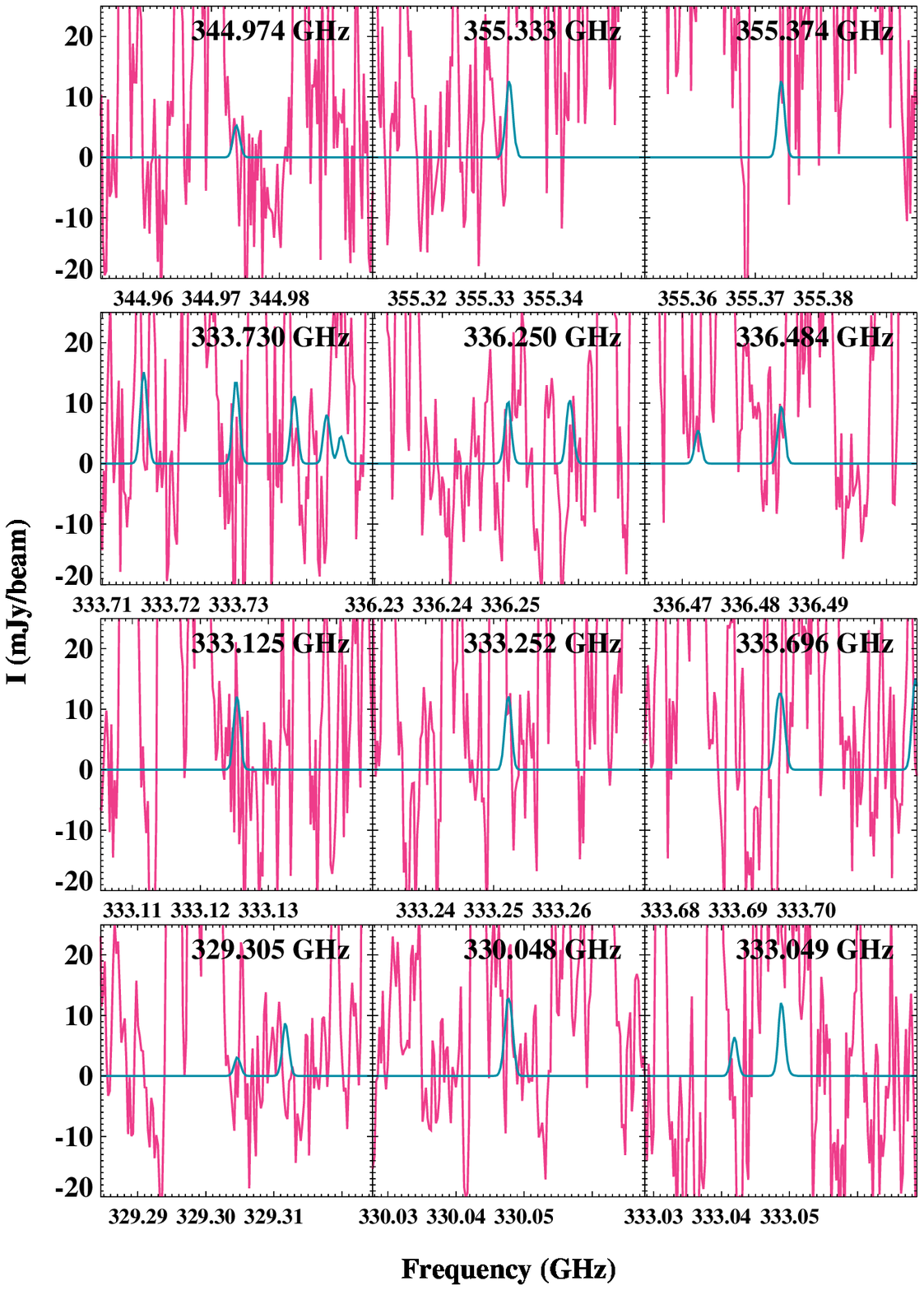}
 \caption{Twelve selected lines of S$_{2}$O in the $\varv=0$ state. Idem Fig.~\ref{fgr:SO2v=0}.}
 \label{fgr:S2Ov=0}
\end{figure*}

\begin{figure*}
 \centering
 \includegraphics[width=0.95\textwidth,height=0.8\textheight,keepaspectratio]{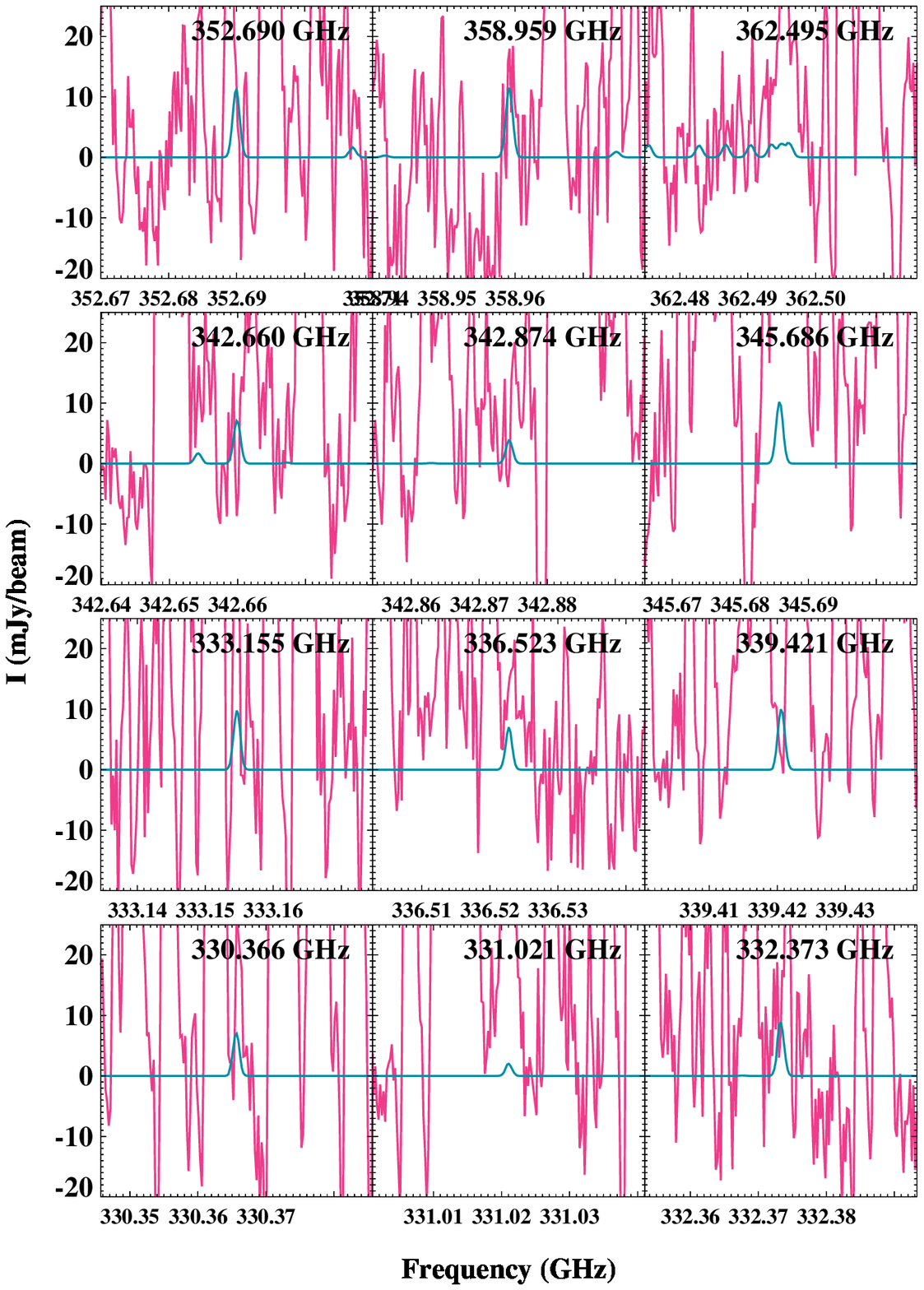}
 \caption{Twelve selected lines of cis-S$_{2}$O$_{2}$. Idem Fig.~\ref{fgr:SO2v=0}.}
 \label{fgr:cis-S2O2}
\end{figure*}

\begin{figure*}
 \centering
 \includegraphics[width=0.95\textwidth,height=0.2\textheight,keepaspectratio]{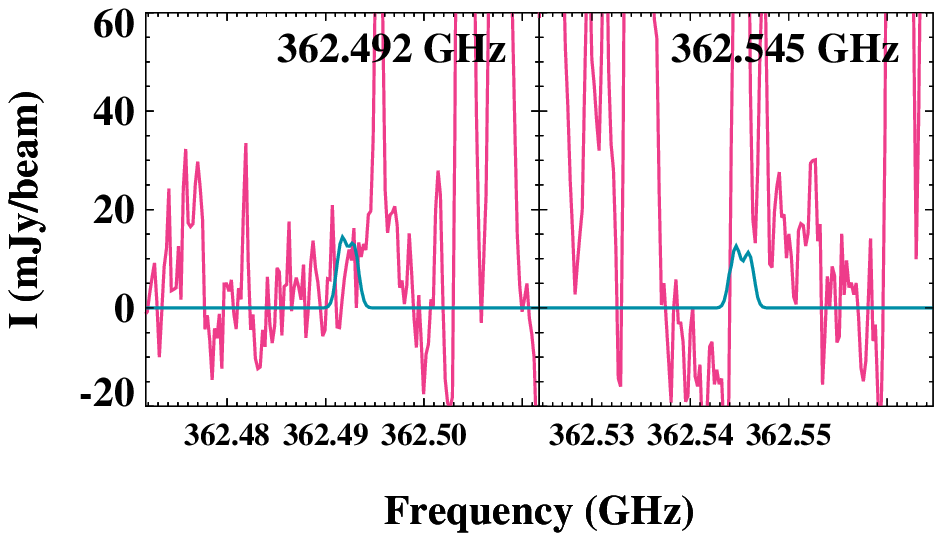}
 \caption{Two selected lines of HCS. Idem Fig.~\ref{fgr:SO2v=0}.}
 \label{fgr:HCS}
\end{figure*}

\begin{figure*}
 \centering
 \includegraphics[width=0.95\textwidth,height=0.8\textheight,keepaspectratio]{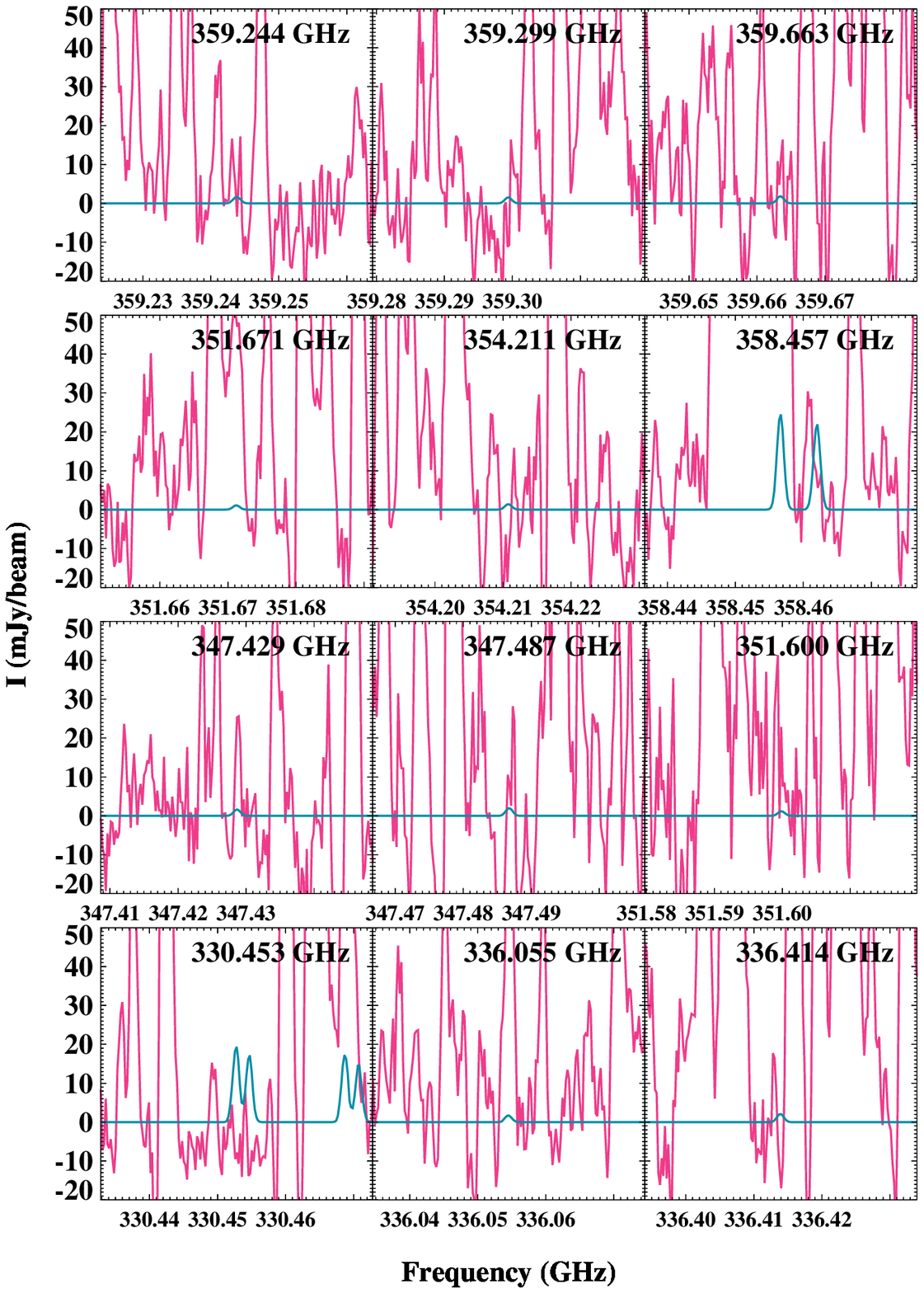}
 \caption{Twelve selected lines of HSC. Idem Fig.~\ref{fgr:SO2v=0}.}
 \label{fgr:HSC}
\end{figure*}

\begin{figure*}
 \centering
 \includegraphics[width=0.95\textwidth,height=0.2\textheight,keepaspectratio]{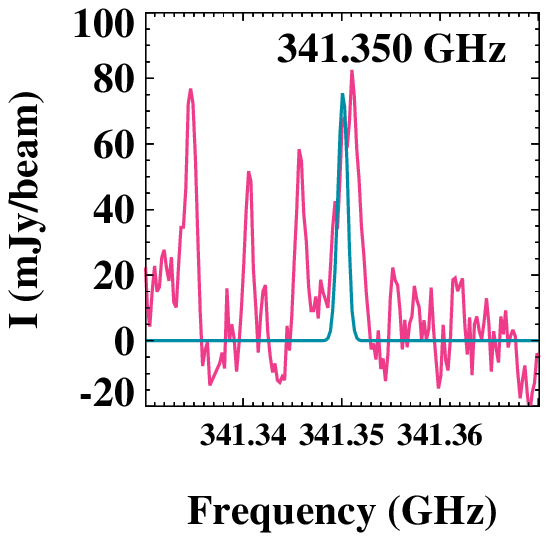}
 \caption{A line of HCS$^{+}$. Idem Fig.~\ref{fgr:SO2v=0}.}
 \label{fgr:HCS+}
\end{figure*}

\begin{figure*}
 \centering
 \includegraphics[width=0.95\textwidth,height=0.2\textheight,keepaspectratio]{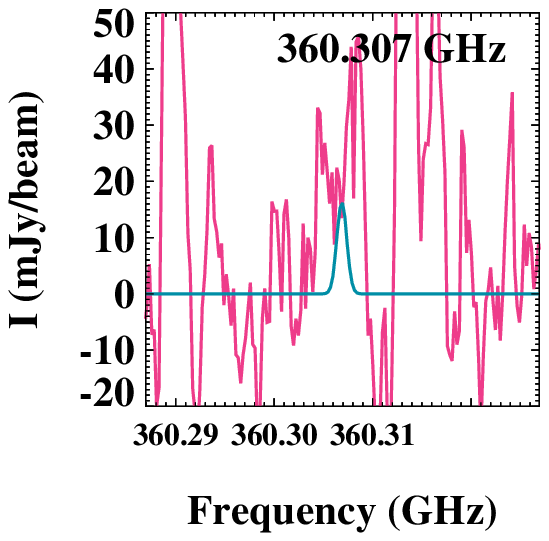}
 \caption{A line of DCS$^{+}$. Idem Fig.~\ref{fgr:SO2v=0}.}
 \label{fgr:DCS+}
\end{figure*}

\begin{figure*}
 \centering
 \includegraphics[width=0.95\textwidth,height=0.2\textheight,keepaspectratio]{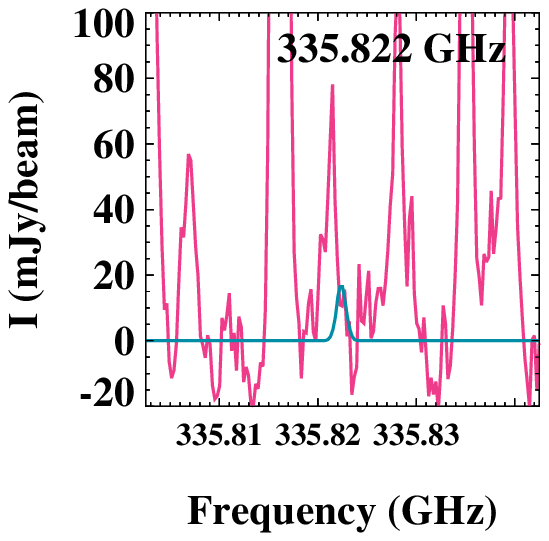}
 \caption{A line of HC$^{34}$S$^{+}$. Idem Fig.~\ref{fgr:SO2v=0}.}
 \label{fgr:HCS-34+}
\end{figure*}

\begin{figure*}
 \centering
 \includegraphics[width=0.95\textwidth,height=0.8\textheight,keepaspectratio]{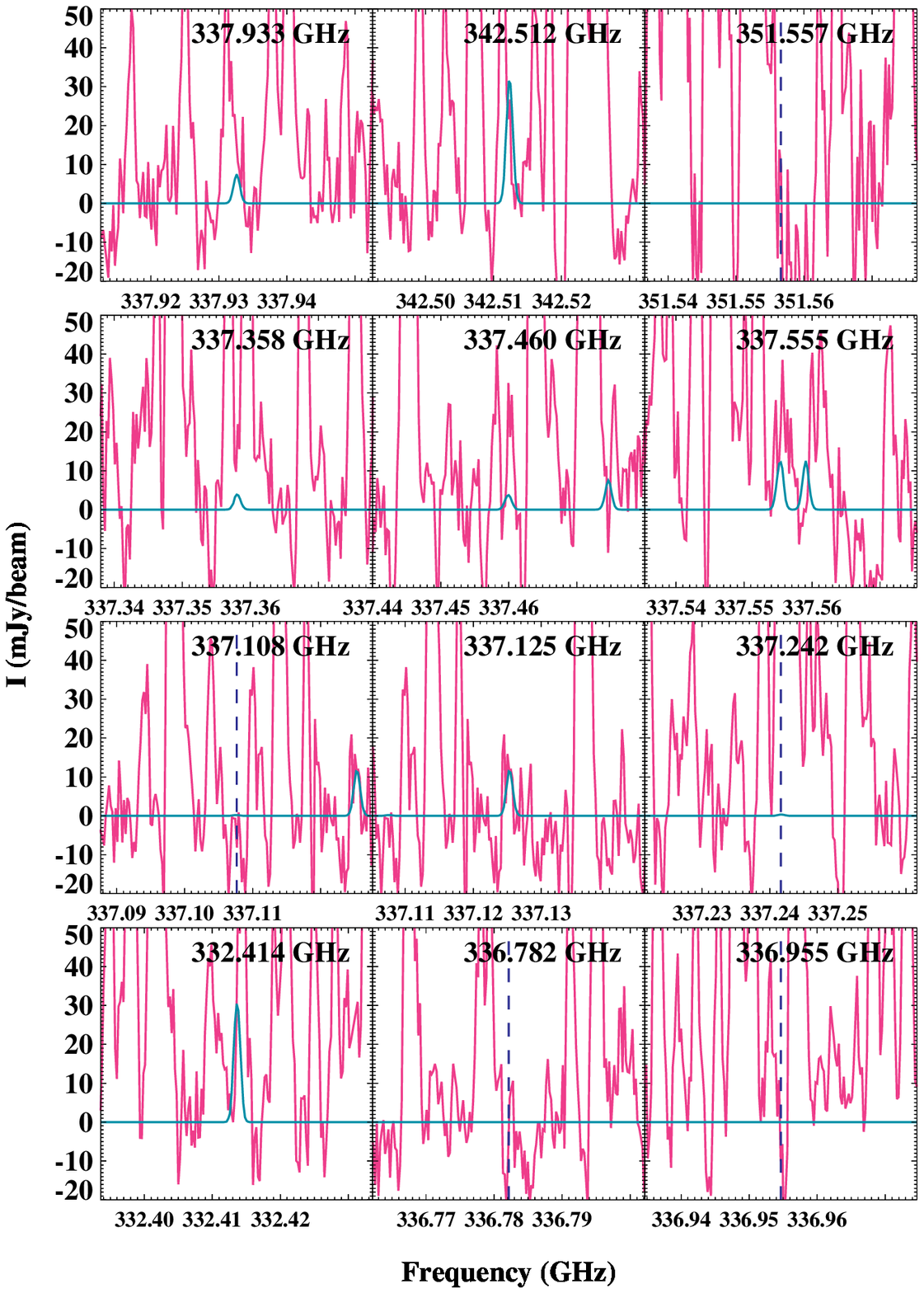}
 \caption{Fourteen selected lines of H$_{2}$C$^{34}$S. Idem Fig.~\ref{fgr:SO2v=0}. The vertical blue dashed line in one of the panels indicates the position of the line that is too weak to generate an emission line under the assumed conditions.}
 \label{fgr:H2CS-34}
\end{figure*}

\begin{figure*}
 \centering
 \includegraphics[width=0.95\textwidth,height=0.8\textheight,keepaspectratio]{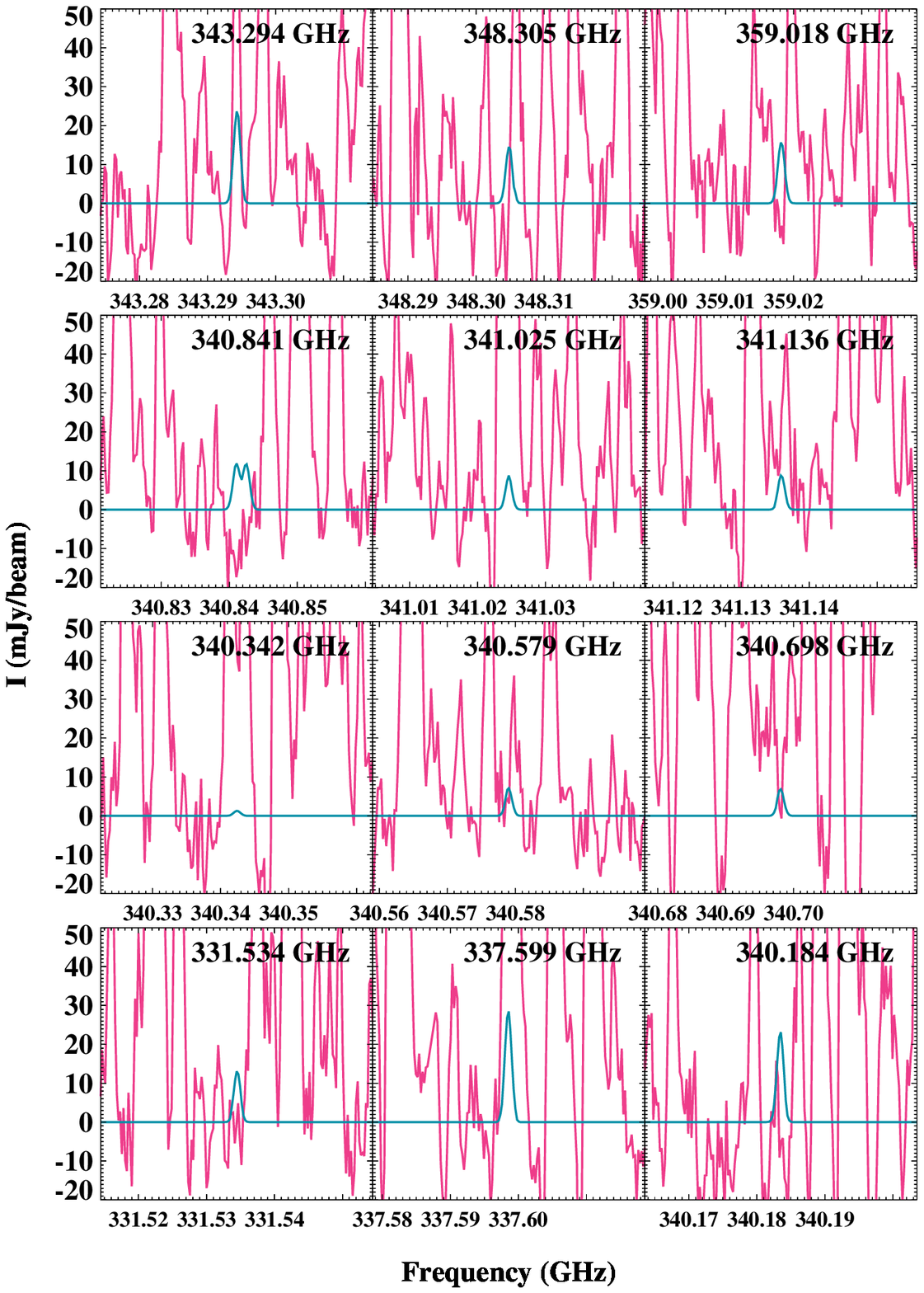}
 \caption{Twelve selected lines of D$_{2}$CS. Idem Fig.~\ref{fgr:SO2v=0}.}
 \label{fgr:D2CS}
\end{figure*}

\begin{figure*}
 \centering
 \includegraphics[width=0.95\textwidth,height=0.6\textheight,keepaspectratio]{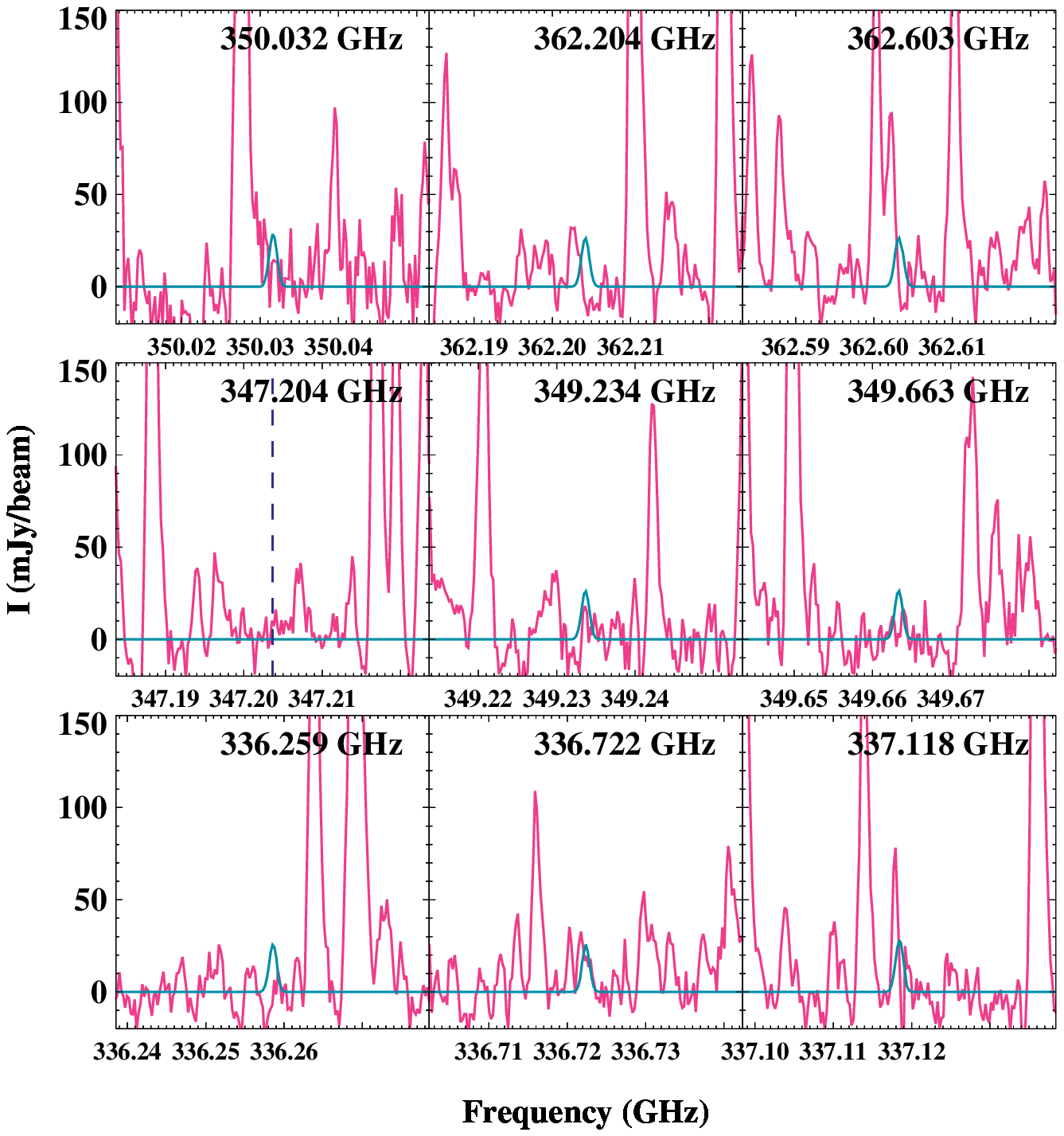}
 \caption{Nine selected lines of CCS. Idem Fig.~\ref{fgr:SO2v=0}. The vertical blue dashed line in one of the panels indicates the position of the line that is too weak to generate an emission line under the assumed conditions.}
 \label{fgr:CCS}
\end{figure*}

\begin{figure*}
 \centering
 \includegraphics[width=0.95\textwidth,height=0.8\textheight,keepaspectratio]{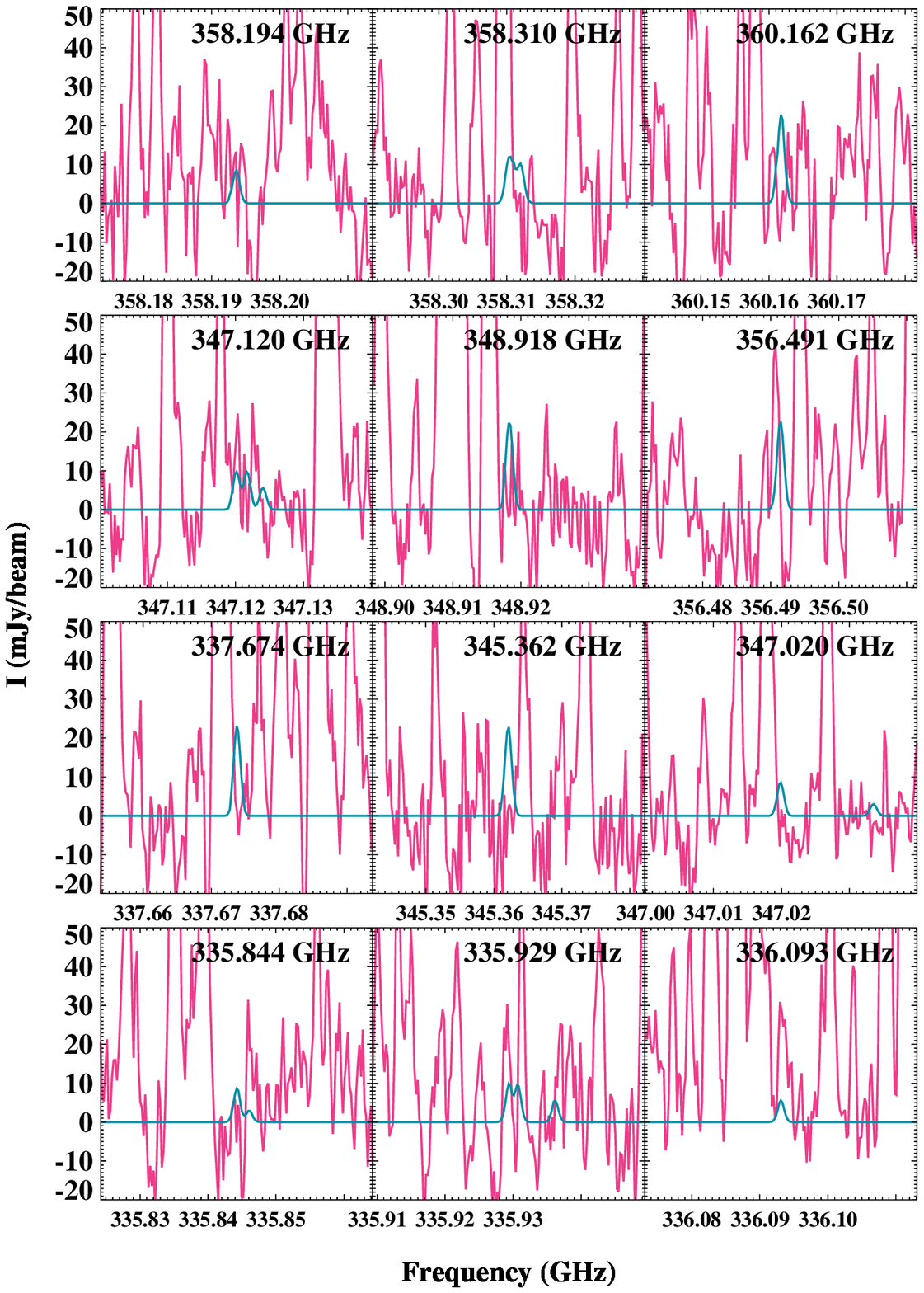}
 \caption{Twelve selected lines of H$_{2}$C$_{2}$S. Idem Fig.~\ref{fgr:SO2v=0}.}
 \label{fgr:H2C2S}
\end{figure*}

\begin{figure*}
 \centering
 \includegraphics[width=0.95\textwidth,height=0.8\textheight,keepaspectratio]{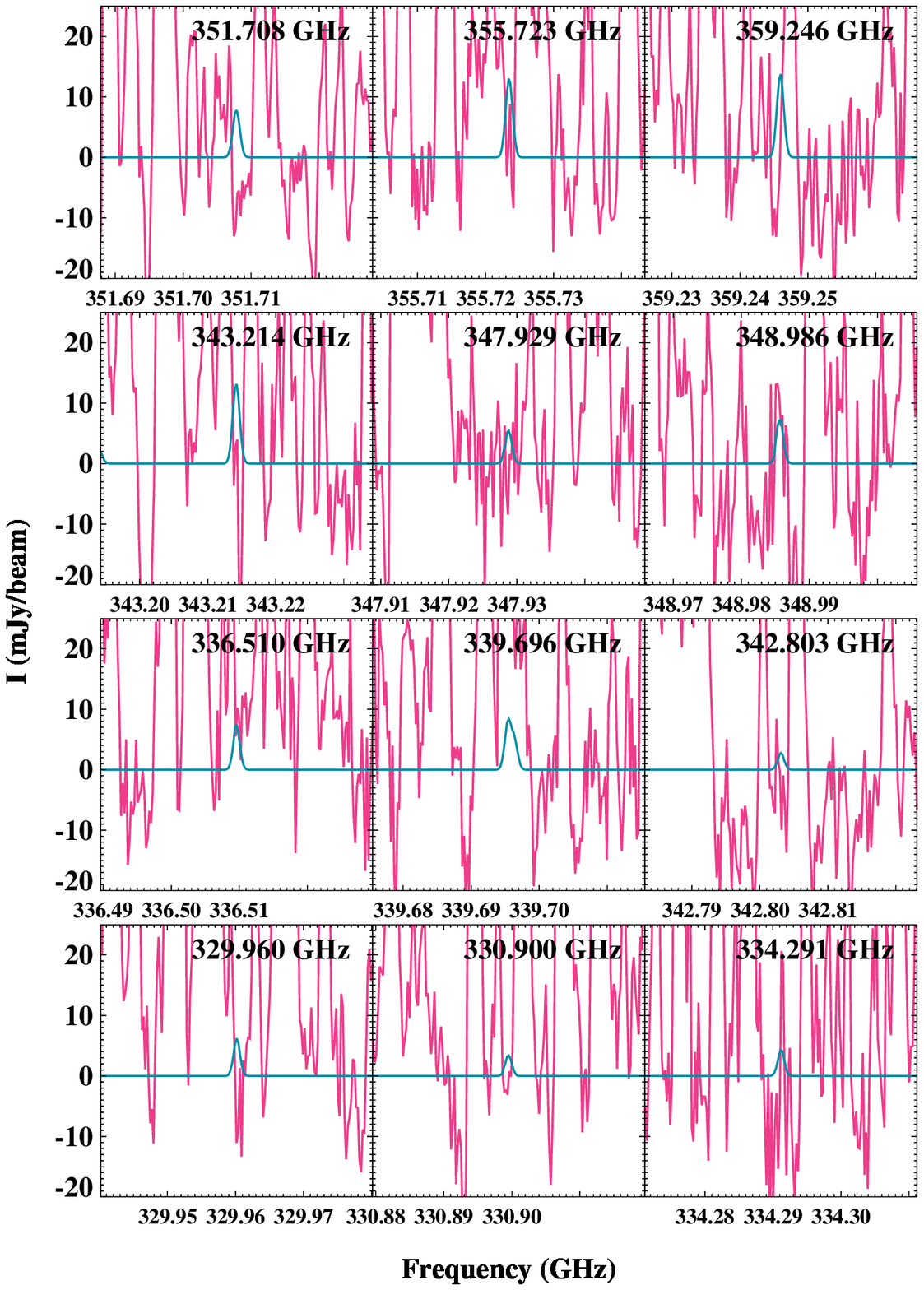}
 \caption{Twelve selected lines of c-C$_{2}$H$_{4}$S. Idem Fig.~\ref{fgr:SO2v=0}.}
 \label{fgr:c-C2H4S}
\end{figure*}

\begin{figure*}
 \centering
 \includegraphics[width=0.95\textwidth,height=0.2\textheight,keepaspectratio]{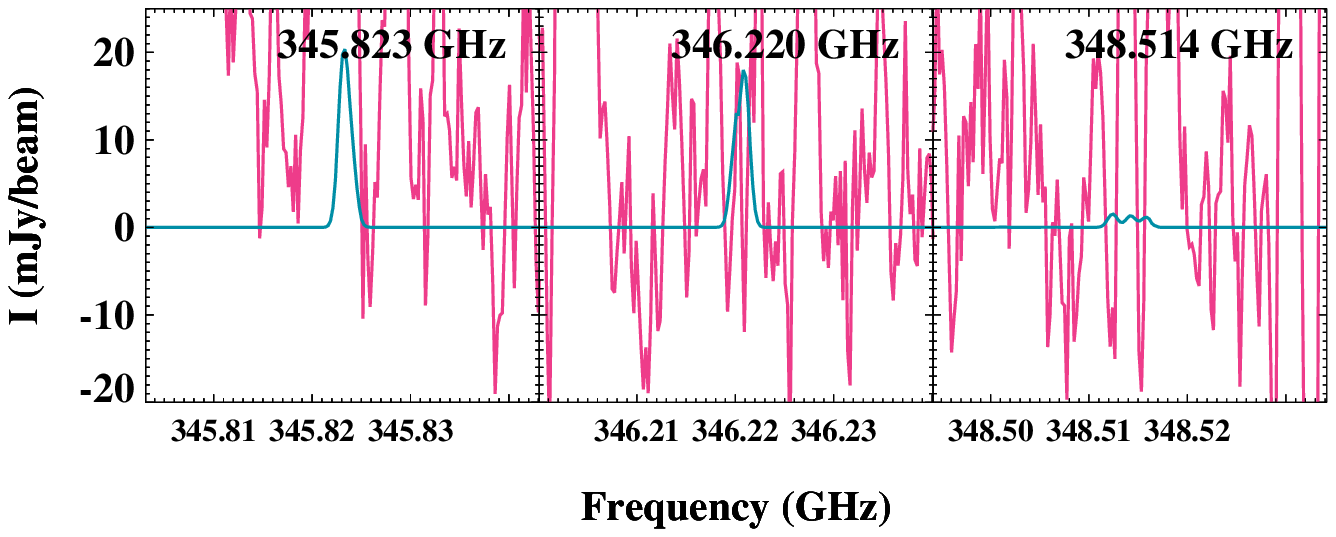}
 \caption{Three selected lines of NS in the $\varv=0$ state. Idem Fig.~\ref{fgr:SO2v=0}.}
 \label{fgr:NSv=0}
\end{figure*}

\begin{figure*}
 \centering
 \includegraphics[width=0.95\textwidth,height=0.2\textheight,keepaspectratio]{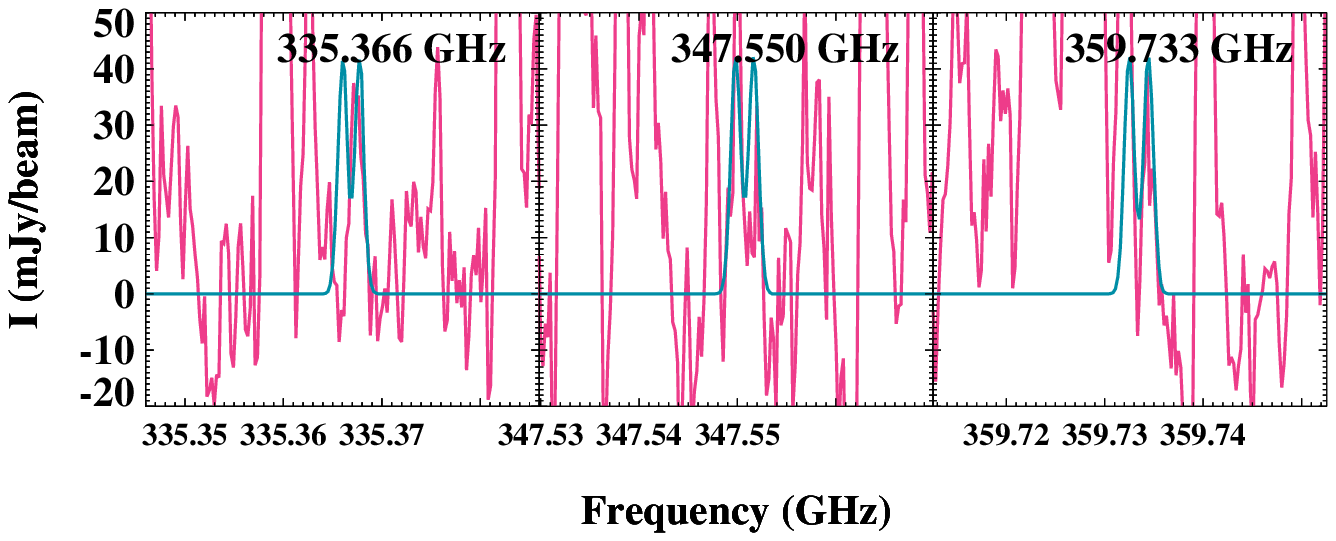}
 \caption{Three lines of NCS. Idem Fig.~\ref{fgr:SO2v=0}.}
 \label{fgr:NCS}
\end{figure*}

\begin{figure*}
 \centering
 \includegraphics[width=0.95\textwidth,height=0.4\textheight,keepaspectratio]{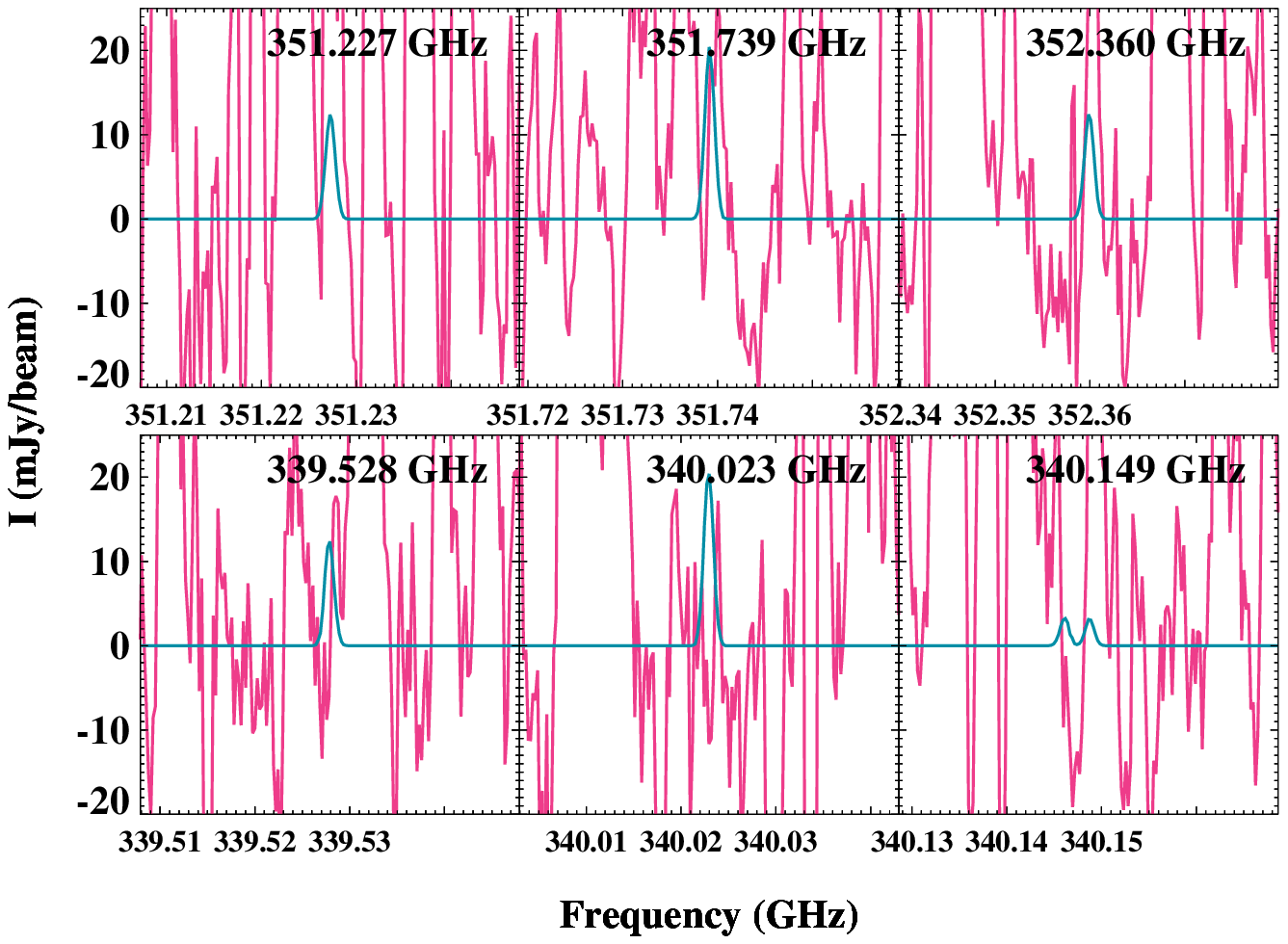}
 \caption{Nine lines of HNCS of the a-type. Idem Fig.~\ref{fgr:SO2v=0}.}
 \label{fgr:HNCSa-type}
\end{figure*}

\begin{figure*}
 \centering
 \includegraphics[width=0.95\textwidth,height=0.8\textheight,keepaspectratio]{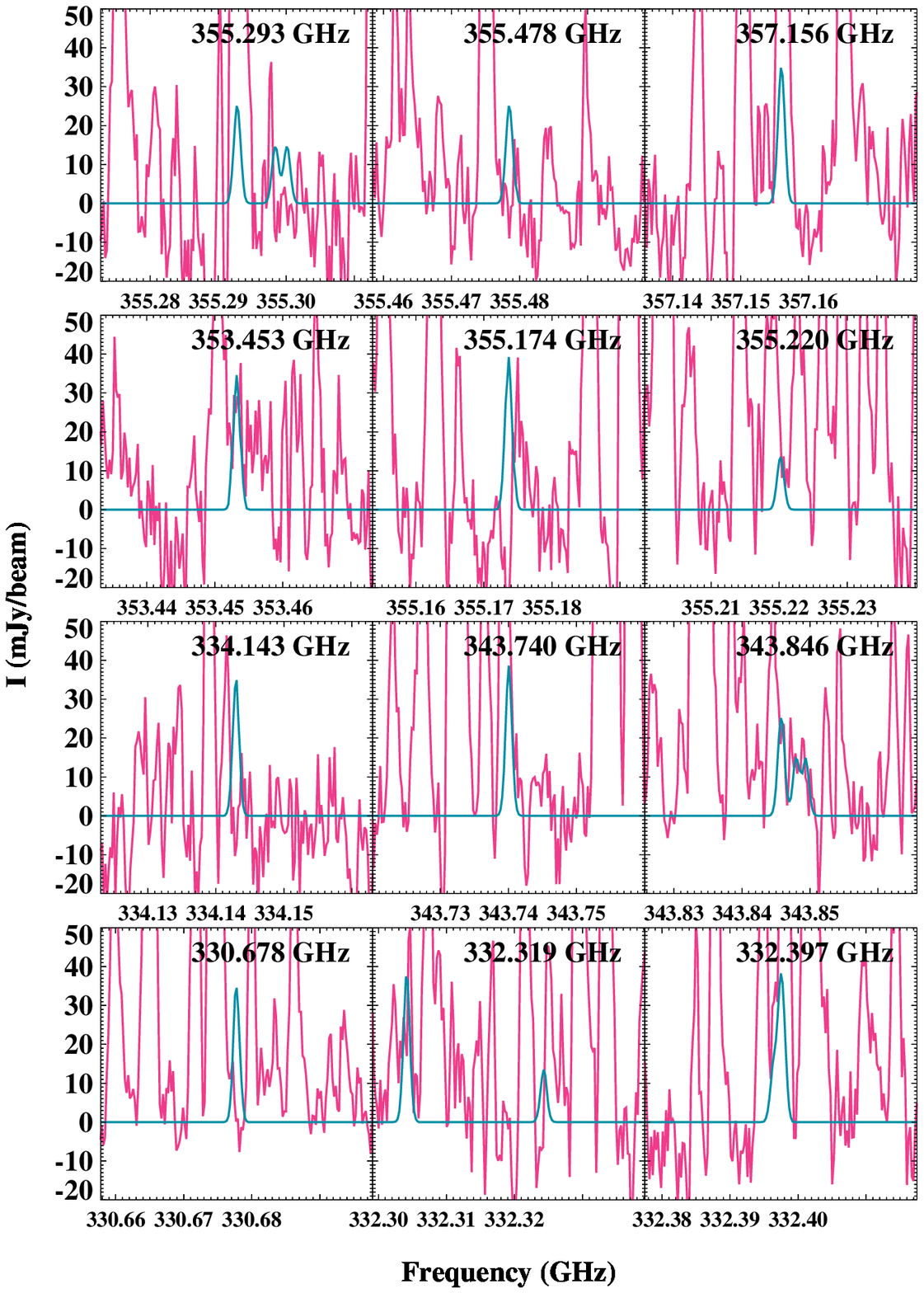}
 \caption{Twelve selected lines of HSCN. Idem Fig.~\ref{fgr:SO2v=0}.}
 \label{fgr:HSCN}
\end{figure*}

\clearpage
\newpage
\section{Spectroscopic laboratory information}
\label{labinfo}

The database sources, CDMS or JPL catalogues with their entry numbers, have been provided in the main text for all the species that have been searched for. In addition, the primary references on which these entries are based along with references for the dipole moments for all detected or possibly detected species are given below. Furthermore, the references with laboratory measurements in the range of the PILS survey in cases, in which the primary reference does not cover such data or if these data are an important contribution to the line list, are also included.

\subsection{SO$_{2}$}

The $\varv = 0$ and $\varv _2 = 1$ entries for the main isotopic species are based on \citet{MullerBrunken2005}, those for $^{34}$SO$_2$ -- on \citet{Belov1998}. The dipole moment of SO$_2$ in several vibrational states was determined by \citet{Patel1979}.

\subsection{SO}

The SO entry is based on \citet{Bogey1997}. Its dipole moment was measured by \citet{PowellLide1964}.

\subsection{OCS}

The main sources for the OCS entries are: \citet{Golubiatnikov2005} for OCS, $\varv = 0$; \citet{Morino2000} for OCS, $\varv _2 = 1$; \citet{Dubrulle1980} for O$^{13}$CS, OC$^{34}$S and OC$^{33}$S. Dipole moment values were determined for OCS in various vibrational states and for several isotopic species by \citet{Tanaka1985}.

\subsection{CS}

The CS entries are based on \citet{Muller2005}, and the main sources of laboratory data are \citet{Bogey1982} and \citet{AhrensWinnewisser1999}. The dipole moments 
of CS in $\varv = 0$ and 1 were measured by \citet{WinnewisserCook1968}.

\subsection{H$_{2}$CS}

The H$_2$CS data are largely from \citet{Maeda2008}, those of HDCS are from \citet{Minowa1997}. \citet{Fabricant1977} determined the dipole moment of H$_2$CS.

\subsection{H$_{2}$S}

The H$_2$S entry is based to a considerable extent on \citet{Belov1995}. The entries of HDS and HD$^{34}$S are based on \citet{Camy-Peyret1985}. The HDS transition frequencies with microwave accuracy were summarized by \citet{Helminger1971}. \citet{HillgerStrandberg1951} reported a small number of HD$^{34}$S and HDS transition frequencies. \citet{ViswanathanDyke1984} determined dipole moments of H$_2$S, HDS and D$_2$S.

\subsection{CH$_{3}$SH}

The CH$_{3}$SH entry is based on \citet{Xu2012} with transition frequencies in the range of our survey from \citet{Bettens1999}. The information on the dipole moment components was provided by \citet{Tsunekawa1989}.

\section{All the non-detected species}
\label{nondetspecies}

The one beam offset position of the PILS Band~$7$ dataset was searched for all the sulphur-bearing molecules available in the CDMS catalogue. All those detected are given in Table~\ref{tbl:bestfit}. All those not detected, but with a derived upper limit are given in Table~\ref{tbl:upperlim}. The following is a list of all other non-detected species at the $1\sigma$ level ($\sigma=10$~mJy beam$^{-1}$ channel$^{-1}$ or $5$~mJy~beam$^{-1}$~km~s$^{-1}$) and their corresponding CDMS entries in brackets: SO$_{2}$ $\varv_{2}=1$ (64503), $^{33}$SO$_{2}$ (65501), S$^{18}$OO (66502), S$^{17}$OO (65502), SO $\varv=1$ (48502), $^{34}$SO (50501), $^{33}$SO (49501), $^{36}$SO (52502), S$^{18}$O (50502), S$^{17}$O (49502), SO$^{+}$ (48010), $^{17}$OCS (61504), OC$^{36}$S (64510), $^{18}$OC$^{34}$S (64511), $^{18}$O$^{13}$CS (63503), O$^{13}$C$^{34}$S (63502), O$^{13}$C$^{33}$S (62507), CS $\varv=0-4$ (44501), CS$^{+}$ (44512), $^{13}$C$^{36}$S (49508), H$_{2}$C$^{34}$S (48508), H$_{2}$C$^{33}$S (47506), H$_{2}^{13}$CS (47505), H$_{2}$S (34502), D$_{2}$S (36503), D$_{2}^{34}$S (38507), cis-HOSO$^{+}$ (65510), Si$^{34}$S $\varv=0-2$ (62508), Si$^{33}$S (61508), Si$^{36}$S (64514), $^{29}$SiS $\varv=0-2$ (61506), $^{30}$SiS $\varv=0-2$ (62510), $^{29}$Si$^{34}$S $\varv=0,1$ (63504), $^{29}$Si$^{33}$S (62512), $^{29}$Si$^{36}$S (65507), $^{30}$Si$^{34}$S $\varv=0,1$ (64513), $^{30}$Si$^{33}$S (63505), $^{30}$Si$^{36}$S (66505), SiS $\varv=0-5$ (60506), HSiS (61512), H$_{2}$SiS (62513), OSiS (76517), HOCS$^{+}$ (61510), HSCO$^{+}$ (61509), t-HC(O)SH (62515), c-HC(O)SH (62516), SH$^{+}$ (33505), HSO (49512), NS $\varv=1$ (46516), N$^{34}$S (48509), N$^{33}$S (47509), N$^{36}$S (50516), $^{15}$NS (47510), DNCS a-type (60510), DNCS b-type (60511), HN$^{13}$CS a-type (60512),H$^{15}$NCS a-type (60513), HNC$^{34}$S a-type (61519), C$_{3}$S $\varv=0$ (68503), C$_{3}$S $\varv_{5}=1$ (68505), C$^{13}$CCS (69502), $^{13}$CCCS (69503), CC$^{13}$CS (69507), C$_{3}^{34}$S (70502), CC$^{13}$C$^{34}$S (71503), C$^{13}$CC$^{34}$S (71504), $^{13}$CCC$^{34}$S (71505), C$^{13}$C$^{13}$CS (70506), $^{13}$CC$^{13}$CS (70507), OC$_{3}$S (84502), H$_{2}$C$_{3}$S (70503), C$_{4}$S (80501), C$_{5}$S (92501), $^{13}$CC$_{4}$S (93501), C$^{13}$CC$_{3}$S (93502), C$_{2}^{13}$CC$_{2}$S (93503), C$_{3}^{13}$CCS (93504), C$_{4}^{13}$CS (93505), C$_{5}^{34}$S (94502), CaS $\varv=0,1$ (72501), ScS (77509), $^{46}$TiS (78503), TiS (80505), $^{50}$TiS (82503), YS (121502).

The following species have not been detected, because they do not have lines in the frequency range surveyed with PILS Band~$7$ data:  CS $\varv=1-0, 2-1$ (44510), CS $\varv=2-0$ (44511), C$^{34}$S $\varv=1-0$ (46510); $^{13}$CS $\varv=0,1$ (45501); $^{13}$CS $\varv=1-0$ (45509); $^{13}$C$^{34}$S (47501), $^{13}$C$^{33}$S (47501), H$_{2}^{34}$S (36504), H$_{2}^{33}$S (35503), Si$^{34}$S $\varv=1-0$ (62509), $^{29}$SiS $\varv=1-0$ (61507), $^{30}$SiS $\varv=1-0$ (62511), SiS $\varv=1-0,2-1$ (60507), SiS $\varv=2-0$ (60508), S (32511), SH$^{-}$ (33504), NS $\varv=1-0$ (46517), $^{15}$N$^{34}$S (49511), HS $\varv=0$ (33508), HS $\varv=1$ (33509), H$^{13}$CS$^{+}$ (46504), HSCH$_{2}$CN (73503), HCNS (59510), HSNC (59511), HS$^{13}$CN (60514), DSCN (60515), HSC$^{15}$N (60516), H$^{34}$SCN (61520), C$_{7}$S (116501).

Finally, the following is a list of all other non-detected species at the $1\sigma$ level ($\sigma=10$~mJy beam$^{-1}$ channel$^{-1}$ or $5$~mJy~beam$^{-1}$~km~s$^{-1}$) that are available only in the JPL catalogue and their corresponding JPL entries in brackets: MgS (56009), $^{13}$CCS (57001), C$^{13}$CS (57002), CC$^{34}$S (58001), PS (63007), SO$_{2}$ $\varv=2$ (64005), H$_{2}$SO$_{4}$ (98001). SD (34005) does not have lines in the frequency range surveyed with PILS Band~$7$ data.

\section{Comparison with a model of hot cores}
\label{app:compmodels}

Efforts on chemical modelling of sulphur networks have recently been revived by \citet{Woods2015}. The authors computed the abundances of sulphur-bearing species for hot core conditions upon the inclusion of recent experimental and theoretical data into their chemical network, including a refractory sulphur residue. The modelled molecular ratios (relative to either H$_{2}$S or OCS; as tabulated in table~$9$ of \citet{Woods2015} for the standard model with the interstellar cosmic ray ionization rate of $1.3\times10^{-17}$~s$^{-1}$) can be compared to those derived in this work for IRAS~16293--2422~B. The exact values are given in Table~\ref{tbl:compmodels}. An agreement within one order of magnitude is found for OCS and CS relative to H$_{2}$S, and for H$_{2}$S relative to OCS. The modelled ratios for H$_{2}$CS are three orders of magnitude higher, which may be explained by the fact that the grain-surface network used in the models is fairly small and the grain-surface chemistry is not accounted for in full. H$_{2}$CS is expected to be involved in many grain-surface reactions and will likely be used up for synthesis of larger sulphur-bearing species, such as CH$_{3}$SH (analogous to the sequential hydrogenation of CO leading to CH$_{3}$OH with H$_{2}$CO as an intermediate).

For SO and SO$_{2}$, the differences are many orders of magnitude between the observed and modelled values of \citet{Woods2015}. This likely has to do with the fact that the models have been run for a very long time of $10^{7}$~yr, which implies that many gaseous species are driven into SO and SO$_{2}$, leading to their overproduction. At earlier times (as seen in fig.~$6$ of \citealt{Woods2015}), the model results have a closer agreement for SO with observations. SO$_{2}$ remains too high nevertheless, which may have to do with thermal desorption being calculated via an efficiency factor. It seems that the modelling approach of \citet{Woods2015} is insufficient at reproducing the sulphur chemistry towards IRAS~16293--2422~B. The differences with cometary ROSINA measurements are just as large. The development of a full gas-grain chemical network for sulphur, including grain-surface chemistry, is the topic of future research. For the case of dark prestellar core, \citet{Vidal2017} have been able to successfully reproduce observed abundances of sulphur-bearing species without the need for artificial sulphur depletion.

A comparison with observations of hot cores, such as Sgr~B2(N), Sgr B2(M) (e.g., \citealt{Belloche2013}) and Orion KL (e.g., \citealt{Esplugues2014}) is beyond the scope of this paper due to the mandatory discussion on emitting regions and structures sampled with observations. Future high spatial resolution ALMA observations will mediate this problem.

\ctable[
 caption = {Molecular ratios relative to H$_{2}$S and OCS as measured with these interferometric ALMA observations at the one beam offset position from source B of IRAS~16293--2422$^{\text{q}}$ in comparison to hot core models$^{\text{r}}$},
 label = {tbl:compmodels},
 star = 1,
 ]{@{\extracolsep{\fill}}lrrrr}{
\tnote[q]{values taken from Table~\ref{tbl:SDcomp}}
\tnote[r]{table~$9$ of \citet{Woods2015} for the standard model with the interstellar cosmic ray ionization rate of $1.3\times10^{-17}$~s$^{-1}$}
 }{
 \hline
 \multirow{2}{*}{Species} & \multicolumn{2}{c}{Molecular ratios relative to H$_{2}$S (\%)} & \multicolumn{2}{c}{Molecular ratios relative to OCS (\%)}\T\\
 & ALMA B & model & ALMA B & model\B\\
 \hline
 H$_{2}$S  & $100$      & $100$             & $68-679$ & $29$\T\\
 OCS       & $147-15$   & $341$             & $100$    & $100$\\
 SO        & $0.3-0.03$ & $58$              & $0.2$    & $17$\\
 SO$_{2}$  & $0.8-0.08$ & $1.5\times10^{4}$ & $0.5$    & $4.5\times10^{3}$\\
 CS        & $2-0.2$    & $1.2$             & $2$      & $0.3$\\
 H$_{2}$CS & $0.8$      & $129$             & $0.5$    & $38$\B\\
\hline}

\bsp 

\label{lastpage}

\end{document}